\documentclass[notitlepage,superscriptaddress,nofootinbib,amsmath,amssymb,aps,11pt]{revtex4-1}
\usepackage[]{inputenc}
\usepackage[dvips]{graphicx}
\usepackage{color}
\usepackage{bm}
\usepackage{mathpazo}
\usepackage[ps2pdf,bookmarks=true,colorlinks,linkcolor=blue,urlcolor=green,citecolor=red]{hyperref}
\renewcommand\a{\alpha}
\renewcommand\b{\beta}
\renewcommand\d{\delta}
\renewcommand\k{\kappa}
\renewcommand\l{\lambda}
\renewcommand\r{\rho}

\renewcommand\t{\tau}

\renewcommand\c{\chi}
\renewcommand\j{\psi}
\renewcommand\o{\omega}
\newcommand\e{\epsilon}
\newcommand\g{\gamma}
\newcommand\z{\zeta}
\newcommand\m{\mu}
\newcommand\n{\nu}

\newcommand\p{\pi}
\newcommand\h{\theta}
\newcommand\s{\sigma}
\newcommand\f{\phi}
\newcommand\w{\eta}
\newcommand\ve{\varepsilon}

\renewcommand\L{\Lambda}

\renewcommand\S{\Sigma}
\renewcommand\O{\Omega}

\newcommand\D{\Delta}
\newcommand\G{\Gamma}

\newcommand\J{\Psi}

\newcommand{\fig}[1]{Fig.~\ref{#1}}
\newcommand{\eq}[1]{Eq.~(\ref{#1})}
\newcommand{\sect}[1]{Sec.~\ref{#1}}
\newcommand{\eqs}[2]{Eqs.~(\ref{#1})-(\ref{#2})}

\newcommand\lb{\left(}
\newcommand\rb{\right)}
\newcommand\ls{\left[}
\newcommand\rs{\right]}
\newcommand\lc{\left\{}
\newcommand\rc{\right\}}
\newcommand{\lan}{\langle}
\newcommand{\ran}{\rangle}

\newcommand\ra{\rightarrow}

\newcommand{\non}{\nonumber\\}
\newcommand\pt{\partial}

\newcommand{\ie}{\emph{i.e.}}

\newcommand{\etal}{\emph{et al.}}
\newcommand{\cl}{{\cal L}}

\newcommand{\re}{{\rm{Re}}}
\newcommand{\im}{{\rm{Im}}}
\newcommand{\bx}{{\vec x}}
\newcommand{\br}{{\vec r}}
\newcommand{\bp}{{\vec p}}
\newcommand{\bk}{{\vec k}}

\newcommand{\bv}{{\vec v}}

\newcommand{\bB}{{\vec B}}
\newcommand{\bE}{{\vec E}}
\newcommand{\bJ}{{\vec J}}

\newcommand{\jb}{{\bar \j}}

\newcommand{\rp}{{\rm RP}}

\renewcommand{\part}{{\rm part}}

\renewcommand{\vec}{\boldsymbol}
\newcommand{\be}{\begin{equation}}
\newcommand{\ee}{\end{equation}}
\newcommand{\bear}{\begin{eqnarray}}
\newcommand{\eear}{\end{eqnarray}}
\newcommand{\ba}{\begin{array}}
\newcommand{\ea}{\end{array}}

\begin{document}
\title{Electromagnetic fields and anomalous transports in heavy-ion collisions\\ --- A pedagogical review}
\author{\normalsize{Xu-Guang Huang}}
\email{huangxuguang@fudan.edu.cn}
\affiliation{Physics Department and Center for Particle Physics and Field Theory, Fudan University, Shanghai 200433, China.}

\date{\today}

\begin{abstract}
The hot and dense matter generated in heavy-ion collisions may contain domains which are not invariant under P and CP transformations. Moreover, heavy-ion collisions can generate extremely strong magnetic fields as well as electric fields. The interplay between the electromagnetic field and triangle anomaly leads to a number of macroscopic quantum phenomena in these P- and CP-odd domains known as the anomalous transports. The purpose of the article is to give a pedagogical review of various properties of the electromagnetic fields, the anomalous transports phenomena, and their experimental signatures in heavy-ion collisions.
\end{abstract}
\maketitle
\tableofcontents

\section {Introduction}\label{sec:intro}
As is well known, the strong interaction provides the mechanism that binds the quarks and gluons together to form the hadrons such as the proton and neutron. Our contemporary understanding of strong interaction is described by quantum chromodynamics (QCD) --- a quantum gauge field theory based on color $SU(3)$ gauge symmetry. Despite its simple form, QCD possesses a number of remarkable properties among which the most mysterious one may be the color confinement: in vacuum, the quarks and gluons, as being colorful particles, are always confined in colorless hadrons. The color confinement property forbids us to observe isolated quark or gluon. However, at high temperature and/or high quark chemical potentials, the normal hadronic matter is expected to transform to deconfined quark-gluon matter. When the temperature is high such quark-gluon matter is usually called the quark-gluon plasma (QGP). Lattice QCD simulations which are the first-principle computations that solve QCD directly show that the ``transition temperature" from hadronic matter to QGP is about $200$ MeV (at zero quark chemical potentials). It is believed that such a high temperature was once realized in the universe at a few microsecond after the Big Bang and the QGP was created at that time. In laboratory, up to now, the only method to achieve such a high temperature is to use the high-energy heavy-ion collisions. Such collisions have been carried out in the Relativistic Heavy Ion Collider (RHIC) at Brookhaven National Laboratory (BNL) and in the Large Hadron Collider (LHC) at the European Organization for Nuclear Research (CERN). The top center-of-mass energy per nucleon pair at RHIC Au + Au collisions is $\sqrt{s}=200$ GeV and at LHC Pb + Pb collisions is $\sqrt{s}=2.76$ TeV and will be upgraded to $\sqrt{s}=5.5$ TeV soon. In these colliders, two nuclei are accelerated to velocity very close to the speed of light and then collide, the energy deposition in the reaction region can be large enough to create the hot and dense environment in which the deconfinement condition is reached. Measurements performed at RHIC and LHC have collected many signals supporting the generation of the QGP and have also revealed a variety of unusual properties of the QGP, e.g., its very small shear viscosity comparing to the entropy density and its high opacity for energetic jets.

Heavy-ion collisions can generate electromagnetic (EM) fields as well~\cite{Rafelski:1975rf}. Recent numerical simulations found that the magnitude of the magnetic field in RHIC Au + Au collisions at $\sqrt{s}=200$ GeV can be at the order of $10^{18}-10^{19}$ Gauss~\footnote{In the relevant literature, people usually use $m_\p^2$ or MeV$^2$ as the unit of $e\bB$ where $e$ is the electron charge magnitude and $m_\p\approx 140$ MeV is the pion mass. In converting to the SI or Gaussian units, it is helpful to note the following relation: $1$ MeV$^2=e\cdot1.6904\times10^{14}$ Gauss (if $\hbar=c=1$, otherwise the right-hand side should be multiplied by $\hbar c^2$).} and in LHC Pb + Pb collisions at $\sqrt{s}=2.76$ TeV can reach the order of $10^{20}$ Gauss~\cite{Kharzeev:2007jp,Skokov:2009qp,Voronyuk:2011jd,Bzdak:2011yy,Ou:2011fm,Deng:2012pc,Bloczynski:2012en,Bloczynski:2013mca,Zhong:2014cda,Zhong:2014sua}. The electric filed can also be generated owing to event-by-event fluctuations~\cite{Bzdak:2011yy,Deng:2012pc,Bloczynski:2012en,Bloczynski:2013mca} (we will explain the physical meaning of such fluctuations in \sect{sec:fields}) or in asymmetric collisions like Cu + Au collision~\cite{Hirono:2012rt,Deng:2014uja,Voronyuk:2014rna}, and its strength is roughly of the same order as the magnetic field. Thus heavy-ion collisions provide a unique terrestrial environment to study QCD matter in strong EM fields. In particular, recently it was proposed that the magnetic field can convert topological fluctuations in the QCD vacuum into global electric charge separation along the direction of the magnetic field. The underlying mechanism is the so-called chiral magnetic effect (CME)~\cite{Kharzeev:2007jp,Fukushima:2008xe}. Some relatives of the CME were also proposed, including the chiral separation effect (CSE)~\cite{Son:2004tq,Metlitski:2005pr}, chiral electric separation effect (CESE)~\cite{Huang:2013iia}, chiral magnetic waves (CMW)~\cite{Kharzeev:2010gd}, chiral vortical effect (CVE)~\cite{Erdmenger:2008rm,Banerjee:2008th,Son:2009tf}, chiral vortical wave~\cite{Jiang:2015cva}, chiral heat wave~\cite{Chernodub:2015gxa}, chiral Alfven wave~\cite{Yamamoto:2015ria}, etc. They all represent special transport phenomena that are closely related to chiral anomaly~\footnote{We will not distinguish the terms ``chiral anomaly", ``axial anomaly", and ``triangle anomaly" in this article.} and thus are called anomalous transports. The experimental searches of the anomalous transports have been carried out at RHIC and LHC and the measurements indeed offered signals consistent with the predictions of the CME, CMW, and CVE; see the discussions in \sect{sec:exper}.

The purpose of this paper is to give a pedagogical review of recent progresses on the study of EM fields and the anomalous transport phenomena induced by EM fields in heavy-ion collisions. We will keep all the discussions as intuitive as possible and lead the readers who wish to understand more technical details to proper literature. To access this paper, the readers do not need to have expert knowledge of QCD; only elementary knowledge of quantum field theory and heavy-ion collisions are needed. (Perhaps the only exception is \sect{sec:cme} (3) where some knowledge of topology and gauge field theory is needed; we thus give more thorough discussion and put necessary references there so that the readers can easily trace the relevant literature.) Thus this paper will be particularly useful for graduate students who have finished their first-year courses and wish to enter the exciting research area of anomalous transport phenomena. In this aspect, the present review is complementary to existing excellent reviews, e.g., Refs.~\cite{Tuchin:2013ie,Kharzeev:2009fn,Kharzeev:2013jha,Kharzeev:2013ffa,Kharzeev:2015kna,Kharzeev:2015znc} in which more advanced materials can be found.

We organize the paper as follows. In \sect{sec:fields}, we shall discuss some general properties of the EM fields in heavy-ion collisions. In \sect{sec:anoma}, we shall give an elementary introduction to the anomalous transports in parity-odd (P-odd) and/or charge-conjugation-odd (C-odd) medium. The experimental implications of the anomalous transports and the current status of their detection in heavy-ion collisions will be reviewed in \sect{sec:exper}. Some discussions will be presented in \sect{sec:discu}.

In addition to the anomalous transports, the EM fields can drive a range of other intriguing phenomena including, for example, the magnetic catalysis of chiral symmetry breaking~\cite{Gusynin:1994re,Gusynin:1995nb,Shovkovy:2012zn}, the inverse magnetic catalysis or magnetic inhibition at finite temperature and density~\cite{Preis:2010cq,Bali:2011qj,Bali:2012zg,Bruckmann:2013oba,Fukushima:2012xw,Fukushima:2012kc,Kojo:2012js,Chao:2013qpa,Yu:2014sla,Feng:2014bpa,Yu:2014xoa,Cao:2014uva,Ferrer:2014qka,Mueller:2015fka,Chen:2015hfc}, the possible $\r$ meson condensation in strong magnetic field~\cite{Chernodub:2010qx,Chernodub:2011mc,Hidaka:2012mz,Liu:2014uwa,Liu:2015pna}, the neutral pion condensation in vaccum~\cite{Cao:2015cka}, the anisotropic viscosities in hydrodynamic equations~\cite{Braginskii:1965,Lifshitz:1981,Huang:2009ue,Huang:2011dc,Tuchin:2011jw}, and the early-stage phenomena in heavy-ion collisions like the EM-field induced particle production~\cite{Tuchin:2010vs,Tuchin:2010gx,Tuchin:2012mf,Tuchin:2013ie,Tuchin:2014nda,Tuchin:2014pka,Basar:2012bp} and the dissociation of heavy-flavor mesons~\cite{Marasinghe:2011bt,Machado:2013rta,Alford:2013jva,Liu:2014ixa,Guo:2015nsa}. These topics will not be the main focus of this article. Some of them are nicely reviewed in Refs.~\cite{Kharzeev:2013jha,Andersen:2014xxa,Miransky:2015ava}.

\section {Properties of electromagnetic fields in heavy-ion collisions}\label{sec:fields}
\begin{figure}
\begin{center}
\includegraphics[width=6.5cm]{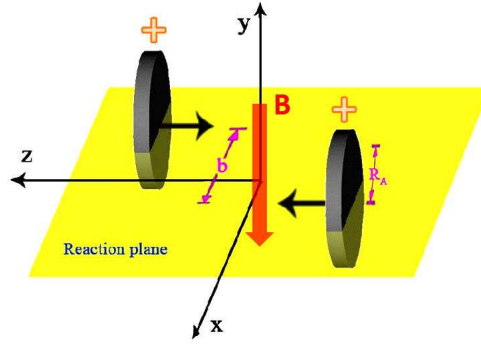}
\caption{The geometrical illustration of the noncentral heavy-ion collision. Here $b$ is the impact parameter and $R_A$ is the radius of the nucleus. The magnetic field $\bB$ is expected to be perpendicular to the reaction plane due to the left-right symmetry of the collision geometry.}
\label{coll_geo}
\end{center}
\vspace{-0.7cm}
\end{figure}

The reason why heavy-ion collisions can generate magnetic fields is simple: nuclei are positively charged and when they move they generate electric currents which in turn induce the magnetic fields. In a noncentral heavy-ion collision, two counter-propagating nuclei collide at a finite impact parameter $b$; one can easily imagine that the magnetic field at the center of the overlapping region will be perpendicular to the reaction plane owing to the left-right symmetry of the collision geometry (see \fig{coll_geo} for illustration). However, in a real collision event, this left-right symmetry may be lost because the nucleon distribution of one nucleus would not be identical to another. We will come to this point later, but first let us estimate how strong the magnetic field can be.

Let us consider Au + Au collisions at fixed impact parameter $b=10$ fm and at RHIC energy $\sqrt{s}=200$ GeV as an example. If we approximate the problem by assuming that all the protons are located at the center of the nucleus, then by naively applying the Biot-Savart law we obtain
\begin{eqnarray}
\label{estimate}
-eB_y&\sim& 2Z_{\rm Au}\g\frac{e^2}{4\p} v_z\lb\frac{2}{b}\rb^2\approx 10 m_\p^2\approx 10^{19}\; {\rm Gauss},
\end{eqnarray}
where $v_z=\sqrt{1-(2m_N/\sqrt{s})^2}\approx 0.99995$ ($m_N$ is the nucleon mass) is the velocity of the nucleus, $\g=1/\sqrt{1-v_z^2}\approx100$ is the Lorentz gamma factor, and $Z_{\rm Au}=79$ is the charge number of gold nucleus. The minus sign on the left-hand side is because the magnetic field in pointing to the $-\hat{\bf y}$ direction in the setup shown in \fig{coll_geo}.

This is really a huge magnetic field. It is much larger than the masses squared of electron, $m_e^2$, and light quarks, $m_u^2, m_d^2$, and thus is capable of inducing significant quantum effects. It is also larger than the magnetic fields of neutron stars including the magnetars which may have surface magnetic fields of the order of $10^{14}-10^{15}$ Gauss~\cite{Olausen:2013bpa,Turolla:2015mwa}. Therefore the magnetic fields generated in high-energy heavy-ion collisions are among the strongest ones that we have ever known in current universe. (In the early universe, there was a possibility to generate an even stronger magnetic field through the electro-weak transition, see Ref.~\cite{Grasso:2000wj} for review.) One can expect that such a huge magnetic field may have important consequences on the dynamics of the quark-gluon matter produced in heavy-ion collisions. We will discuss several such consequences, namely, the anomalous transport phenomena in Sec.~\ref{sec:anoma}. In this section we will focus on the fields themselves.

\subsection {Computations of the electromagnetic fields}\label{sec:setup}
The estimation given in \eq{estimate} is too simplified, in order to have a more reliable simulation for the electromagnetic (EM) fields in heavy-ion collisions, the following issues need to be taken into account. (1) We need the knowledge of the proton and neutron distributions in a given nucleus. For this purpose, we can choose the well established Woods-Saxon distribution to use. (2) In real heavy-ion collisions, because the proton distribution varies from one nucleus to another, the generated EM fields vary from event to event. It is thus important to study the event-by-event fluctuation of the EM fields~\cite{Bzdak:2011yy,Deng:2012pc,Bloczynski:2012en,Bloczynski:2013mca}. (3) We need to replace the Biot-Savart law by the full relativistic Li\'enard-Wiechert potentials which contain the retardation effect,
\begin{eqnarray}
\label{LWE}
e\bE(t,\br)&=&\frac{e^2}{4\p}\sum_n Z_n\frac{{\vec R}_n-R_n\bv_n}{(R_n-{\vec R}_n\cdot\bv_n)^3}(1-v_n^2),\\
\label{LWB}
e\bB(t,\br)&=&\frac{e^2}{4\p}\sum_n Z_n\frac{\bv_n\times{\vec R}_n}{(R_n-{\vec R}_n\cdot\bv_n)^3}(1-v_n^2),
\end{eqnarray}
where the summation is over all the charged particles, $Z_n$ is the charge number of the $n$th particle, ${\vec R}_n=\br-\br_n$ is the relative
position of the field point $\br$ to the source point $\br_n$ of the $n$th particle, $\bv_n$ is the velocity of $n$th particle at the retarded time
$t_n=t-|\br-\br_n|$. Note that \eqs{LWE}{LWB} have singularities at $R_n=0$; in practical
calculations a variety of regularization schemes have been used and consistent results are obtained after taking the event average~\cite{Skokov:2009qp,Voronyuk:2011jd,Bzdak:2011yy,Deng:2012pc,Bloczynski:2012en,Bloczynski:2013mca,Deng:2014uja}.

As the EM fields in heavy-ion collisions can be much larger than the electron and light quark masses squared, one may worry about the possible quantum electrodynamics (QED) correction to the otherwise classical Maxwell field equations (the Li\'enard-Wiechert potentials are the solutions of the Maxwell equations). So let us make a magnitude estimate of such QED correction by using the one-loop Euler-Heisenberg effective lagrangian for soft photons (see Ref.~\cite{Dunne:2004nc} for review):
\begin{eqnarray}
\label{euler}
\cl_{\rm EH}=-A_\m J^\m-\frac{1}{4}F^{\m\n}F_{\m\n}-\frac{e^2}{32\p^2}\int_0^\infty\frac{ds}{s}e^{-sm_e^2}\ls\frac{\re \cosh(esX)}{\im \cosh(esX)}F_{\m\n}\tilde{F}^{\m\n}-\frac{4}{e^2s^2}-\frac{2}{3}F_{\m\n}F^{\m\n}\rs,
\end{eqnarray}
where $J^\m$ is the electric current, $A^\m$ is the EM potential, $F^{\m\n}$ is the strength tensor, $\tilde{F}^{\m\n}=\frac{1}{2}\e^{\m\n\a\b}F_{\a\b}$, and $X=\sqrt{\frac{1}{2}F_{\m\n}F^{\m\n}+\frac{i}{2}F_{\m\n}\tilde{F}^{\m\n}}$. At strong-field limit, the asymptotic form of \eq{euler} behaves like~\cite{Dunne:2004nc}
\begin{eqnarray}
\cl_{\rm EH}\sim -A_\m J^\m-\frac{1}{4}\ls1-\frac{e^2}{24\p^2}\ln\frac{e^2|F^2|}{m_e^4}\rs F^{\m\n}F_{\m\n},
\end{eqnarray}
where $F^2=F^{\a\b}F_{\a\b}$.
The field equations derived from this lagrangian can be regarded as the Maxwell equations but with a renormalized charge (keeping only the leading-log term because $|eF|\gg m_e^2$)
\begin{eqnarray}
e\rightarrow \tilde{e}\approx e\ls1-\frac{e^2}{24\p^2}\ln\frac{e^2|F^2|}{m_e^4}\rs^{-1}
\end{eqnarray}
at leading-log order.
Thus, we can find that even for very strong EM field, e.g., $|eF|\sim 100 m_\p^2$, the quantum correction can only amend the final restuls by a few percent. This justifies the applicability of \eqs{LWE}{LWB}.

In the following subsections we will review the recent results of the EM fields in heavy-ion collisions obtained by using the Li\'enard-Wiechert potentials on event-by-event basis. We will mainly focus on Au + Au collisions at RHIC and Pb + Pb collisions at LHC; other collision systems will be briefly discussed in Sec.~\ref{sec:other}.

\subsection {Impact parameter dependence}\label{sec:impac}
\begin{figure}[!htb]
\begin{center}
\includegraphics[width=7.0cm]{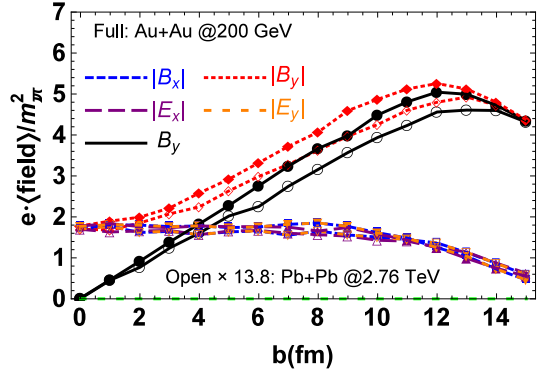}
\caption{The EM fields at $t=0$ and $\br={\bf 0}$
as functions of the impact parameter $b$. Figures are modified from Ref.~\cite{Deng:2012pc}.}
\label{bdepe}
\end{center}
\end{figure}

We first show in \fig{bdepe} the impact parameter dependence of the EM fields at $\br={\bf 0}$ and $t=0$ where the initial time $t=0$ is set to be the time when the two colliding nuclei completely overlap.
The curves with full dots are for Au + Au collision at RHIC energy $\sqrt{s}=200$ GeV and the curves with open dots are for fields scaled by a factor $13.8=\sqrt{s_{\rm LHC}}/\sqrt{s_{\rm RHIC}}$ for Pb + Pb collision at LHC energy $\sqrt{s}=2.76$ TeV. In these figures (and also in the figures hereafter) $\lan \cdots\ran$ represents the average over events.
One can find that:\\ (1) The event averaged EM fields have only one nonzero component, $\lan B_y\ran\neq0$; all other components vanish: $\lan B_{x}\ran=\lan B_{z}\ran=\lan \bE\ran=0$. However, owing to the fluctuations of the positions of protons in the nuclei, their magnitudes in each event can be large (except for the $z$-components which are always small). This is reflected in the averaged absolute values of the fields and is most evident for central collisions~\cite{Bzdak:2011yy,Deng:2012pc}.\\ (2) When $b< 2 R_A$ with $R_A$ the nucleus radius, the event-averaged field $e\lan B_y\ran$ is proportional to $b$ and it reaches its maximum value around $2R_A$. The fluctuation-induced fields are not sensitive to $b$ when $b<2R_A$.

\subsection {Collision energy dependence}\label{sec:energ}
As investigated in Ref.~\cite{Bzdak:2011yy,Deng:2012pc}, to high precision, the magnitudes of EM fields linearly depend on the collision energy $\sqrt{s}$.
Actually the absolute values of the EM fields satisfy very well the following scaling law, $e|{\rm Field}|\propto\sqrt{s}f(b/R_A)$ where $f(b/R_A)$ is a universal function which has the shapes for $\lan e|B_{x,y}|\ran$ and
$\lan e|E_{x,y}|\ran$ as shown in \fig{bdepe}. For the event-averaged magnetic field, $e\lan B_y\ran$, the following formula approximately expresses its impact parameter $b$, collision energy $\sqrt{s}$, charge number $Z$ and atomic number $A$ dependence:
\begin{eqnarray}
e\lan B_y\ran\propto\frac{\sqrt{s}}{2m_N}\frac{Z}{A^{2/3}}\frac{b}{2R_A} m_\p^2,\;\;\;\; {\rm for\;\;} b<2R_A.
\end{eqnarray}
Note that the prefactor $\sqrt{s}/(2m_N)$ is nothing but the Lorentz gamma factor.

\subsection {Spatial distributions}\label{sec:spat}
\begin{figure*}[!htb]
\begin{center}
\includegraphics[width=7.5cm]{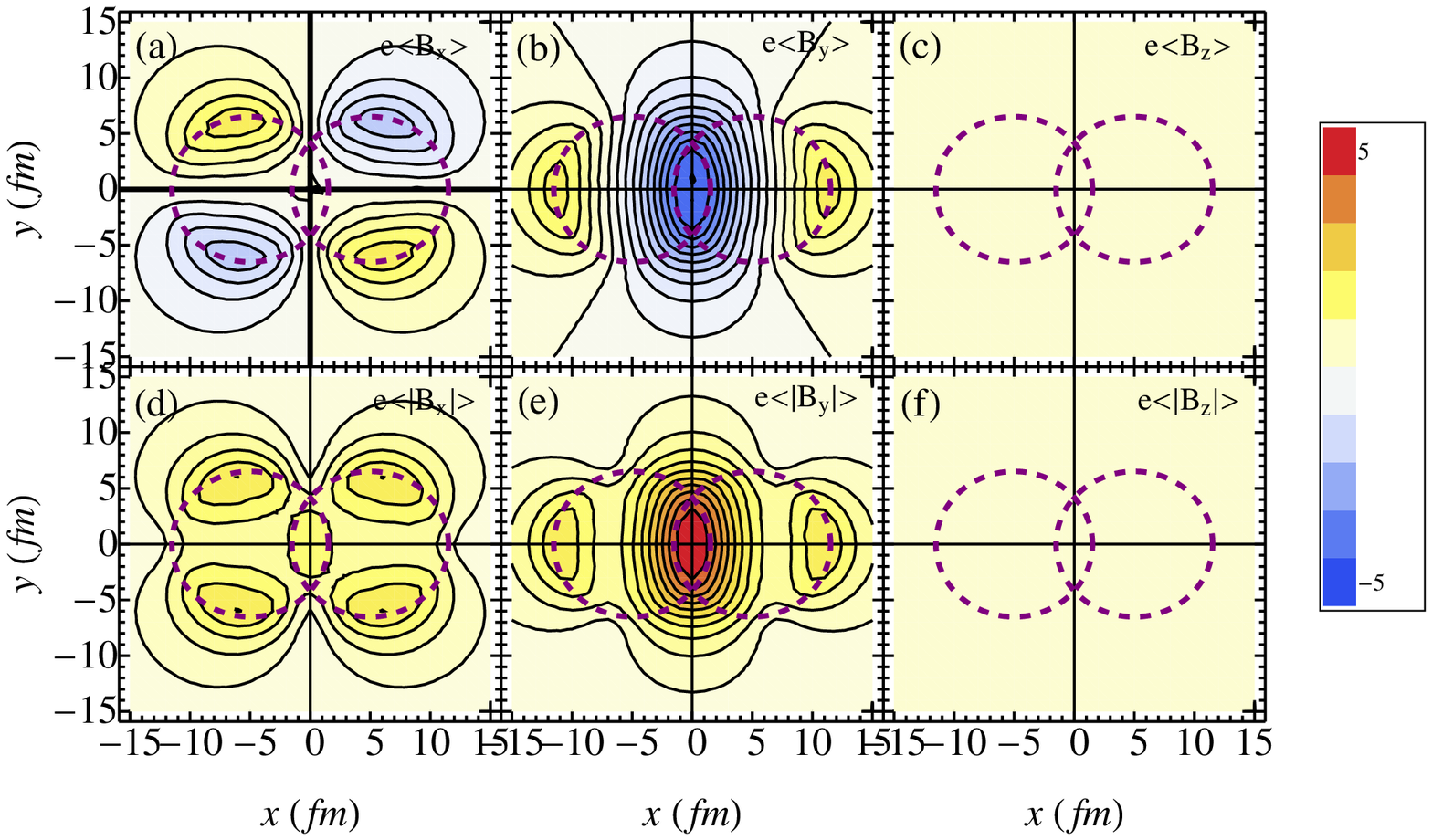}
\includegraphics[width=7.5cm]{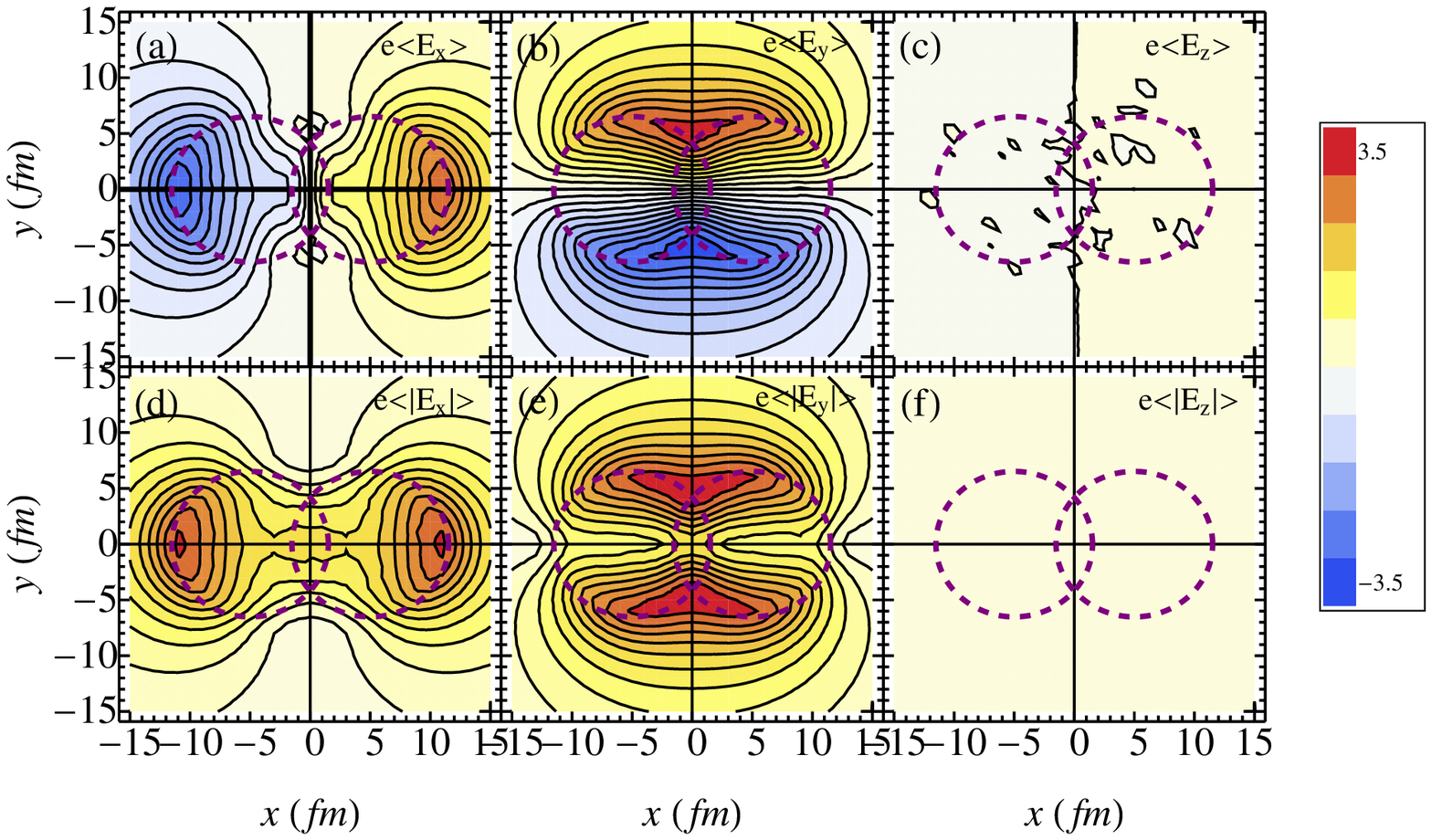}
\caption{(Color online) The spatial distributions of the EM fields in the transverse plane at $t=0$ for
$b=10$fm at RHIC energy. The unit is $m_\p^2$. The dashed circles indicate the two colliding nuclei. Figures are from Ref.~\cite{Deng:2012pc}.}
\label{sdist}
\end{center}
\end{figure*}
The spatial distributions of the EM fields are evidently inhomogeneous.
The contour plots for $\lan eB_{x,y,z}\ran$, $\lan eE_{x,y,z}\ran$,
$\lan e|B_{x,y,z}|\ran$, and $\lan e|E_{x,y,z}|\ran$ in the transverse plane at impact parameter $b=10$ fm and at $t=0$ for RHIC energy are shown in \fig{sdist}.
The distributions of the fields for LHC energy is merely the same but with
$2760/200=13.8$ times larger magnitudes according to Sec. \ref{sec:energ}.

One may notice that for noncentral collision, the $y$-component of
the electric field is very large along the $y$-direction, reflecting the fact that at $t=0$ a large amount
of net charges stays temporally in the center of the ``almond''-shaped overlapping region. This strong, out-of-plane electric field may drive positive
(negative) charges to move outward (toward) the reaction
plane, and thus induce an electric quadrupole moment in the produced quark-gluon matter. Such
an electric quadrupole moment, as argued in Ref.~\cite{Burnier:2011bf},
may lead to an elliptic flow imbalance between $\p^+$ and $\p^-$; see Refs.~\cite{Deng:2012pc,Stephanov:2013tga} for the detail.

\subsection {Azimuthal correlation with the participant planes}\label{sec:azimu}
\begin{figure*}[!htb]
\begin{center}
\includegraphics[width=6.5cm]{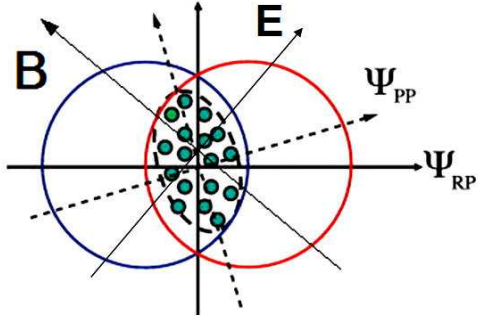}
\caption{Illustration of the azimuthal angles of the participant plane $\Psi_{\rm PP}$ and electric and magnetic fields.}
\label{inifluct}
\end{center}
\end{figure*}

\begin{figure*}
\begin{center}
\includegraphics[width=4.cm]{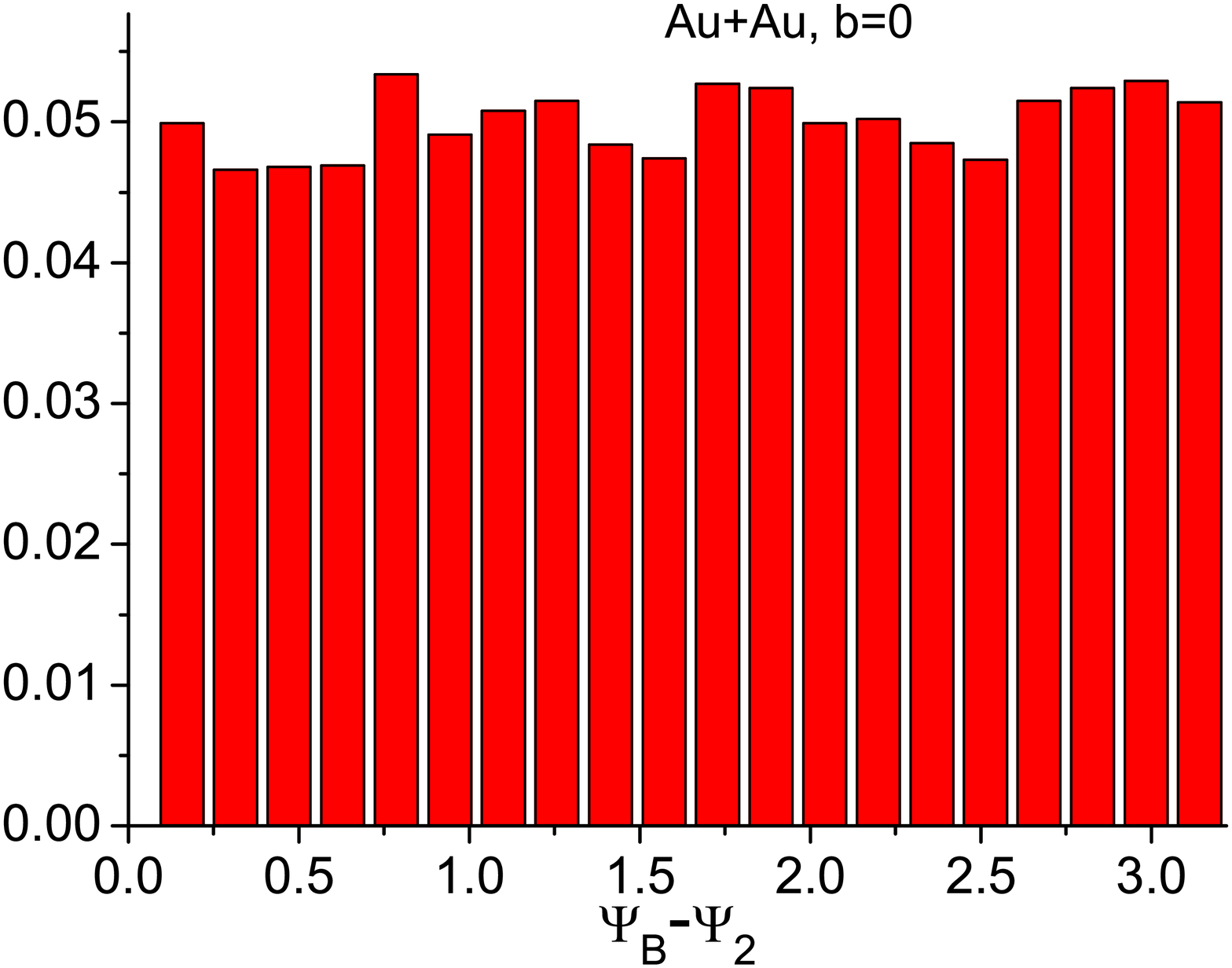}
\includegraphics[width=4.cm]{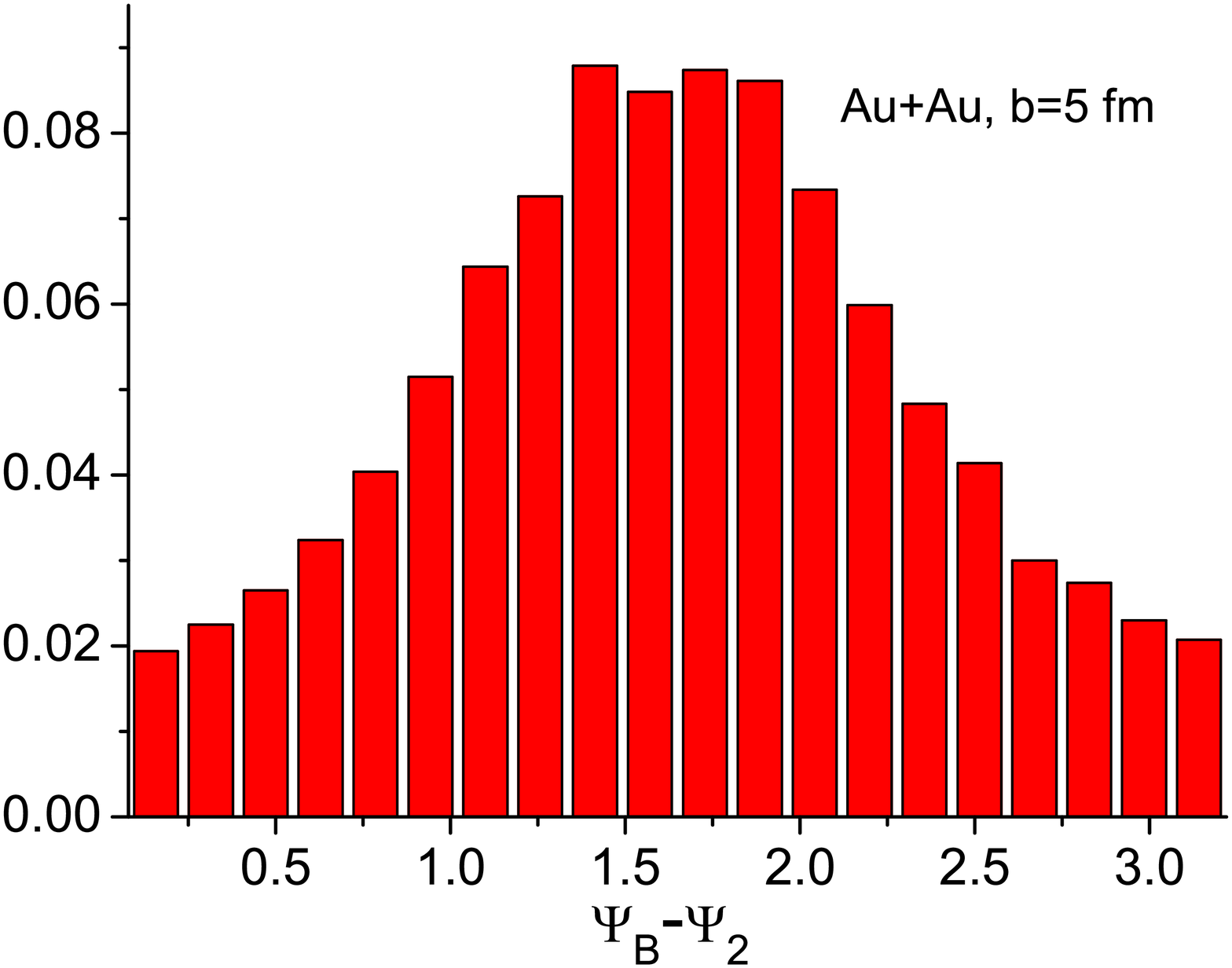}
\includegraphics[width=4.cm]{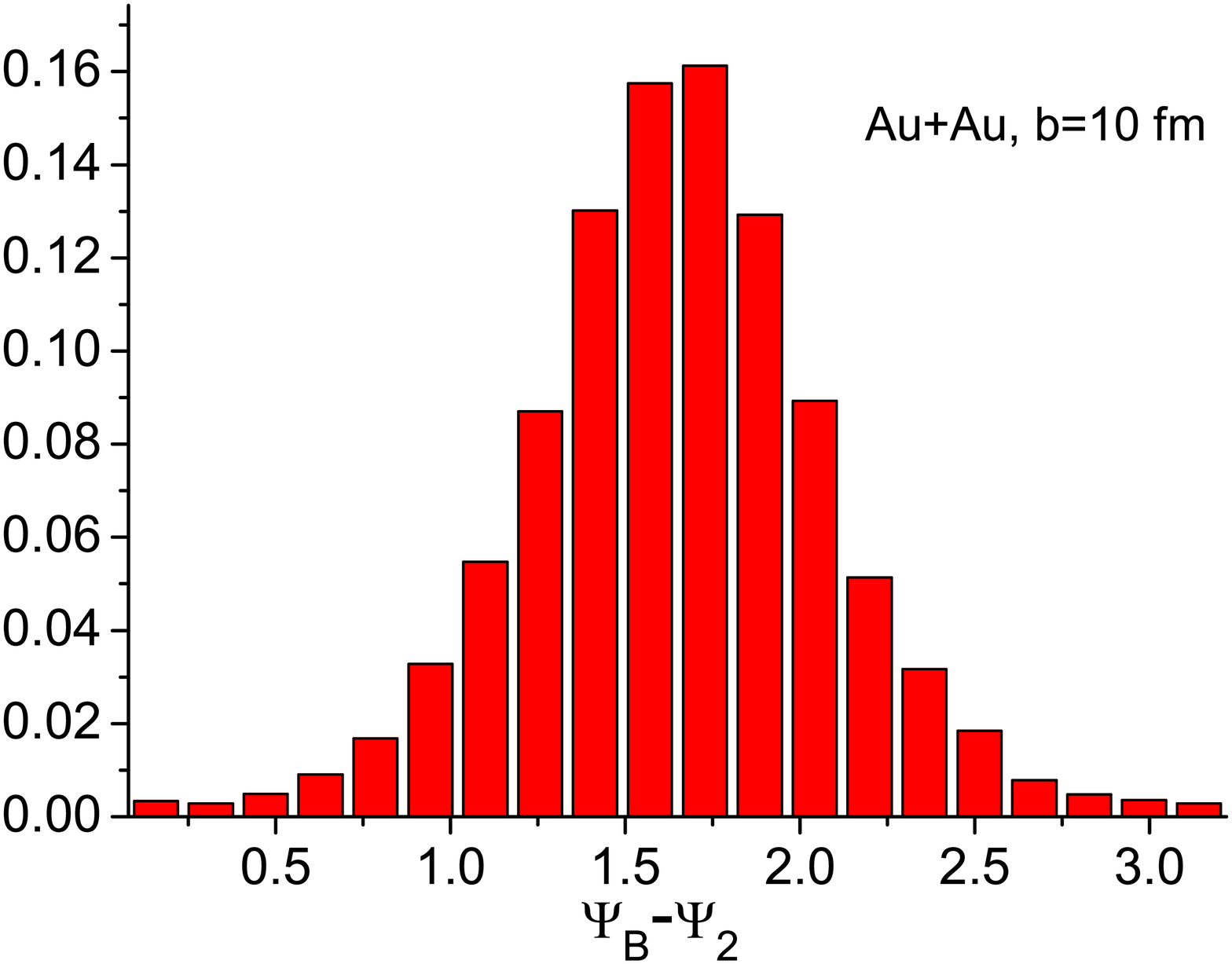}
\includegraphics[width=4.cm]{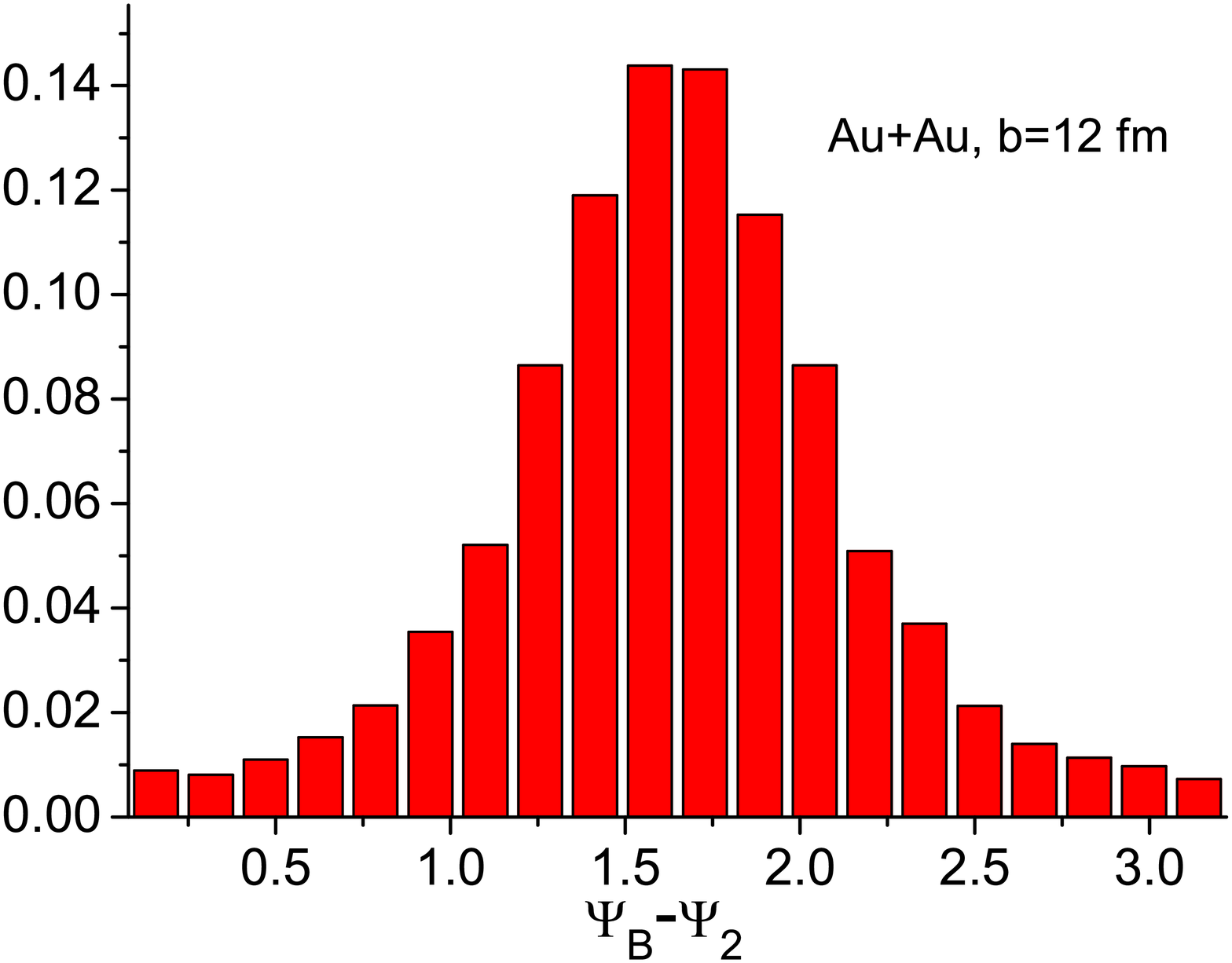}
\caption{The event-by-event histograms of $\J_\bB-\J_2$ at impact parameters $b=0, 5, 10, 12$ fm for Au + Au collision at RHIC energy. Here $\J_\bB$ is the azimuthal direction of $\bB$ field (at $t=0$ and $\br=(0,0,0)$) and
$\J_2$ is the second harmonic participant plane. This figure is from Ref.~\cite{Bloczynski:2012en}.}
\label{his}
\end{center}
\vspace{-0.2cm}
\end{figure*}

We have seen that the event-by-event fluctuations of the nuclear distribution can strongly modify the magnitudes of the EM fields, one then may ask how these event-by-event fluctuations affect the azimuthal orientations of the EM fields (see the illustrating \fig{inifluct}). In fact, as revealed recently~\cite{Bloczynski:2012en}, the event-by-event fluctuations generally make the magnetic fields unaligned with the normal direction of the reaction plane. As a consequence of the event-by-event fluctuations of nuclear distribution, the distribution of the participants (the nucleons that participate in the collision) in the overlapping region varies from event to event as well. Thus for each event, the overlapping region is not perfectly almond-shaped and its short-axis may be rendered away from the impact-parameter direction. Such shape and direction variations can be captured by the so-called eccentricity parameters $\e_n$ and harmonic angles $\J_n$, $n=1,2,3,\cdots$. Mathematically, they are defined as
 \begin{eqnarray}
\label{part:def}
\e_1e^{i\J_1} &=&-\frac{\int d^2\br\r(\br) r^3 e^{i\f}}{\int d^2\br\r(\br) r^3},\\
\e_n e^{in\J_n}&=&-\frac{\int d^2\br\r(\br) r^n e^{in\f}}{\int d^2\br\r(\br) r^n},\;\; n>1,
\end{eqnarray}
where $\r(\br)$ is the transverse distribution function of the participants. If there were no event-by-event fluctuations, $\J_2$ for each event should be equal to $\J_\rp$. As we will see, the azimuthal orientation between $\psi_{\bB}$ of $\bB$ (and $\psi_{\bE}$ of $\bE$) and $\Psi_{2}$ (the second harmonic angle of the participants) fluctuates with sizable spread in their relative angle $\psi_{\bB}-\Psi_2$ about the expected value $\p/2$. (Note that this would imply that $\psi_{\bB}-\Psi_{\rm RP}$ and $\psi_{\bB}-\Psi_{\rm E}$ with $\Psi_{\rm RP}$ and $\Psi_{\rm E}$ the reaction plane angle and event plane angle also fluctuate.) This can be clearly seen from \fig{his} which shows the histograms of $\psi_{\bB}-\Psi_2$ over events for different $b$. For the most central case the events are uniformly distributed indicating negligible correlation between $\j_\bB$ and $\J_2$; while for noncentral collisions the event distribution behaves like a Gaussian peaking at $\psi_{\bB}-\Psi_2=\p/2$ indicating correlation between $\J_\bB$ and $\J_2$.
For the following reason, such fluctuations in the correlation between $\j_\bB$ (as well as $\j_\bE$) and the participant planes may have important impacts on the experimentally measured quantities.

Let us consider the chiral magnetic effect (CME, see \sect{sec:cme}) as a concrete example, but the analysis can be extended to the observables of other EM-field-induced transport phenomena. The CME contributes to the single particle distribution for charged hadrons a component $f_q\propto qeB \cos(\f-\j_\bB)$ with $q=\pm$ the charge which in turn contributes to the two-particle distribution the following term,
\begin{eqnarray}
f_{q_\a q_\b} \propto q_\a q_\b (e\bB)^2\cos(\phi_\a-\psi_{\bB}) \, \cos(\phi_\b-\psi_{\bB}),
\end{eqnarray}
where $\f_{\a,\b}$ are the azimuthal angles of the hadrons $\a$ and $\b$. We therefore can extract the CME contribution to the two-particle correlation $\gamma_{\a\b}=\lan\cos(\phi_\a+\phi_\b-2\Psi_{2})\ran$ (which was used by STAR and ALICE collaborations to detect the charge separation with respect to the reaction plane, see \sect{sec:corrcme}) as
\begin{eqnarray}
\gamma_{\a\b}\propto  q_\a q_\b \lan (e\bB)^2 \cos[2(\psi_{\bB}-\Psi_2)]\ran.
\end{eqnarray}
If the $\bB$-direction were always perpendicular to the reaction plane while $\Psi_2$ always coincide with $\Psi_{\rm RP}$ (which we set to be zero here), then we simply have $\gamma_{\a\b}\propto  - q_\a q_\b\lan (eB)^2\ran$. But the fluctuations in magnetic field as well as in participant planes will blur the relative angle between the two and modify the signal by a factor $\sim\lan \cos[2(\psi_{\bB}-\Psi_2)]\ran$. (Here we note that the magnitude of the magnetic field has no noticeable correlation to its azimuthal direction~\cite{Bloczynski:2012en}.) Similarly, if one measures the two-particle correlation with respect to higher harmonic participant plane, for example, the fourth harmonic plane $\Psi_4$, $\lan\cos[2(\phi_\a+\phi_\b-2\Psi_{4})]\ran$, the azimuthal fluctuations of $\bB$ will again contribute a modification factor $\sim\lan\cos[4(\psi_{\bB}-\J_4)]\ran$ to it.

\begin{figure*}
\begin{center}
\includegraphics[width=7.0cm]{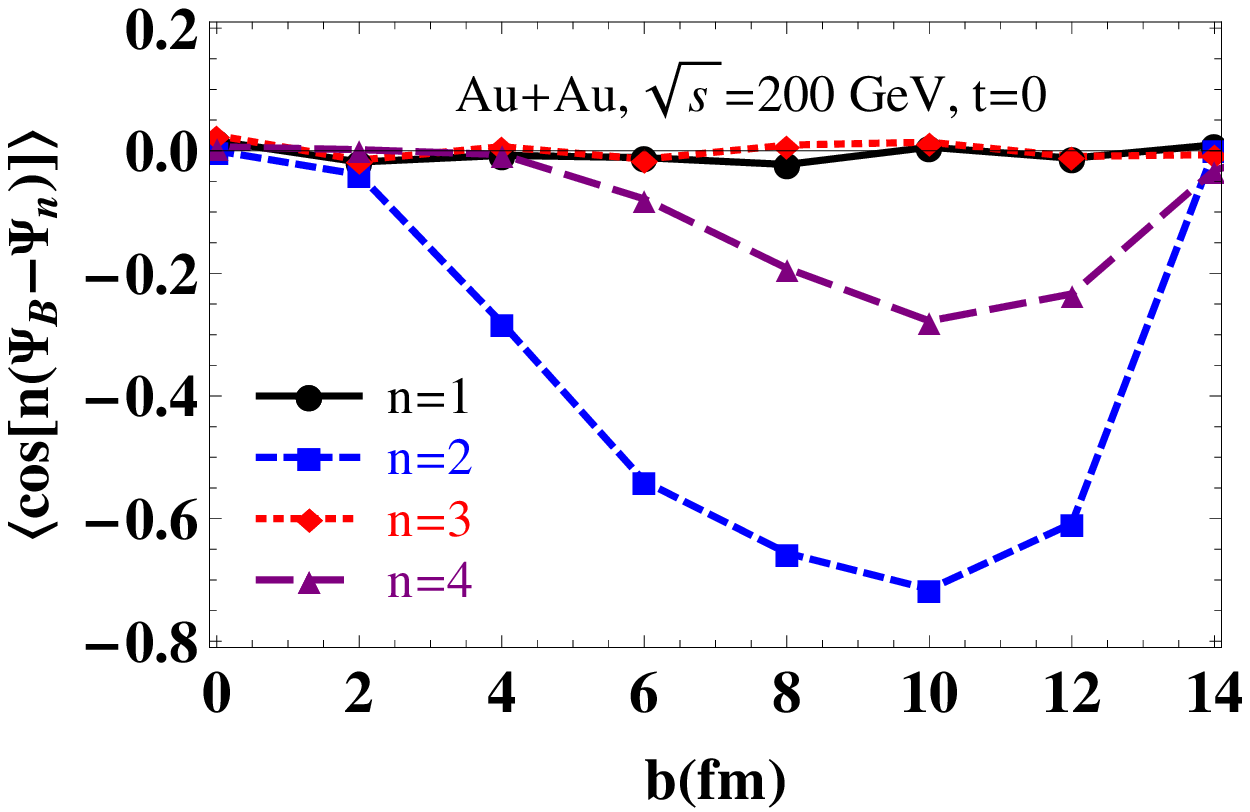}
\includegraphics[width=7.0cm]{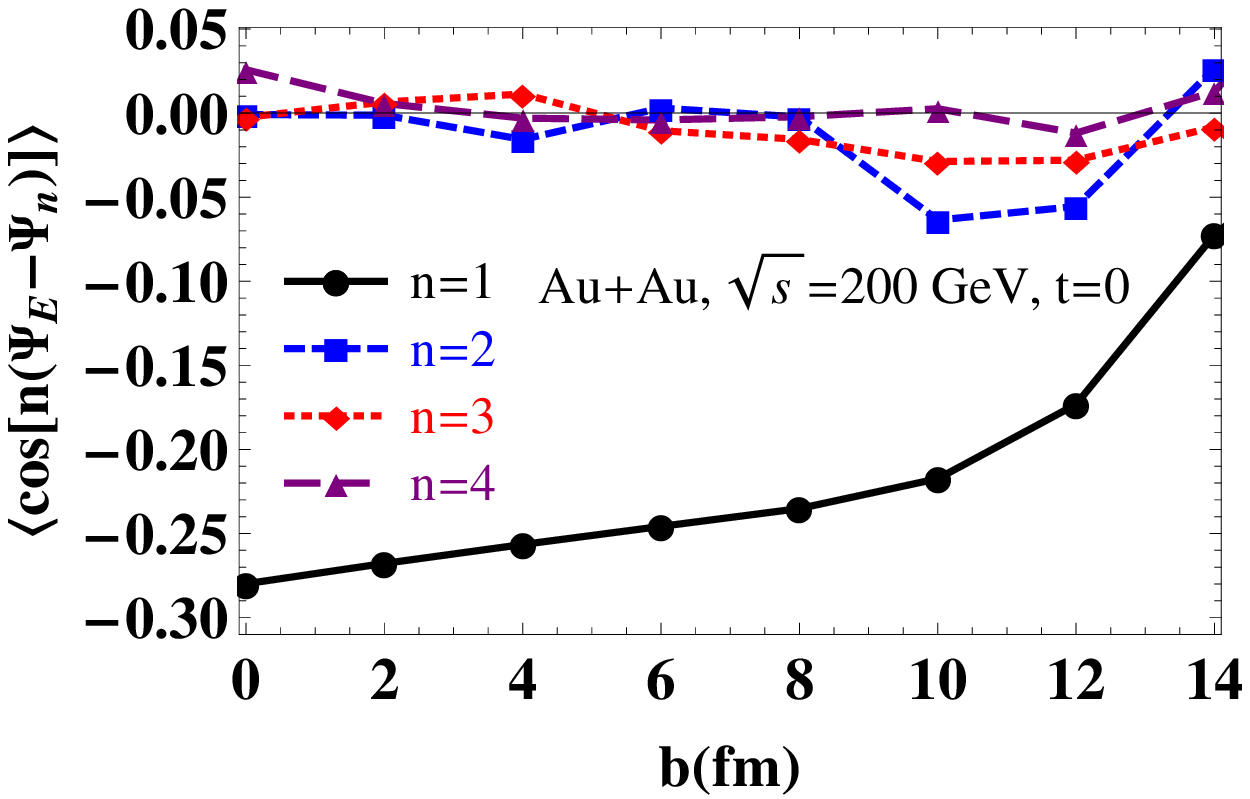}
\caption{The correlations $\lan\cos[n(\psi_{\bB}-\Psi_n)]\ran$ and $\lan\cos[n(\psi_{\bE}-\Psi_n)]\ran$ for $\bB$- and $\bE$-fields at the center of the overlapping region, $\br={\vec 0}$, as functions of impact parameter. Figures are modified from Ref.~\cite{Bloczynski:2012en}.}
\label{icos2}
\end{center}
\end{figure*}
In Fig.~\ref{icos2} (left panel) one can find the computed average values of  $\lan\cos[n(\psi_{\bB}-\Psi_n)]\ran$ as functions of the impact parameter from the event-by-event determination of the $\bB$-field direction $\psi_{\bB}$ at the collision center $\br={\vec0}$ and the participants harmonics, $\Psi_n$, $n=1,2,3,4$. The plots suggest: \\
(1) The correlations between $\j_\bB$ and the odd harmonics $\J_1, \J_3$ are practically zero (as a consequence of parity invariance), while the correlations of $\j_\bB$ with even harmonics $\J_2, \J_4$ are nonzero but suppressed comparing to the non-fluctuating case. \\
(2) The centrality dependence of $\lan\cos[2(\psi_{\bB}-\Psi_2)]\ran$ agrees with the patterns shown in the histograms Fig.~\ref{his}: it is significantly suppressed in the most central and most peripheral collisions indicating no correlations between $\j_\bB$ and $\J_2$ while is still sizable for moderate values of $b$. \\
(3) As checked in Ref.~\cite{Bloczynski:2012en}, there is no visible difference between the $(e\bB)^2$-weighted correlation $\lan (e\bB)^2\cos[n(\psi_{\bB}-\Psi_n)]\ran/\lan(e\bB)^2\ran$ and the unweighted correlation $\lan \cos[n(\psi_{\bB}-\Psi_n)]\ran$ for $n=1,2,3,4$. This indicates
that the magnitude of the magnetic field does not noticeably correlate to its azimuthal direction.

The event-by-event fluctuations also bring modification to the correlations between $\bE$-field orientation and the participant planes.
In parallel to the $\bB$-field case, in Fig.~\ref{icos2} (right panel), we show the correlations $\lan\cos[n(\psi_{\bE}-\Psi_n)]\ran$, $n=1,2,3,4$, as functions of the impact parameter. It is seen that:\\
(1) There is a sizable negative correlation (i.e., back-to-back) between $\j_\bE$ and $\Psi_1$ which is strongest in the central collisions. This is simply because the pole of $\Psi_1$ with more matter will concurrently have more positive charges from protons which induce the $\bE$-field pointing opposite to $\J_1$.\\
(2) There is also a weak correlation between $\j_\bE$ and $\J_3$.\\
(3) Similar to the $\bB$-field case, $(e\bE)^2$-weighted correlations
$\lan(e\bE)^2\cos[n(\psi_{\bE}-\Psi_n)]\ran/\lan(e\bE)^2\ran$ have no visible difference from the unweighted correlations $\lan\cos[n(\psi_{\bE}-\Psi_n)]\ran$ indicating no correlation between $\bE$-field magnitude and orientation.

\subsection {Early-stage time evolution}\label{sec:early}
\begin{figure*}[!htb]
\begin{center}
\includegraphics[height=4.5cm]{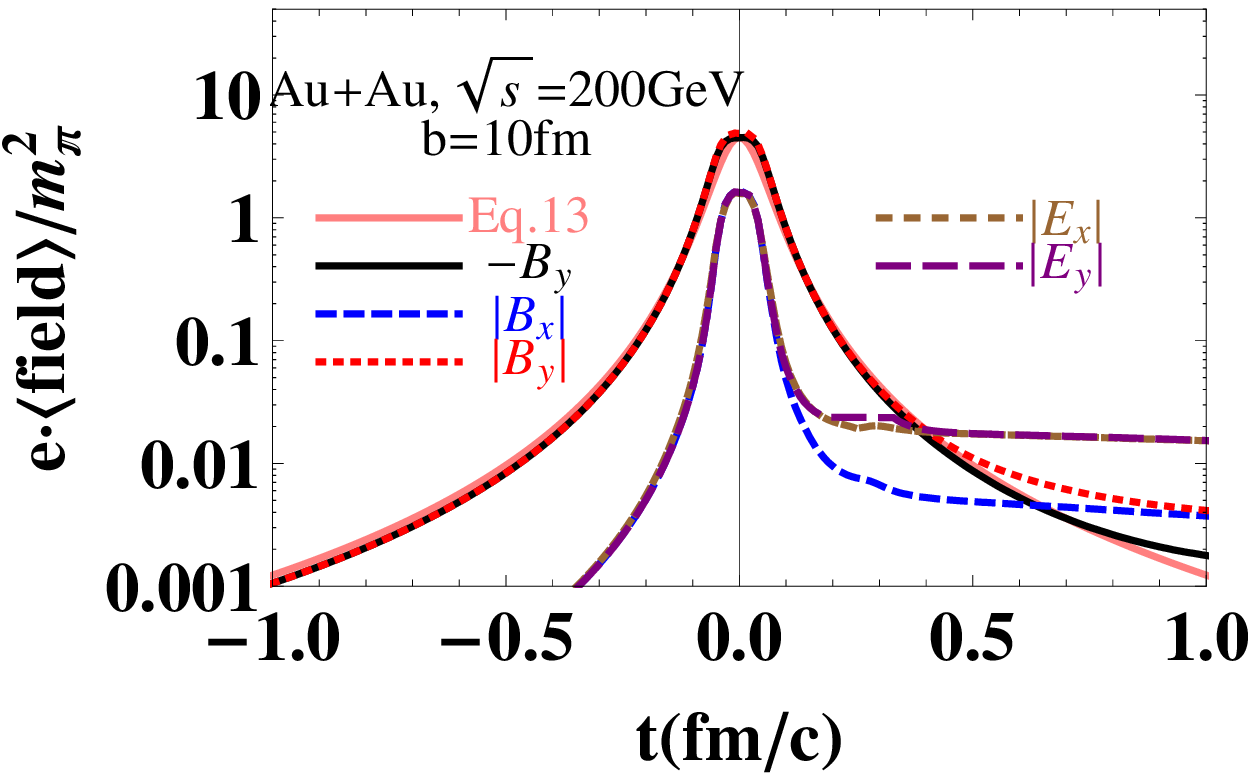}
\includegraphics[height=4.5cm]{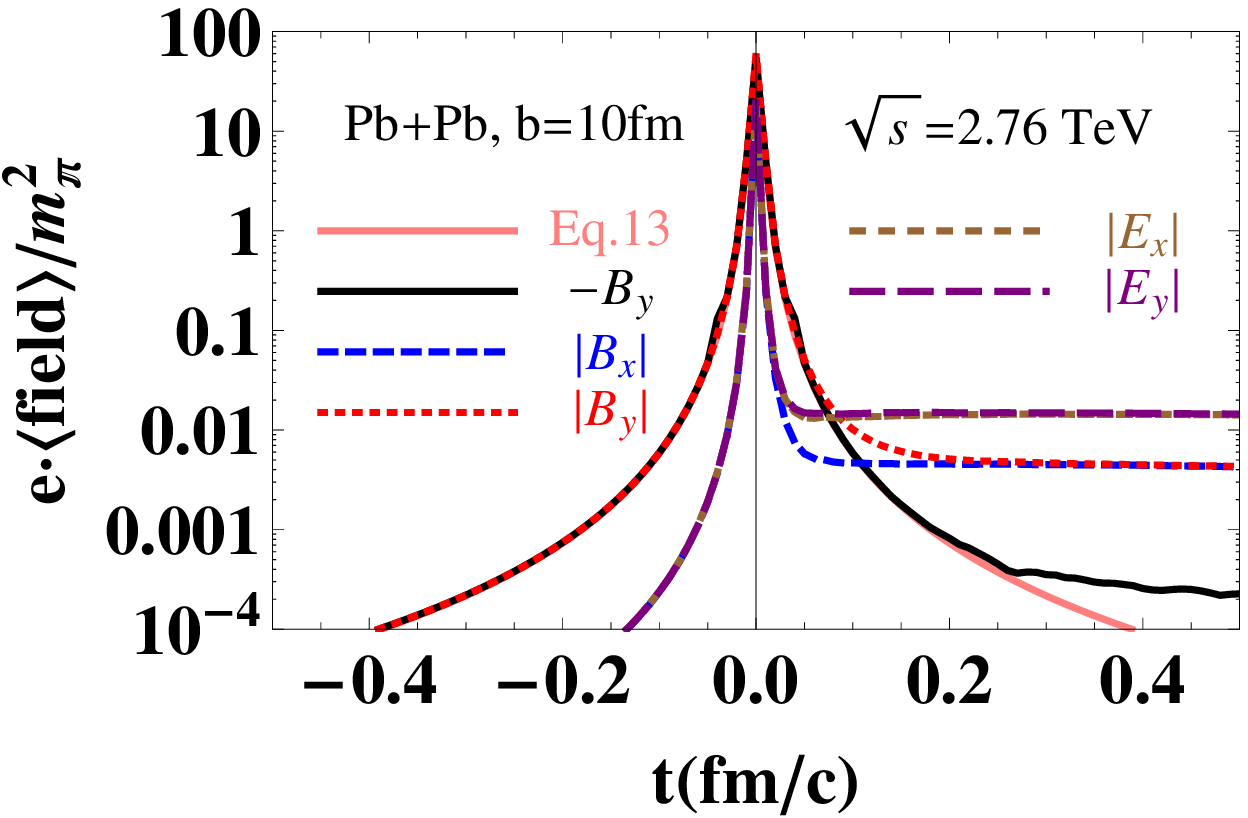}
\caption{The early-stage time evolution of the electromagnetic fields at $\br=0$
with impact parameter $b=10$ for Au + Au collision at $\sqrt{s}=200$ GeV and
Pb + Pb collision at $\sqrt{s}=2.76$ TeV. For $t\lesssim 10 t_B$, \eq{evofit} fit the curve for $\lan eB_y\ran$ very well. After that, the remnants essentially
slow down the decays of the transverse fields and \eq{evofit} does not work well any more. Figures are modified from Ref.~\cite{Deng:2012pc}.}
\label{tdepe}
\end{center}
\end{figure*}
In a high-energy heavy-ion collision, right after the collision, the produced partonic matter is mainly consist of gluons and is in a far-from-equilibrium state. This partonic matter subsequently evolves toward thermal equilibrium and a large number of quarks and anti-quarks are excited during this thermalization process. Although so far we still lack a theory to quantitatively understand the thermalization problem, the phenomenological studies revealed that the time scale of the completion of the thermalization is very short comparing to the total lifetime of the thermalized quark-gluon plasma (QGP). The relevant information can be found in the review articles, Refs.~\cite{Arnold:2007pg,Gelis:2012ri,Huang:2014iwa,Berges:2015kfa}. Once the thermalization is locally achieved, the bulk evolution of the system can be well described by hydrodynamics. One of the transport coefficients of the hydrodynamics, namely, the electric conductivity $\s$ has been numerically simulated by using lattice QCD recently and it was found that $\s$ for QGP is very large (see next subsection). A large $\s$ makes the QGP sensitive to the variation of the EM fields and which in turn strongly influence the time evolution of the EM fields themselves. Thus the time evolution of the EM fields in the QGP stage need special treatments and we leave this issue to next subsection. In this subsection we will focus on the stage before the thermalization is achieved (we call this stage the ``early stage"). The quark-gluon matter in the early stage is expected to be much less conducting than that in the QGP stage and we will simply assume it is insulating and thus ignore the response of the matter to the EM fields.

In \fig{tdepe} one can find the numerical results of the early-stage time evolutions of the EM fields at $\br={\bf0}$ in collisions with $b=10$ fm for Au + Au collision at $\sqrt{s}=200$ GeV and for Pb + Pb collision at $\sqrt{s}=2.76$ TeV. The results are from Ref.~\cite{Deng:2012pc}.
The contributions to the EM fields come from the charged particles in spectators, participants, and remnants. We can see that the transverse fields decay very fast after the collision reflecting the fact that the spectators are leaving the collision region very fast. Once the spectators are all far away from the collision region, the remnants which moves much slower than the spectators become important and they essentially slow down the decays of the transverse fields.
The lifetime of the magnetic fields due to spectators can be estimated as
\begin{eqnarray}
t_B\approx R_A/(\g v_z) \approx\frac{2m_N}{\sqrt{s}} R_A,
\end{eqnarray}
which is just half the time that one proton needs to pass through the nucleus freely. The lifetime $t_B$ is very short for large $\sqrt{s}$: for Au + Au collision at 200 GeV, $t_B\approx 0.065$ fm, while for Pb + Pb collision at 2.76 TeV, $t_B\approx 0.005$ fm. Within the time period $(0,t_B)$ the fields decay slowly and when $t\gtrsim t_B$, we can approximate the early-stage time evolution of the event-averaged magnetic field, that is, $\lan eB_y\ran$ as
\begin{eqnarray}
\label{evofit}
\lan eB_y(t)\ran\approx\frac{\lan eB_y(0)\ran}{(1+t^2/t^2_B)^{3/2}}.
\end{eqnarray}
This formula works better for larger impact parameter $b$ and larger $\sqrt{s}$. As seen from \fig{tdepe}, \eq{evofit} fit the simulation results for $\lan eB_y(t)\ran$ very well for time $t\lesssim 10 t_B$; after that the remnants dominate and the separation between the curves from \eq{evofit} and from the simulations are visible. The \eq{evofit} shows that for $t> t_B$ the magnetic field decays fast, $\lan eB_y(t)\ran \sim t_B^3/t^3$~\footnote{One should note that the magnitude of the magnetic field is still very large even at $t>10 t_B$; for example, $\lan eB_y\ran\sim 40$ MeV$^2$ for RHIC Au + Au collisions at $t=1$ fm which is still comparable to the light quark mass squared, $m_{u,d}^2$.}. However, if at that time $t_B$ the QGP has been already formed, its EM response will significantly modify the time evolution of the fields.

\subsection {Late-stage (QGP-stage) time evolution}\label{sec:late}
The discussions and simulations presented in the last subsection are based on the assumption that the produced matter
is ideally insulting. This assumption is adoptable only in the early stage where the system is gluon-dominated but becomes less and less justified as the system evolves and more and more quarks and anti-quarks emerge. As a matter of fact, the QGP is a good conductor according to the theoretical and lattice QCD studies.
At very high temperature the perturbative study gives that the electric conductivity of QGP is $\s\approx 6T/e^2$~\cite{Arnold:2003zc}. An old lattice calculation with $N_f=0$ found that $\s\approx 7 C_{\rm EM}T$~\cite{Gupta:2003zh} at $1.5T_c<T<3 T_c$ with $T_c$ the deconfinement temperature. Another quenched lattice simulation using staggered fermions found that $\s\approx 0.4 C_{\rm EM}T$~\cite{Aarts:2007wj}. Recent quenched lattice studies using Wilson fermions obtained that $\s\approx (1/3) C_{\rm EM}T-C_{\rm EM}T$~\cite{Ding:2010ga,Francis:2011bt,Ding:2014dua}
for temperature $T_c<T<3 T_c$. The lattice calculation with $N_f=2$ dynamical Wilson fermions found that $\s\approx 0.4 C_{\rm EM}T$ at $T\sim 250$ MeV~\cite{Brandt:2012jc}. Another new lattice simulation using $N_f=2+1$ fermions obtained that $\s\approx 0.1 C_{\rm EM} T- 0.3 C_{\rm EM} T$ for temperature $T_c<T<2 T_c$~\cite{Amato:2013naa,Aarts:2014nba}. In these results, the EM vertex parameter $C_{\rm EM}\equiv \sum_f q_f^2$ and $q_f$ is the charge of quark with flavor $f$; for example, $C_{\rm EM}=(5/9)e^2$ if $u, d$ quarks are considered while $C_{\rm EM}=(2/3)e^2$ if $u, d, s$ quarks are considered. Note that the deconfinement temperature $T_c$ is different in $N_f=0$ and $N_f\neq 0$ cases; for example, if $N_f=0$ we have $T_c\sim 270$ MeV while if $N_f=2+1$ we have $T_c\sim 170$ MeV.

At $T=350$ MeV and choosing $\s\approx 0.3 C_{\rm EM}T$ with $u,d,s$ quarks contributing to $C_{\rm EM}$, one can find that the resulted $\s$ is about $10^3$ times larger than that of copper at room temperature ($\s_{\rm Cu}\approx 4.43\times 10^{-3}$ MeV at $T=20 ^oC$).

Now let us analyze how the large $\s$ influences the time evolution of the EM fields in the QGP stage which we refer to as ``late stage".

Our discussion will be based on magnetohydrodynamics. We first write down
the Maxwell's equations,
\begin{eqnarray}
\label{maxwell}
&\displaystyle\nabla\times\bE=-\frac{\pt\bB}{\pt t},&\\
&\displaystyle\nabla\times\bB=\frac{\pt\bE}{\pt t}+\bJ,&
\end{eqnarray}
where $\bJ$ is the electric current. We treat the QGP as being locally charge neutral but conducting, thus $\bJ$ is the sum of the external one and the one determined by the Ohm's law,
\begin{eqnarray}
\label{ohm}
\bJ=\s\lb\bE+\bv\times\bB\rb+\bJ_{\rm ext},
\end{eqnarray}
where $\bv$ is the flow velocity of QGP and $\bJ_{\rm ext}$ is the current due to the motion of unwounded protons (most are in spectators). Using \eq{ohm}, we can
rewrite the Maxwell's equations as magnetohydrodynamic equations
\begin{eqnarray}
\label{induce1}
&\displaystyle\frac{\pt\bB}{\pt t}=\nabla\times(\bv\times\bB)
+\frac{1}{\s}\lb\nabla^2\bB-\frac{\pt^2\bB}{\pt t^2}+\nabla\times\bJ_{\rm ext}\rb,&\\
\label{induce2}
&\displaystyle\frac{\pt\bE}{\pt t}+\frac{\pt\bv}{\pt t}\times\bB=\bv\times(\nabla\times\bE)
+\frac{1}{\s}\lb\nabla^2\bE-\frac{\pt^2\bE}{\pt t^2}-\frac{\bJ_{\rm ext}}{\pt t}\rb,&
\end{eqnarray}
where we have used the Gauss laws $\nabla\cdot\bB=0$ and $\nabla\cdot\bE=\r=0$.
Equation (\ref{induce1}) is the {\it induction equation}, which plays a central role in describing
the dynamo mechanism of stellar magnetic field generation. The first terms on the right-hand sides
of \eqs{induce1}{induce2}
are the convection terms, while the remained terms are called ``diffusion terms" although they are not exactly in the diffusion-equation type.
Let us discuss some outcomes of these magnetohydrodynamic equations.

(1) If $\bv=\vec0$, that is, if the QGP does not flow, the \eq{induce1} reduces to
\begin{eqnarray}
\label{induce12}
\displaystyle\frac{\pt\bB}{\pt t}=\frac{1}{\s}\lb\nabla^2\bB-\frac{\pt^2\bB}{\pt t^2}+\nabla\times\bJ_{\rm ext}\rb.
\end{eqnarray}
This equation can be solved by using the method of Green's function, the details can be found in Refs.~\cite{Tuchin:2013ie,Tuchin:2013apa,Tuchin:2014iua,Tuchin:2014hza,Gursoy:2014aka,Zakharov:2014dia} in which the authors studied how the spectators induced magnetic field evolve in QGP phase (assuming the system is already in the QGP phase at the initial time). The main information from these studies are that the presence of the conducting matter can significantly delay the decay of the magnetic field. This is easily understood as the consequence of the Faraday induction: a fast decaying external magnetic field induces a circular electric current in the medium which in turn causes a magnetic field that compensates the decaying external magnetic field.

For late times, the external current $\bJ_{\rm ext}$ from the spectators can be neglected (we will always assume this case in this subsection hereafter). If $\s\gg 1/t_c$ with $t_c$ the characteristic time scale over which the field strongly varies, one can neglect the second-order time derivative term and render \eq{induce12} a diffusion equation:
\begin{eqnarray}
\displaystyle\frac{\pt\bB}{\pt t}=\frac{1}{\s}\nabla^2\bB.
\end{eqnarray}
This case was studied in Ref.~\cite{Tuchin:2010vs}. This equation describes the decay of the field due to diffusion, and the diffusion time of the magnetic field is given by
\begin{eqnarray}
t_D=L^2\s,
\end{eqnarray}
with $L$ a characteristic length scale of the system over which the magnetic field varies strongly.
Upon setting $L\sim 10$ fm and $\s\approx 0.3C_{\rm EM}T\approx 6$ MeV at $T=300$ MeV, the diffusion time is about $t_D\sim 3$ fm. However, as argued by Mclerran and Skokov~\cite{McLerran:2013hla}, in this case the condition $\s\gg 1/t_c\sim1/t_D$ is not satisfied, so it is more realistic to solve \eq{induce12} instead its diffusion-type simplification.

(2) If $\bv\neq\vec0$ and the magnetic Renolds number $R_m=LU\s\gg 1$ (the magnetic Renolds number quantifies the ratio of the convection term over the ``diffusion term"), we can approximately keep only the convection terms in \eqs{induce1}{induce2}. This corresponds to the {\it ideally conducting} limit. The equations such obtained are
\begin{eqnarray}
\label{induce1new}
&\displaystyle\frac{\pt\bB}{\pt t}=\nabla\times(\bv\times\bB),&\\
\label{induce2new}
&\bE=-\bv\times\bB.&
\end{eqnarray}
It is well-known that \eq{induce1new} leads to the {\it frozen-in theorem} for ideally conducting plasma,
\ie, the magnetic lines are frozen in the
plasma elements or more precisely the magnetic flux through a closed loop defined by plasma elements keeps
constant. Thus the decay of the fields are totaly due to the expansion of the QGP. To see the consequence of \eqs{induce1new}{induce2new}, we
assume for simplicity a initial Gaussian transverse entropy density profile
\begin{eqnarray}
s(x,y)=s_0\exp{\lb-\frac{x^2}{2a^2_x}-\frac{y^2}{2a^2_y}\rb},
\end{eqnarray}
where $a_{x,y}$ are the root-mean-square widths of the transverse distribution. They are of order of the nuclei radii if the
impact parameter is not large. For example, for Au + Au collision at RHIC,
$a_x\sim a_y\sim3$ fm for $b=0$, and $a_x\sim2$ fm, $a_y\sim3$ fm for $b=10$ fm.
By assuming the Bjorken longitudinal expansion,
\begin{eqnarray}
v_z=\frac{z}{t},
\end{eqnarray}
one can solve the ideal hydrodynamic equations for transverse expansion and obtain~\cite{Ollitrault:2007du},
\begin{eqnarray}
v_x&=&\frac{c_s^2}{a^2_x}xt,\\
v_y&=&\frac{c_s^2}{a^2_y}yt,
\end{eqnarray}
where $c_s=\sqrt{\pt P/\pt\ve}$ is the speed of sound.  Substituting the velocity fields into \eq{induce1new}, we can solve out $\bB(t)$ analytically. For example, the $B_y(t)$ at $\br=\vec0$ is give by
\begin{eqnarray}
\label{byevo}
B_y(t, {\bf 0})&=&\frac{t_0}{t}e^{-\frac{c_s^2}{2a_x^2}(t^2-t_0^2)}B_y(t_0, {\bf 0}).
\end{eqnarray}
This is just manifestation of the frozen-in theorem, because the areas of the cross section of the QGP expands according to $\frac{t}{t_0}\exp{{\frac{c_s^2}{2a_x^2}(t^2-t_0^2)}}$ in $x-z$ plane, thus the total flux across the $x-z$ plane is a constant.
Setting $a_x\sim a_y\sim3$ fm and $c_s^2\sim1/3$, we see from \eq{byevo}
that for $t\lesssim5$ fm $B_y(t)$ decays approximately as $B_y(t)\propto (t_0/t) B_y(t_0)$ --- much slower than the $1/t^3$-type decay in the insulating case discussed in last subsection.

So far, we discussed two special cases of \eqs{induce1}{induce2} which permit analytical treatments. It is desirable to solve the most general equations in accompanying with the hydrodynamic or kinetic-equation simulation for the, i.e. the fluid velocity and temperature, of the fireball. But up to today, this is not done yet.

\subsection {Electromagnetic fields in other collision systems} \label{sec:other}
In recent years, RHIC has also run heavy-ion collisions of nuclei other than gold, for example, the Cu + Au and U + U collisions. The EM fields in these collision systems were also studied~\cite{Hirono:2012rt,Bloczynski:2013mca,Deng:2014uja,Voronyuk:2014rna,Chatterjee:2014sea}. We here give a brief summary of these studies.

\begin{figure*}[!htb]
\begin{center}
\includegraphics[height=4.2cm]{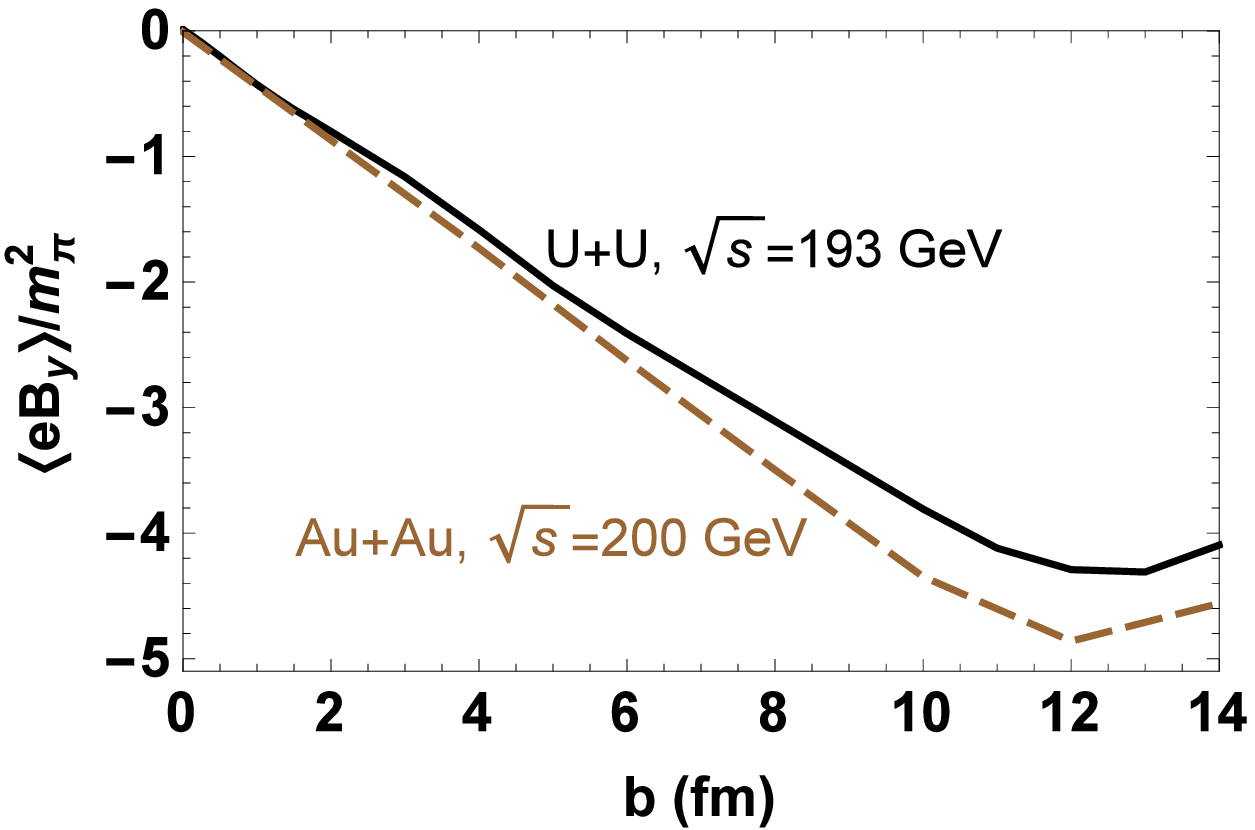}\;\;\;
\includegraphics[height=4.2cm]{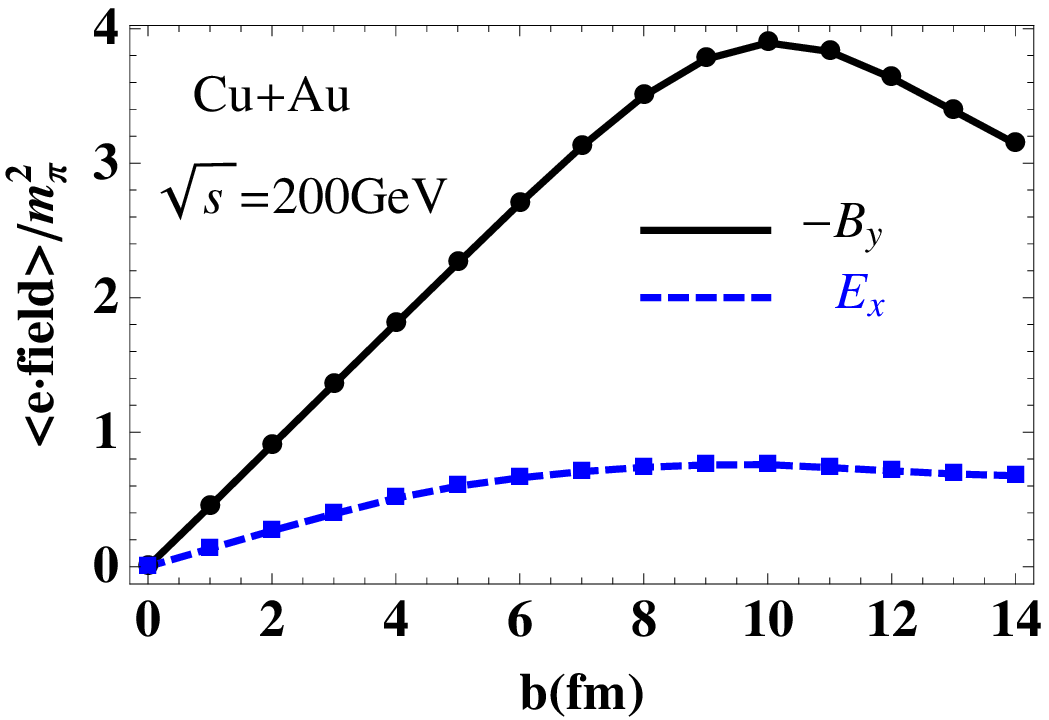}
\caption{The event-averaged EM fields in U + U and Cu + Au collisions. Figures are taken from Refs.~\cite{Bloczynski:2013mca,Deng:2014uja}.}
\label{figuuca}
\end{center}
\end{figure*}

The EM fields in U + U collisions were studied thoroughly in Ref.~\cite{Bloczynski:2013mca}. The uranium $^{238}$U nucleus, unlike the Au or Pb nucleus, has a highly deformed prolate shape. But this shape deformation does not bring much effect to the event-averaged magnetic fields. As simulated in Ref.~\cite{Bloczynski:2013mca}, the U + U collisions at $\sqrt{s}=193$ GeV produce an event-averaged magnetic field just lightly smaller than that in Au + Au collisions at $\sqrt{s}=200$ GeV, see the left panel of \fig{figuuca}. The readers can find more information, especially those related to the event-by-event fluctuation, in Ref.~\cite{Bloczynski:2013mca}.

The Cu + Au collisions are geometrically asymmetric: both the charge number and the total size of the gold nucleus are much larger than that of copper nucleus. Thus the Cu + Au collisions may be able to produce nonzero electric fields along the in-plane Au-to-Cu direction even after the event average.
The numerical simulation presented in Ref.~\cite{Deng:2014uja} found that:\\ (1) The strength of the event averaged magnetic field (which is along the $-y$ direction) in Cu + Au collisions at 200 GeV is comparable to that in Au + Au collisions.\\ (2) There is a strong event averaged electric field pointing from the Au nucleus to Cu nucleus which is at the order of one $m_\p^2$, see the right panel of \fig{figuuca}.\\ (3) The azimuthal angle of the electric field, $\j_\bE$, has a strong back-to-back correlation with $\j_1$, the first harmonic angle of the participants; this is the same as that in Au + Au collisions~\cite{Bloczynski:2012en}. The new feature is that in noncentral Cu + Au collisions there is a clear positive correlation between $\j_\bE$ and $\j_2$ signaling a persistent in-plane electric field. More details can be found in Ref.~\cite{Deng:2014uja}.

Very recently, there were interesting proposals for the novel effects of this in-plane $\bE$ field in Cu + Au collisions, for example: the $\bE$ field can lead to a directed flow $v_1$ splitting between positively and negatively charged hadrons~\cite{Hirono:2012rt,Voronyuk:2014rna}, the presence of the in-plane $\bE$ field may strongly suppress or even reverse the sign of the charge-dependent correlation $\g_{\a\b}$ (see \eq{gamma} for its definition)~\cite{Deng:2014uja}, and Cu + Au may serve to test the chiral electric separation effect~\cite{Ma:2015isa}; more relevant discussions are given in \sect{sec:exper}.

\section {Anomalous transports in P- and C-odd quark-gluon plasma}\label{sec:anoma}
The strong EM fields may induce a variety of novel effects to the quark-gluon plasma, among which we will focus in this section on the ones that are deeply related to the topology and symmetry of QCD and QED. It was found that, in addition to the normal electric current driven by $\bE$ field, there can emerge three new currents in P- and C-odd regions in QGP as responses to the applied EM fields. They are the chiral magnetic effect (CME)~\cite{Kharzeev:2007jp,Fukushima:2008xe}, the chiral separation effect (CSE)~\cite{Son:2004tq,Metlitski:2005pr}, and the chiral electric separation effect (CESE)~\cite{Huang:2013iia}. Thus the complete response of the P- and C-odd QGP to external EM field can be expressed as:
\begin{eqnarray}
\begin{pmatrix}
\bJ_V \\
\bJ_A
\end{pmatrix}
=
\begin{pmatrix}
\s_{VV} & \s_{VA} \\
\s_{AV} & \s_{AA}
\end{pmatrix}
\begin{pmatrix}
\bE \\
\bB
\end{pmatrix},
\end{eqnarray}
where $\bJ_V$ and $\bJ_A$ represent vector and axial currents and $\s$'s are corresponding conductivities. The Ohm's law and the conductivity $\s_{VV}$~\footnote{We will specifically refer to the vector current $\bJ_V$ as the $U(1)$ electric current $\bJ$ unless otherwise stated. Then $\s_{VV}$ is then just the usual electric conductivity which we denoted by $\s$ in last section.} is physically well understood, so we will not discuss them. In this section, we will focus on the CME, CSE, and CESE which are anomalous in the sense that their appearances are closely related to the topologically nontrivial vacuum structure of QCD and the axial anomaly. We will discuss their experimental consequences in next section.

\subsection {Chiral magnetic effect}\label{sec:cme}
{\bf (1) What is the chiral magnetic effect? ---} The CME is the generation of vector current by external magnetic field in chirality-imbalanced (P-odd) medium. Historically, the CME has been studied through different theoretical approaches in a number of contexts ranging from astrophysics~\cite{Vilenkin:1980fu}, condensed matter systems~\cite{Nielsen:1983rb,Alekseev:1998ds,Volovik:2003fe,Zyuzin:2012tv,Kharzeev:2012dc,Chen:2013mea,Basar:2013iaa,Landsteiner:2013sja,Huang:2015mga}, QCD physics~\cite{Kharzeev:2004ey,Kharzeev:2007tn,Kharzeev:2007jp,Fukushima:2008xe,Kharzeev:2009pj,Fukushima:2009ft,Asakawa:2010bu,Fukushima:2010vw,Fukushima:2010zza,Brits:2010pw,Hou:2011ze,Kharzeev:2011ds} , to holographic models~\cite{Newman:2005hd,Yee:2009vw,Rebhan:2009vc,D'Hoker:2009bc,Rubakov:2010qi,Gorsky:2010xu,Gynther:2010ed,Hoyos:2011us,Amado:2011zx,Nair:2011mk,Kalaydzhyan:2011vx,Loganayagam:2012pz,Lin:2013sga}. The recent reviews of CME are~\cite{Kharzeev:2009fn,Kharzeev:2013ffa,Fukushima:2012vr,Kharzeev:2015kna}.

The CME can be neatly expressed as
\begin{eqnarray}
\label{cme}
\bJ_V &=&
\s_{VA}
\bB,\\
\label{cmecond}\s_{VA}&=&\frac{e^2}{2\p^2}\m_A,
\end{eqnarray}
for each specie of massless fermions with charge $e$, where the current $\bJ_V$ is defined by $J^\m_V=e\lan\jb \g^\m\j\ran$ and $\m_A$ is a parameter that characterizes the chirality imbalance of the medium. The $\m_A$ is commonly called aixal or chiral chemical potential although it actually does not conjugate to any conserved charges of the fermions; we will discuss its meaning later. For QGP, the total CME current is obtained by adding up all the light quark's contributions and the CME conductivity should be $\s_{VA}=N_C\m_A\sum_f q_f^2/(2\p^2)$ with $q_f$ the charge of quark of flavor $f$ and $N_c=3$ the number of color.
\begin{figure}[!htb]
\begin{center}
\includegraphics[width=7cm]{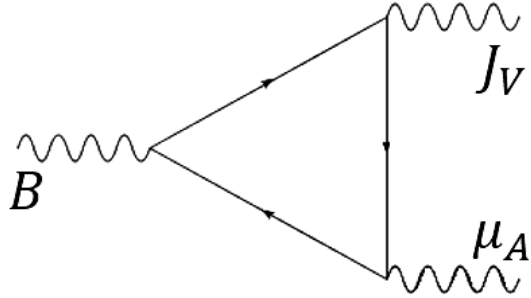}
\caption{The triangle diagram for CME. It is the same diagram that produces the axial anomaly.}
\label{diagcme}
\end{center}
\end{figure}

From \eq{cme}, we can first recognize that CME is P odd because $\bJ_V$ (a P-odd quantity) and $\bB$ (a P-even quantity) transform differently under parity. Thus CME can occur only in P-odd medium characterized by finite $\m_A$. Second, the CME is C even as $\bJ_V$ and $\bB$ are both C-odd. Third, the CME is T-even as both $\bJ_V$ and $\bB$ are T-odd~\footnote{Here, ``T" stands for ``time reversal".}; this is also evident from the fact that the CME conductivity $\s_{VA}$, as expressed in \eq{cmecond}, is temperature independent. The time-reversal-evenness of the CME conductivity indicates that the emergence the the CME current is a non-dissipative phenomenon~\cite{Kharzeev:2011ds,Kharzeev:2013ffa}. (Note that the usual electric conductivity $\s_{VV}$ is T-odd and thus can generate entropy.)

To get an intuitive understanding of \eq{cme}, let us consider a system with unequal numbers of right-handed (RH) and left-handed (LH) $u$ quarks (for example, consider that $N_R>N_L$ with $N_{R/L}$ the total number of RH$/$LH quarks) subject to uniform magnetic field. We know that the Landau quantization has the property that the lowest Landau level permits only one spin polarization which minimizes the excitation energy of quarks but is highly degenerate with a degeneracy factor proportional to the total magnetic flux. Thus if the magnetic field is strong enough so that this degeneracy factor is larger than $N_R$ or $N_L$, all the $u$ quarks are confined to the lowest Landau level on which their spins are totally polarized to be along the direction of the magnetic field. Now the RH $u$ quarks will prefer to move along their spin direction that is the direction of the magnetic field; while the LH $u$ quarks will prefer to move opposite to their spin direction which is opposite to the direction of the magnetic field. Because the number of RH quarks are larger than the number of LH quarks, the overall effect of the motion of quarks will be to generate a net current of $u$ along the magnetic field. For $\bar u$ quark, similar argument leads to a net current of $\bar u$ moving opposite to the magnetic field or a net electric current along the direction of the magnetic field~\footnote{One should be noticed that the antipartile of a RH-chirality (RH-helicity) massless fermion is of LH-chirality (RH-helicity)}. This is the intuitive picture of the CME. Note that if the magnetic field is not strong enough so that the higher Landau levels are also occupied, then each higher Landau level will contain equal numbers of spin-up and spin-down quarks and the CME current of spin-up quarks will exactly cancel the CME current of spin-down quarks for each higher Landau level. Therefore the higher Landau levels do not contribute to the total current --- only the lowest Landau level is responsible to the CME.

{\bf (2) How to derive the CME? ---} The emergence of the CME, \eq{cme}, is due to the axial anomaly in QED sector which couples the vector current $\bJ_V$ to the magnetic field $\bB$ and the axial chemical potential $\m_A$, see \fig{diagcme}. There are a variety of methods to derive \eq{cme} from microscopic quantum field theory~\cite{Vilenkin:1980fu,Nielsen:1983rb,Alekseev:1998ds,Kharzeev:2007jp,Fukushima:2008xe,Fukushima:2009ft,Fukushima:2010vw,Brits:2010pw,Hou:2011ze,Zahed:2012yu,Warringa:2012bq}, mesoscopic kinetic theory~\cite{Son:2012wh,Stephanov:2012ki,Gao:2012ix,Chen:2012ca,Son:2012zy,Chen:2013dca,Dwivedi:2013dea,Chen:2013iga,Chen:2014cla,Akamatsu:2014yza,Manuel:2013zaa,Manuel:2014dza,Duval:2014ppa,Chen:2015gta,Gao:2015zka}, to macroscopic hydrodynamic approach~\cite{Son:2009tf,Sadofyev:2010is,Sadofyev:2010pr,Zakharov:2012vv,Banerjee:2012iz,Neiman:2010zi,Jensen:2012jy}. Here we pick up one of these derivations given by Fukushima, Kharzeev, and Warringa~\cite{Fukushima:2009ft} because it is elementary and easy to see the relation between CME and the lowest Landau level and the axial anomaly.

Let $\O$ be the thermodynamical potential of fermions of charge $e$ in a magnetic field $\bB=B\hat{\vec z}$ at finite vector and axial chemical potentials, $\m_V=(\m_R+\m_L)/2, \m_A=(\m_R-\m_L)/2$. In the noninteracting limit, $\O$ can be written as
\begin{eqnarray}
\O=\frac{|eB|}{2\p}\sum_{s=\pm}\sum_{n=0}^\infty\a_{n,s}\int\frac{dp_z}{2\p}\lc E_{n,s}+T\ln\ls1+e^{-\b(E_{n,s}-\m_V)}\rs\ls1+e^{-\b(E_{n,s}+\m_V)}\rs\rc,
\end{eqnarray}
where $n$ runs over all the Landau levels and $s$ is over spins, $E_{n,s}$ is the dispersion relation of the fermions
\begin{eqnarray}
E_{n,s}=\sqrt{[{\rm sgn}(p_z)(p_z^2+2n|eB|)^{1/2}+s\m_A]^2+m^2},
\end{eqnarray}
and $\a_{n,s}$ is a degenerate constant given by $\a_{n,s}=1-\d_{0n}\d_{s,-{\rm sgn}(eB)}$ which accounts the fact that only one spin state occupies the lowest Landau level. The current $J_V^z$ can be obtained through differentiation of $\O$ with respect to vector potential $A_z$, $J_V^z=\pt\O/\pt A_z|_{A_z=0}$, which owing to gauge invariance is equivalent to $J_V^z=e\pt\O/\pt p_z$ in the integrand,
\begin{eqnarray}
J_V^z=\frac{e|eB|}{2\p}\sum_{s=\pm}\sum_{n=0}^\infty\a_{n,s}\int_{-\L}^\L\frac{dp_z}{2\p}\frac{\pt E_{n,s}}{\pt p_z}\ls1-n_F(E_{n,s}+\m_V)-n_F(E_{n,s}-\m_V)\rs,
\end{eqnarray}
where $\L$ is a ultraviolet cutoff that guarantees the finiteness of the calculation at the intermediate steps and goes to infinity at the end of the calculation, and $n_F(x)=(\exp{(x/T)}+1)^{-1}$ is the Fermi-Dirac distribution. It is then straightforward to find that
\begin{eqnarray}
J_V^z&=&\frac{e|eB|}{(2\p)^2}\sum_{s=\pm}\sum_{n=0}^\infty\a_{n,s}\ls E_{n,s}(p_z=\L)-E_{n,s}(p_z=-\L)\rs\non
&=&\frac{e|eB|}{(2\p)^2}\sum_{s=\pm}\sum_{n=0}^\infty\a_{n,s}\ls (\L^2+2n|eB|)^{1/2}+s\m_A-((\L^2+2n|eB|)^{1/2}-s\m_A)\rs\non
&=&\frac{e^2\m_A}{2\p^2}B.
\end{eqnarray}
This is just \eq{cme}. This derivation shows that: (1) The CME is due to the ultraviolet surface integral and is unaffected by infrared parameters, like $T$, $\m_V$, $m$, etc. (2) Only the lowest Landau level contributes to CME, reflecting the fact that only the lowest Landau level permits a touching node of opposite chirality. This touching node is known to be responsible for the $U_A(1)$ axial anomaly~\cite{Nielsen:1983rb,Ambjorn:1983hp}: the applied $\bE$ and $\bB$ fields pump the fermions at LH-chirality branch to RH-chirality branch at the touching node at a rate $\sim [e^2/(2\p^2)]\bE\cdot\bB$.

We emphasize again that although the derivation here is for non-interacting system and relies on Landau quantization picture, the CME conductivity $\s_{VA}=e^2\m_A/(2\p^2)$ is actually fixed by the axial anomaly equation and thus universal no matter how strong the interaction between fermions is (Recall that the axial anomaly equation itself is universal in the sense that it does not receive perturbative correction from scattering between fermions, a result usually referred to as Adler-Bardeen theorem). This is particularly supported by the derivation of CME based on holographic models~\cite{Newman:2005hd,Yee:2009vw,Rebhan:2009vc,D'Hoker:2009bc,Rubakov:2010qi,Gorsky:2010xu,Gynther:2010ed,Hoyos:2011us,Amado:2011zx,Nair:2011mk,Kalaydzhyan:2011vx,Loganayagam:2012pz,Lin:2013sga} which intrinsically describe strongly coupled system, where although the whole setup is very different from the above derivation and other calculations based on perturbation theory, the CME conductivity is shown to be given by the same universal result.

{\bf (3) How can QGP be chiral? ---} The appearance of CME requires a nonzero $\m_A$ which characterizes the strength of the chirality imbalance of the system. Then the question is: how can the QGP generate a net chirality imbalance? To answer this question, following the argument in Ref.~\cite{Kharzeev:2007jp},
let us first consider the vacuum state of the $SU(3)$ gauge theory. To make the energy minimized, the vacuum must satisfy the condition $G^a_{\m\n}=0$ ($G^a_{\m\n}$ is the field strength tensor) which requires the gauge field to be pure gauge: $\mathcal{A}_\m(x)=ig^{-1}U^{-1}(x)\pt_\m U(x)$ with $U(x)\in SU(3)$. Working in temporal gauge $\mathcal{A}_0(x)=0$ and by noting that for any time-independent gauge transformation, $\pt_0 U(\bx)=0$, the temporal gange fixing condition is unchanged, $0=\mathcal{A}_0(x)\ra U^{-1}(\bx) \mathcal{A}_0(x) U(\bx)+ig^{-1}U^{-1}(\bx)\pt_0 U(\bx)=0$, one can realize that the vacuum is described by a time-independent $\mathcal{A}_i(\bx)$ which is a pure gauge potential
\begin{eqnarray}
\mathcal{A}_i(\bx)=ig^{-1}U^{-1}(\bx)\pt_i U(\bx).
\end{eqnarray}
If we impose the boundary condition $U(\bx)\ra $ constant at $|\bx|\ra\infty$ (see, for example, Refs.~\cite{Srednicki:2007qs,Weinberg:1996kr,Cheng:1985bj,Nair:2005iw} for relevant discussions about the boundary condition), the gauge transformation $U(\bx)$ defines a map from $S^3$ ($\mathbb{R}^3$ with the infinity identified as an ordinary point) to $SU(3)$ which is characterized by a winding number $n_w\in \p_3(SU(3))=\mathbb{Z}$,
\begin{eqnarray}
n_w=\frac{1}{24\p^2}\int d^3\bx \e^{ijk}{\rm tr}[(U^{-1}\pt_i U)(U^{-1}\pt_j U)(U^{-1}\pt_k U)].
\end{eqnarray}
This winding number is a topological invariant as can be checked by smoothly deforming $U(\bx)$.
Thus the $U(\bx)$'s corresponding to different $n_w$ are topologically distinct in the sense that then cannot be smoothly deformed into each other without passing through gauge field configurations whose field strengths are nonzero. In other words, the $U(\bx)$'s of different $n_w$ define multiple degenerate vacua (called the $\h$-vacua) separated by finite energy barriers.

On the other hand, one can categorize all the gauge field configurations in to topologically distinct classes characterized by different values of the following topological invariant,
\begin{eqnarray}
q=\frac{g^2}{32\p^2}\int d^4 x G^a_{\m\n}\tilde{G}^{\m\n}_a,
\end{eqnarray}
where $\tilde{G}^{\m\n}_a=\frac{1}{2}\e^{\m\n\r\s}G_{\r\s}^a$.
This $q$ is called the (second) Chern number of configuration $\mathcal{A}$ and is always an integer~\cite{Nakahara:2003nw,Nash:1983cq}. It is straightforward to show that under the condition $G_{\m\n}^a\ra0$ at infinity ($t\ra\pm\infty$ or $|\bx|\ra\infty$),
\begin{eqnarray}
q=\frac{1}{24\p^2}\int d\S_\m\e^{\m\n\r\s}{\rm tr}[(U^{-1}\pt_\n U)(U^{-1}\pt_\r U)(U^{-1}\pt_\s U)],
\end{eqnarray}
where $\S^\m$ is a surface at infinity in four-dimensional spacetime and $U(x)\in SU(3)$ satisfies $U(t,\bx)\ra U_\pm(\bx)$ for $t\ra\pm\infty$ and $U(t,\bx)\ra $ constant for $|\bx|\ra\infty$~\cite{Srednicki:2007qs,Weinberg:1996kr,Cheng:1985bj,Nair:2005iw}. Thus, after several steps of manipulations,
\begin{eqnarray}
q=n_w(t=\infty)-n_w(t=-\infty).
\end{eqnarray}
This means that the gauge field configuration which goes to pure gauge at infinity and has finite $q$ can induce a transition from vacuum of winding number $n_w(t=-\infty)$ to another vacuum of winding number $n_w(t=\infty)$. At zero temperature, such gauge field configurations are called instantons~\cite{Belavin:1975fg} and they are responsible for the quantum tunneling through the energy barrier between vacua of different winding numbers~\cite{tHooft:1976up,tHooft:1976fv,Jackiw:1976pf,Callan:1976je}.

The high energy barrier ($\sim\L_{\rm QCD}\sim 200$ MeV) between two vacua suppresses the instanton transition rate exponentially, but at high-enough temperature, the transition between different vacua can also be induced by another, classical, thermal excitation called sphaleron~\cite{Manton:1983nd,Klinkhamer:1984di} which, instead of tunneling through the barrier, can take the vacuum over the barrier. In electroweak theory, sphaleron transitions cause baryon number violation and may be important for the cosmological baryogenesis~\cite{Kuzmin:1985mm,Rubakov:1996vz}. In QCD, the existence of the sphaleron configurations at finite temperature enormously increases the transition rate~\cite{McLerran:1990de,Arnold:1996dy,Huet:1996sh,Moore:1997im,Moore:1997sn,Bodeker:1998hm,Bodeker:1999gx,Shuryak:2001cp,Ostrovsky:2002cg,Moore:2010jd}. At very high temperature, the perturbative calculation of the sphaleron transition rate gives ~\cite{Kharzeev:2007jp,Arnold:1996dy,Huet:1996sh,Moore:1997im,Moore:1997sn,Bodeker:1998hm,Bodeker:1999gx,Shuryak:2001cp,Ostrovsky:2002cg,Moore:2010jd}
\begin{eqnarray}
\G_{\rm sph}\sim (\a_s N_c)^5 T^4,
\end{eqnarray}
while the strong coupled holographic approach gives an even larger rate $\G_{\rm sph}=(g^2 N_c)^2/(256\p^2)T^4$~\cite{Son:2002sd}. Thus at high temperature, the sphaleron transition rate can be very large. This provides a machinery of generating P and CP odd bubbles in QGP (note that a transition process from a topologically trivial vacuum to a topologically nontrivial vacuum violates P and CP symmetry as is evident from the integrand of $q$).

Now if we integrate the axial anomaly equation in the QCD sector (for massless quarks)
\begin{eqnarray}
\pt_\m {\cal J}_A^\m=\frac{g^2N_f}{16\p^2} G^a_{\m\n}\tilde{G}^{\m\n}_a,
\end{eqnarray}
where ${\cal J}_A^\m=\sum_f\lan\jb_f\g^\m\g_5\j_f\ran$ ($f$ is over all massless flavors) is the axial current, we see that
\begin{eqnarray}
N_A(t=\infty)-N_A(t=-\infty)=2q,
\end{eqnarray}
where $N_A=\int d^3\bx {\cal J}_A^0(x)$ is the total chirality or axial charge. This demonstrates that a topologically nontrivial gauge field configuration can create or annihilate the total chirality of fermions, and thus if the QGP contains a (sufficiently large) domain in which $q$ is finite we would expect that it finally will contain unequal numbers of RH and LH quarks or anti-quarks even if initially $N_A(t=-\infty)=0$. This is how QGP can become chiral. We note here that the probabilities of generating positive chirality and negative chirality are equal which means that over many colliding events in heavy-ion collisions the averaged chirality should vanish. What remains after event average is the chirality fluctuation rather the chirality itself and any measurement of the chirality-imbalance effects should be on the event-by-event basis, see \sect{sec:exper}.

{\bf (4) What is axial chemical potential? ---} There is a conceptual problem in interpreting $\m_A$ as the axial chemical potential for fermions: as when we talk about the chemical potential we always need to associate to it a conserved quantity but we know that the axial charge $\int d^3\bx {\cal J}_A^0$ of fermions is in general not conserved when the fermions are coupled to gauge fields. Furthermore, if $\m_A$ is really a chemical potential conjugate to a conserved axial charge of fermions, then \eq{cme} would imply a persistent electric current even at global equilibrium. But the appearance of CME current in equilibrium violates the gauge invariance in the QED sector as discussed in Ref.~\cite{Kharzeev:2013ffa}, see also Refs.~\cite{Rebhan:2009vc,Rubakov:2010qi,Basar:2013iaa,Fukushima:2012vr} for relevant discussions. Thus the physical meaning of $\m_A$ is actually quite confusing, and there have been a number of relevant discussions in literature~\cite{Frohlich:2000en,Vilenkin:1980ft,Kharzeev:2009fn,Rebhan:2009vc,Rubakov:2010qi,Fukushima:2012vr,Kharzeev:2013ffa,Basar:2013iaa}. According to these studies, the CME should be considered as a non-equilibrium phenomenon which vanishes when the system is in global equilibrium and, correspondingly, $\m_A$ shouldn't be regarded as a fixed chemical potential of the fermions in equilibrium. It is more appropriate to view $\m_A$ as the rate of the time changing of the $\h$ field (see below), $\m_A=\pt_t\h$; it is the $\h$ parameter that describes the state of the system.

To reveal the relation between $\m_A$ and the $\h$ field let us consider the $\h$-vacua of QCD. The effect of the $\h$-vacua can be encoded into a $\h$-term in QCD Lagrangian,
\begin{eqnarray}
\cl_{\rm QCD}&=&-\frac{1}{4}G^a_{\m\n}G_a^{\m\n}+\sum_f\jb_f [i\g^\m(\pt_\m-ig{\mathcal{A}_\m})]\j_f-\h \frac{g^2 N_f}{32\p^2} G^a_{\m\n}\tilde{G}_a^{\m\n},
\end{eqnarray}
where $\mathcal{A}_\m$ is the gluon field. For massless quarks the $\h$-term does not give observable consequence because it can be rotated away by $U_A(1)$ transformation, but if we promote the $\h$ angle to be spacetime dependent (a pseudoscalar `axion' background), then it will show important physical consequences. Perform the path integral over the quarks:
\begin{eqnarray}
Z[\mathcal{A},\h]=\int[d\j][d\jb]e^{iS(\mathcal{A})}.
\end{eqnarray}
The $\h$-term in the action can be eliminated by performing a local $U_A(1)$ transformation to quark fields and by using the Fujikawa method:
\begin{eqnarray}
\j_f(x)&\ra& e^{i\h(x)\g_5}\j_f(x),\non
\jb_f(x)&\ra& \jb_f(x)e^{i\h(x)\g_5}.
\end{eqnarray}
The resulting Lagrangian reads
\begin{eqnarray}
\cl_{\rm QCD}&=&-\frac{1}{4}G^a_{\m\n}G_a^{\m\n}+\sum_f\jb_f [i\g^\m(\pt_\m-ig{\mathcal{A}_\m}+i(\pt_\m\h)\g_5)]\j_f.
\end{eqnarray}
From this Lagrangian we can identify $\m_A=\pt_t\h$, that is, $\m_A$ is the time derivative of the $\h$ field.

Now let us consider the QED sector. Then the following Lagrangian should be added to $\cl_{\rm QCD}$:
\begin{eqnarray}
\cl_{\rm QED}&=&-\frac{1}{4}F_{\m\n}F^{\m\n}-A_\m J_V^\m,
\end{eqnarray}
where $A_\m$ is the photon field and $J_V^\m=\sum_f q_f\jb_f\g^\m\j_f$ with $q_f$ the charge of quark of flavor $f$. The above $U_A(1)$ transformation will then induce a new term in the the final effective Lagrangian
\begin{eqnarray}
\cl_{\rm QCD\times QED}&=&-\frac{1}{4}F_{\m\n}F^{\m\n}-\frac{1}{4}G^a_{\m\n}G_a^{\m\n}+\sum_f\jb_f [i\g^\m(\pt_\m-ig{\mathcal{A}_\m}+i(\pt_\m\h)\g_5)]\j_f-A_\m J_V^\m+\frac{\k}{4}\h F^{\m\n}\tilde{F}_{\m\n},\non
\end{eqnarray}
where
\begin{eqnarray}
\k=N_c\sum_f\frac{q_f^2}{2\p^2}.
\end{eqnarray}
The EM sector of this Lagrangian is the Maxwell-Chern-Simons or axion Lagrangian. The equations of motion derived from this Lagrangian is called the axion electrodynamics~\cite{Wilczek:1987mv}, which read (see e.g. Ref.~\cite{Ozonder:2010zy})
\begin{eqnarray}
&&{\vec\nabla}\cdot\bE=J_V^0+\k{\vec\nabla}\h\cdot\bB,\\
&&{\vec\nabla}\times\bB-\frac{\pt \bE}{\pt t}=\bJ_V-\k(\m_A\bB+{\vec\nabla}\h\times\bE),\\
&&{\vec\nabla}\cdot\bB=0,\\
&&{\vec\nabla}\times\bE+\frac{\pt \bB}{\pt t}=0.
\end{eqnarray}
From the second equation, we can identify that $\k\m_A\bB$ is the CME current which plays the same role as the applied current $\bJ_V$.

We see that the out-of-equilibrium nature the CME is already indicated by the identification $\m_A=\pt_t\h$: if we allow the $\h$ angle to be the correct physical quantity specifying the physical state of the system, then a constant $\m_A$ would already indicate a linearly growing $\h$ in time.
Then what is the counterpart of CME in equilibrium? This can be guessed by integrating both sides of \eq{cme} over time and we can write the result down as (for constant magnetic field):
\begin{eqnarray}
{\vec P}=\h\frac{e^2}{2\p^2}\bB+{\rm constant},
\end{eqnarray}
where $\vec P$ is the electric polarization vector. This equation describes a magnetic-field induced charge polarization through the axion-EM-field coupling, which can appear in insulating matter even in equilibrium and is called the topological magentoelectric effect~\cite{Qi:2008ew,Essin:2008rq}. In particular, recently found time-reversal-invariant topological insulators fulfill the topological magnetoelectric effect with a special value of $\h$, i.e., $\h=\p$.

\subsection {Chiral separation effect}\label{sec:cse}
{\bf (1) What is the chiral separation effect? ---} The chiral separation effect (CSE) is the dual effect to the CME in the sense that the former is expressed from the latter by interchanging the vector and axial quantities~\cite{Son:2004tq,Metlitski:2005pr,Newman:2005as}. In formula, the CSE is given by
\begin{eqnarray}
\label{cse}
\bJ_A &=&
\s_{AA}
\bB,\\
\label{csecond}\s_{AA}&=&\frac{e^2}{2\p^2}\m_V,
\end{eqnarray}
where the axial current is defined by $J_A^\m=e\lan\jb\g^\m\g_5\j\ran$, and $\m_V$ is the vector chemical potential corresponding to the global $U_V(1)$ symmetry (can be considered as baryon chemical potential)~\footnote{It is not necessary to define $J_A^\m$ and $\m_V$ this way. We can also define the axial current as $J_A^\m=\lan\jb\g^\m\g_5\j\ran$ (like ${\cal J}_A^\m$ in last subsection), i.e., without the $e$ factor, and $\m_V$ as the electric chemical potential. Or we can also define the axial current as $J_A^\m=\lan\jb\g^\m\g_5\j\ran$ and $\m_V$ as the baryon chemical potential, but then $\s_{AA}$ will read $\s_{AA}=e\m_V/2\p^2$}.
Thus the CSE represents the generation of the axial current along with the external magnetic field in the presence of finite vector charge density parametrized by the vector chemical potential $\m_V$. Unlike the CME, the CSE is P-even and T-even but C-odd, the appearance of CSE does not require a parity-violating environment. Similar with the CSE, the T-even nature of CSE signals that it is not dissipative.
\begin{figure}[!htb]
\begin{center}
\includegraphics[width=7cm]{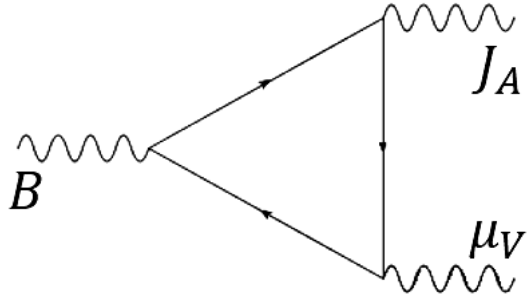}
\caption{The Feynman diagram for CSE.}
\label{diagcse}
\end{center}
\end{figure}

The intuitive picture of the emergence of the CSE at finite vector density is the following. Imagine a fermionic system with more charged massless fermions than anti-fermions. At extremely large magnetic field pointing to up direction, supposing the fermions are positively charged, then all the fermions are confined in the lowest Landau levels and their spins are aligned upward. Then the up-moving fermions carry positive helicity while the down-moving fermions carry negative helicity; because the helicity of a massless fermion is equal to its chirality, we see the fermions contribute an axial current along the direction of the magnetic field. Similarly, it is easy to see that the anti-fermions contribute an axial current opposite to the direction of the magnetic field. Because the system contains more fermions than anti-fermions, a net axial current moving upward is generated.

{\bf (2) How to derive the CSE? ---} Similar with the CME, the generation of CSE is also a consequence of the axial anomaly, see \fig{diagcse}. Comparing \fig{diagcse} to \fig{diagcme} one can again recognize the dual relation between CSE and CME. The expression (\ref{cse}) has been already indicated in the axial anomaly equation,
\begin{eqnarray}
\pt_t J_A^0+{\vec\nabla}\cdot\bJ_A=\frac{e^3}{2\p^2}\bE\cdot\bB.
\end{eqnarray}
Assuming steady state and noticing that we can write $e\bE=\nabla\m_V$, this suggests that $\bJ_A=[e^2/(2\p^2)]\m\bB$ for uniform $\bB$. A more rigorous derivation along this line can be found in, for example, Ref.~\cite{Metlitski:2005pr} in which the surface integral are more appropriately handled. Other derivations and various aspects of the CSE are studied in Refs.~\cite{Son:2004tq,Metlitski:2005pr,Newman:2005as,Bergman:2008qv,Gorbar:2009bm,Basar:2010zd,Gorbar:2010kc,Hong:2010hi,Landsteiner:2011cp,Gorbar:2013upa,Yamamoto:2015fxa,Huang:2015mga}. We here just stress one fact that like the CME, only the lowest Landau level is responsible to the arising of CSE. Intuitively, this is because the higher Landau levels are occupied by equal numbers of spin-up and spin-down fermions (supposing $\bB$ is along the up direction), it is easy to see that they give opposite contributions to $J^z_A$ and thus cancel out (Similar argument applies to anti-fermions as well).

{\bf (3) The chiral Magnetic Wave. ---} The duality between CME and CSE allows the existence of interesting collective modes called chiral magnetic waves (CMWs). This can be seen in the following way. We can express the CME (\ref{cme}) and CSE (\ref{cse}) in terms of RH and LH currents and corresponding chemical potentials:
\begin{eqnarray}
\label{cmecse1}
\bJ_R &=&\frac{e^2}{4\p^2}\m_R\bB,\\
\label{cmecse2}
\bJ_L &=&-\frac{e^2}{4\p^2}\m_L\bB,
\end{eqnarray}
where
\begin{eqnarray}
\bJ_R &=&\frac{1}{2}(\bJ_V+\bJ_A),\non
\bJ_L &=&\frac{1}{2}(\bJ_V-\bJ_A),\non
\m_R &=&\m_V+\m_A,\non
\m_L &=&\m_V-\m_A.
\end{eqnarray}
In terms of RH and LH quantities, the continuation equations read,
\begin{eqnarray}
\label{continuation}
\pt_t J^0_{R/L}+{\vec\nabla}\cdot\bJ_{R/L} &=&0,
\end{eqnarray}
where $J^0_R$ and $J^0_L$ are the densities of RH and LH fermions. Now consider a small departure from equilibrium so that we can write $J^0_{R,L}$ and $\m_{R,L}$ as $\m_{R,L}=\m_0+\d\m_{R,L}$, $J^0_{R,L}=n_0+\d J^0_{R,L}$ where $\d J^0_{R,L}, \d\m_{R,L}$ are all small numbers. Substituting \eqs{cmecse1}{cmecse2} into \eq{continuation} and keeping linear terms in $\d\m$ and $\d J^0$), one can obtain
\begin{eqnarray}
\label{waveequ}
\pt_t\d J^0_R+\frac{e^2}{4\p^2\c_R}\bB\cdot\vec\nabla\d J^0_R &=&0,\\
\label{waveequ2}
\pt_t\d J^0_L-\frac{e^2}{4\p^2\c_L}\bB\cdot\vec\nabla\d J^0_L &=&0,
\end{eqnarray}
where $\c_R=\pt J^0_R/\pt\m_R$ and $\c_L=\pt J^0_L/\pt\m_L$ are susceptibilities for RH and LH chiralities. These are two wave equations expressing collective modes arising from the coupled evolution of CME and CSE. They are just the CMWs~\cite{Kharzeev:2010gd} (see also Ref.~\cite{Stephanov:2014dma} for a derivation of CMW based on the kinetic theory), one propagating along the direction of $\bB$ and another opposite to $\bB$ with wave velocities
\begin{eqnarray}
v_R&=&\frac{e^2}{4\p^2\c_R},\non
v_L&=&\frac{e^2}{4\p^2\c_L}.
\end{eqnarray}
Some comments are in order. (1) If in addition to $\bB$ there is also a finite $\bE$ such that $\bE\cdot\bB\neq 0$, then the continuation equations are anomalous and the right-hand side of Eqs.~(\ref{continuation}) and thus the wave equations (\ref{waveequ}) and (\ref{waveequ2}) should be added terms $\pm [e^3/(2\p^2)]\bE\cdot\bB$. These terms pump chirality into the system and provide sources to the wave equations. The CMW velocities are not affected by the anomalous terms. (2) In general, the microscopic scattering among constitute particles will lead to diffusion of the RH and LH charges. In this case, we should add the diffusion terms, $-D\nabla J^0_{R/L}$, where $D$ are diffusion constants, into Eqs. (\ref{cmecse1}) and (\ref{cmecse2}). The presence of diffusion renders the dispersion relation of CMW to receive a $-i D\bk^2$ term with $\bk$ the wave number of the CMW, and thus dampens the mode of wave number $|\bk|\gtrsim v_{R/L}/D$.

\subsection {Chiral electric separation effect}\label{sec:cese}
{\bf (1) What is the chiral electric separation effect? ---} The CME and CSE are both induced by magnetic field. As we have seen in \sect{sec:fields} that heavy-ion collisions can generate strong electric fields as well. Thus one may wonder whether the electric field can induce an anomalous transport phenomenon.
Such an anomalous transport indeed exists, as first found in Ref.~\cite{Huang:2013iia} in weakly coupled QED plasma and later in weakly coupled QCD plasma~\cite{Jiang:2014ura}, and is called the CESE. It is also confirmed in some strongly-coupled holographic setups~\cite{Pu:2014cwa,Pu:2014fva}.

\begin{figure}[!htb]
\begin{center}
\includegraphics[width=7cm]{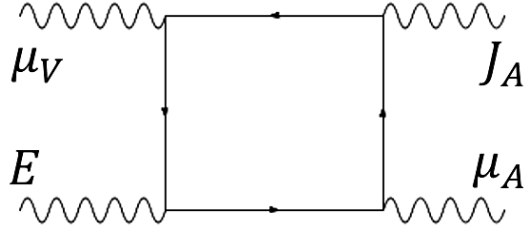}
\caption{The Feynman diagram for CESE.}
\label{diagcese}
\end{center}
\end{figure}

The CESE predicts the generation of axial current in external electric field when the system has both finite vector and axial densities characterized by corresponding chemical potentials $\m_V$ and $\m_A$ respectively,
\begin{eqnarray}
\label{cese}
\bJ_A &=&
\s_{AV}
\bE,\\
\label{cesecond}\s_{AV}&=&\c_e\m_V\m_A,
\end{eqnarray}
where $\c_e$ is not universal and depends on the microscopic scattering processes. Unlike the CME and CSE, the generation of CESE is not directly related to triangle anomaly in the QED sector. Actually, the one-loop diagram for CESE has four legs which is not responsible for the triangle anomaly, see \fig{diagcese}. To understand why $\s_{AV}\propto\m_V\m_A$, let us examine the CPT properties of $\bJ_A$ and $\bE$: $\bJ_A$ is P-even, C-even, and T-odd while $\bE$ is P-odd, C-odd, and T-even. Thus $\s_{AV}$ must be P-odd, C-odd and T-odd. This means that $\s_{AV}$ should be proportional to odd powers of $\m_V$ and $\m_A$; when $\m_V/T$ and $\m_A/T$ are small (as indeed in high-energy heavy-ion collisions) the leading contribution would be proportional to $\m_A\m_V/T^2$. On one hand, the $T$-odd property of $\s_{AV}$ suggests that the CESE may be dissipative, but on the other hand $\s_{AV}$ is not positively definite because the sign of $\m_V$ (and $\m_A$) can be either positive or negative suggesting CESE to be a non-dissipative phenomenon. Detailed study~\cite{Jiang:2014ura} revealed that the dissipation property of CESE is very subtle: CESE can affect the entropy production and in this sense it is dissipative in nature, but the CESE always appears in accompany with the usual electric conduction, it alone does not generate entropy. This is why $\s_{AV}$ can be either positive or negative.

{\bf (2) The CESE conductivity. ---} The CESE conductivity $\s_{AV}$ depends on the microscopic interaction. It can be calculated along the same line as the computation of the usual electric conductivity $\s_{VV}$. In Ref.~\cite{Huang:2013iia} the CESE conductivity for hot QED plasma was calculated by using the Kubo formula within the Hard Thermal Loop perturbation theory~\cite{Jeon:1994if,ValleBasagoiti:2002ir,Aarts:2002tn,Arnold:2000dr,Arnold:2003zc}. At the leading-log order in the coupling constant $e$ the result reads
\begin{eqnarray}
\s_{AV}^{\rm QED}=20.499\frac{\m_V\m_A}{T^2}\frac{T}{e^2\ln(1/e)},
\end{eqnarray}
where one should note that the axial current $\bJ_A$ in the present paper is $e$ times the axial current defined in Ref.~\cite{Huang:2013iia} so that there is factor $e$ difference between $\s_{AV}$ here and that in Ref.~\cite{Huang:2013iia}. The similar perturbative calculation was extended to hot QCD plasma later, and the result for the two-flavor case that $u$ and $d$ quarks are massless reads~\cite{Jiang:2014ura}
\begin{eqnarray}
\s_{AV}^{\rm QCD,2f}=14.5163\, {\rm Tr}_f (Q_e Q_A)\frac{\m_V\m_A}{T^2}\frac{e^2T}{g^4\ln(1/g)}.
\end{eqnarray}
Note that in this case the axial current is defined as $J_A^\m=e\lan\jb\g^\m\g_5 Q_A\j\ran$ and $J_e^\m=e\lan\jb\g^\m Q_e\j\ran$ with $Q_e={\rm diag}(2/3,-1/3)$ and $Q_A$ a generic matrix in flavor space; while the chemical potentials $\m_V$ and $\m_A$ are corresponding to $U_V(1)$ and $U_A(1)$ currents. For the case with $u, d$, and $s$ quarks massless, the result is~\cite{Jiang:2014ura}
\begin{eqnarray}
\s_{AV}^{\rm QCD,3f}=13.0859\,{\rm Tr}_f (Q_e Q_A)\frac{\m_V\m_A}{T^2}\frac{e^2T}{g^4\ln(1/g)}.
\end{eqnarray}

The CESE conductivity was also studied in certain holographic model~\cite{Pu:2014cwa} and it was found that $\s_{AV}^{\rm holography}\propto N_c^2 g_{\rm YM}^2(\m_V\m_A/T^2) T$ which was shown to be surprisingly robust even for large $\m_V/T$ and $\m_A/T$.

{\bf (3) The collective modes due to CESE. ---} With the CME, CSE, and CESE found, we can write down the complete response of the chiral medium to external EM field as (here we focus on QED plasma for simplicity)
\begin{eqnarray}
\bJ_V&=&\s_{VV}\bE+\s_{VA} \bB = \s_{VV}\bE+\s_5 \m_A\bB,\\
\bJ_A&=&\s_{AV}\bE+\s_{AA} \bB =\c_e\m_V\m_A\bE+\s_5\m_V \bB,
\end{eqnarray}
where $\s_5=e^2/(2\p^2)$ and $\c_e$ is the numerical factor in the CESE conductivity.
Because $\s_{VA}\propto\m_A$, $\s_{AA}\propto\m_V$, and $\s_{AV}\propto \m_V\m_A$, we see that in the presence of external EM field, the vector and axial densities mutually induce each other and get entangled in a nontrivial way. As a consequence, the coupled evolution of small fluctuations of the two densities can induce collective modes in a way similar to that the energy density fluctuation in usual hydrodynamics lead to sound wave. The first such example is the CMW which we have discussed in last subsection. The CESE introduces nonlinearity (in the $\m_V\m_A$ term) in to the equations and make the problem more complex~\cite{Huang:2013iia}.

To see how the collective modes arise, we consider static and homogeneous external $\vec E, \vec B$ fields, and study the coupled evolution of the small fluctuations in vector and axial charge densities. First, we can write the total EM field as $\vec E_{\rm tot}=\vec E +\d\vec E$ and $\vec B_{\rm tot}=\vec B +\d\vec B$ with $\d \vec E\propto J_V^0$ and $\d \vec B\propto \vec J_V$ are induced fields due to the vector density and current fluctuations. Second, taking into account that the chemical potentials are small, we can replace the chemical potentials by the corresponding charge densities: $\mu_{V,A}=\alpha_{V,A} J_{V,A}^0$, where the $\alpha_{V,A}$ are the inverse susceptibilities defined as $\alpha_{V,A} = \partial \mu_{V,A} / \partial J_{V,A}^0$. Third, we expand $\s_{VV}$ as $\s_{VV}=\s_0+\s_2(\mu_V^2+\mu_A^2)$ and omit higher order terms in the chemical potentials. Then by linearizing the continuity equations $\partial_t J_{V,A}^0 + \vec\nabla \cdot \vec J_{V,A}=0$, one can obtain
\begin{eqnarray}
\label{vacoupled}
&&\partial_t \d j_V^0 + \sigma_0 \d j_V^0
+ \sigma_5 \alpha_A (\vec B\cdot \vec\nabla) \d j_A^0 \nonumber \\
&&\quad +2\sigma_2\alpha_V^2 n_V(\vec E\cdot\vec\nabla ) \d j_V^0+2\sigma_2\alpha_A^2 n_A(\vec E\cdot\vec\nabla) \d j_A^0=0\, , \nonumber \\
   &&\partial_t \d j_A^0
+\sigma_5 \alpha_V (\vec B\cdot \vec \nabla ) \d j_V^0 +  \chi_e \alpha_V \alpha_A  n_V( \vec E\cdot \vec \nabla )  \d j_A^0  \nonumber \\
&&\quad +  \chi_e \alpha_V \alpha_A  n_A( \vec E \cdot \vec \nabla )  \d j_V^0
 =0     \, , \quad
\end{eqnarray}
where $n_V$ and $n_A$ are small uniform background vector and axial densities and $\d j_{V,A}^0\ll n_{V,A}$ are fluctuations on top of them.

Without loss of generality, we suppose $\vec B$ is along $z$-axis, i.e., $\vec B = B \hat{z}$ while $\vec E =  E \hat{e}$. The dispersion relation obtained from Eq. (\ref{vacoupled}) is
\begin{eqnarray}
\label{dispersion}
\o=-\frac{1}{2}\ls i\s_0-v_+(\hat{e}\cdot\vec k)\rs\pm
\frac{1}{2}\sqrt{\ls i\s_0-v_-(\hat{e}\cdot\vec k)\rs^2+4{\cal A}_\c(\vec k)},
\end{eqnarray}
where $v_\pm=v_v\pm v_a$ with $v_v=2\sigma_2\alpha_V^2 n_V E$ and $v_a = \chi_e \alpha_V  \alpha_A n_V E$, and
\begin{eqnarray}
{\cal A}_\c(\vec k)=\ls\s_5\a_A B(\hat{z}\cdot\vec k)+2\s_2\a_A^2 n_A E(\hat{e}\cdot\vec k)\rs\times\ls\s_5\a_V B(\hat{z}\cdot\vec k)+\c_e\a_V\a_A n_A E(\hat{e}\cdot\vec k)\rs.
\end{eqnarray}

To reveal the physical content of the above dispersion relation, we consider three special cases.\\
{\it (1)$\vec B = B \hat z$ and $\vec E=0$.}\\ Then \eq{dispersion} reduces to
$\omega = \pm \sqrt{(v_\chi k_z)^2 - (\sigma_0/2)^2 } - i (\sigma_0/2)$ with speed $v_\chi = \sigma_5 \sqrt{\alpha_V \alpha_A} B$. When $v_\chi k_z \gg \sigma_0/2$ it represents two propagating modes $\omega\approx  \pm v_\chi k_z -  i (\sigma_0/2)$: these are generalized CMWs which reduce to the CMWs discussed in last subsection when $\s_0=0$ and $\alpha_V=\alpha_A$. When $v_\chi k_z \le e\sigma_0/2$ the two modes are damped.\\
{\it (2) $\vec E =E \hat z$, $\vec B=0$, and $n_V=0$.} \\
In this case, we find two modes from (\ref{dispersion})
\begin{eqnarray}
\omega = \pm \sqrt{(v_e k_z)^2 - (\sigma_0/2)^2 } - i (\sigma_0/2)
\end{eqnarray}
with $v_e = \a_A n_A \sqrt{2\s_2\c_e\alpha_V \alpha_A} E$. Similar to the CMWs, when $v_e k_z \gg \sigma_0/2$ there are two  well-defined modes $\omega\approx  \pm v_e k_z -  i (\sigma_0/2)$ from CESE that propagate along $\vec E$ field and can be called the Chiral Electric Waves (CEWs).  They become damped when $v_e k_z \le \sigma_0/2$. \\
{\it (2) $\vec E =E \hat z$, $\vec B=0$, and $n_A=0$.} \\
In this case, the vector and axial modes become decoupled, and Eq. (\ref{dispersion}) leads to
\begin{eqnarray}
\o_V(\vec k)&=& v_v k_z - i \s_0 \; ,\non
\o_A(\vec k)&=& v_a k_z.
\end{eqnarray}
They represent a ``vector density wave" (VDW) with speed $v_v=2\sigma_2\alpha_V^2 n_V E$ that transports vector charges along $\vec E$ field but will be damped on a time scale $\sim 1/\s_0$ and a propagating ``axial density wave" (ADW) along $\vec E$ with speed $v_a = \chi_e \alpha_V  \alpha_A n_V E$ without damping.

These collective modes may be able to transport chirality and charges in heavy-ion collisions and can lead to specific charge azimuthal distributions from which the CESE may be detected experimentally. We now turn to discuss the experimental status of the searches of the anomalous transports.

\section {Experimental searches of the anomalous transports}\label{sec:exper}
QCD dynamics does not violate parity globally which means that in heavy-ion collisions after average over many events there will not remain a finite axial chemical potential $\m_A$ --- The $\m_A$ just fluctuates from event to event. As a result, the P-odd anomalous transport phenomena can only be measured on the event-by-event basis. We now discuss the event-by-event observables for CME, CMW, and CESE.

\subsection {The charge-dependent azimuthal correlation and chiral magnetic effect}\label{sec:corrcme}
\begin{figure}[!htb]
\begin{center}
\includegraphics[width=7cm]{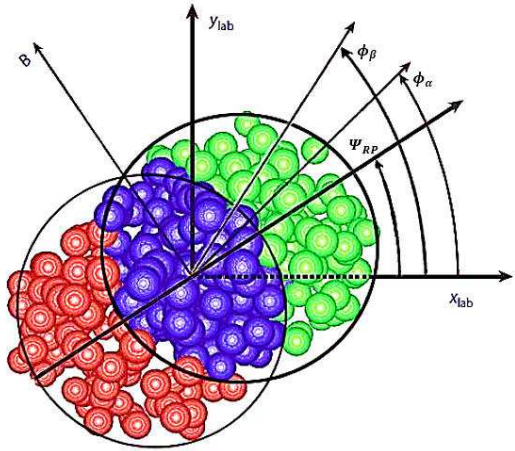}
\caption{Illustration of a typical noncentral collision. Modified from Ref.~\cite{Adamczyk:2014mzf}.}
\label{illcme}
\end{center}
\end{figure}

The CME transports charges along the direction of $\bB$. In noncentral heavy-ion collisions, the $\bB$ direction is, on average, perpendicular to the reaction plane, thus the CME is expected to induce a charge separation with respect to the reaction plane. After the hydrodynamic evolution of the fireball, this out-of-plane charge dipolar configuration in coordinate space can be converted to special distribution of the final charged hadrons in momentum space which can be detected by utilizing specifically designed charge-dependent azimuthal correlations. One such correlation has been proposed by Voloshin~\cite{Voloshin:2004vk} (see also Ref.~\cite{Kharzeev:2004ey}),
\begin{eqnarray}
\label{gamma}
\g_{\a\b}&\equiv&\lan\cos(\f_\a+\f_\b-2\J_\rp)\ran,
\end{eqnarray}
where the indices $\a$ and $\b$ denote the charge of the hadrons, $\f_\a$ and $\f_\b$ are the azimuthal angles of hadrons of charge $\a$ and $\b$ respectively, $\J_\rp$ is the reaction plane angle, and $\lan\cdots\ran$ denotes average over events; See \fig{illcme} for illustration. It is easy to understand that a charge separation with respect to the reaction plane will give a positive $\g_{+-}$ and a negative $\g_{++}$ or $\g_{--}$. In real experiments, the measurements are done with three-particle correlations where the third hadron (of arbitrary charge) is used to reconstruct the reaction plane,
\begin{eqnarray}
\lan\cos(\f_\a+\f_\b-2\f_3)\ran\approx v_2\g_{\a\b},
\end{eqnarray}
where $v_2$ is the elliptic flow.

\begin{figure}[!htb]
\begin{center}
\includegraphics[width=7cm,height=5cm]{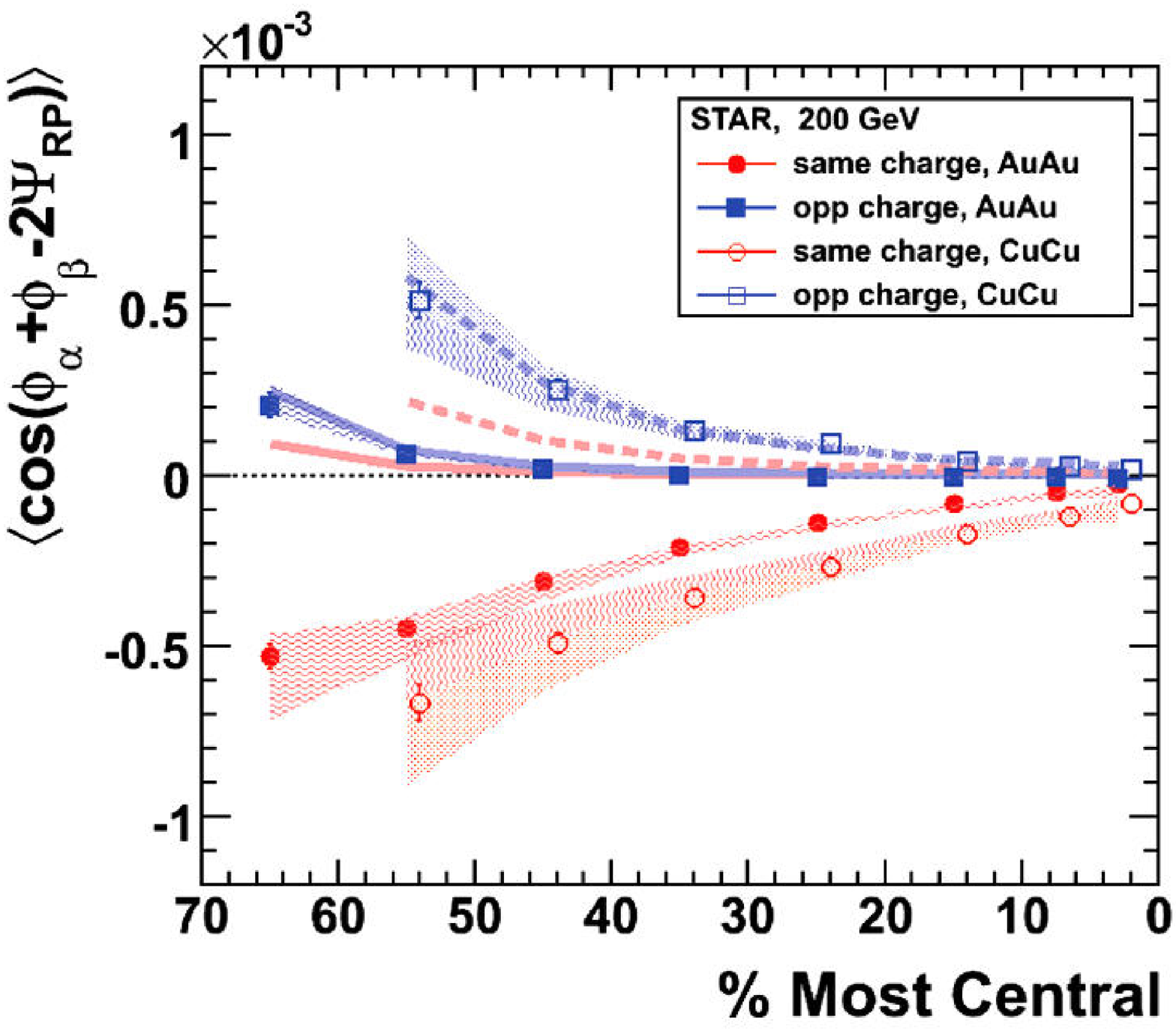}
\caption{The measured charge-dependent correlation $\g_{\a\b}$ by STAR Collaboration. This figure is from Ref.~\cite{Abelev:2009ac}.}
\label{starcme}
\end{center}
\end{figure}

{\bf Experimental results for $\g_{\a\b}$. ---} The first measurement of $\g_{\a\b}$ was officially reported by STAR Collaboration at RHIC for Au + Au and Cu + Cu collisions at $\sqrt{s}=200$ GeV in 2009~\cite{Abelev:2009ac,Abelev:2009ad}, see \fig{starcme} (PHENIX Collaboration at RHIC also reported equivalent measurement~\cite{PHENIX:2010,Ajitanand:2010rc}). From several aspects, these experimental results are consistent with the expectation of the CME: (1) Away from the central collision region, the data shows a clear positive opposite-sign correlation and negative same-sign correlation as the CME predicted. The magnitude of the measurement is in agreement with the theoretical estimate by Kharzeev \etal~\cite{Kharzeev:2004ey,Kharzeev:2007jp}. (2) The observed correlation increases from zero at the most central collisions to the peripheral collisions, consistent with the fact that the magnetic field increases as the centrality grows. (3) The date shows that the opposite-sign correlation is smaller that the same-sign correlation at a fixed centrality, in agreement with the fact that the opposite-sign particles are emitted in opposite directions and thus their correlation suffers from quenching because at least one of the two correlated particles needs to propagate through the opaque medium. Recently, the STAR Collaboration officially reported the measurement of $\g_{\a\b}$ at different beam energies~\cite{Wang:2012qs,Adamczyk:2014mzf}. The main message of the new measurement is that the CME-induced signal is reduced at lower energy; in particular, for energy smaller than $20$ GeV, the same-sign and opposite-sign correlations become to overlap with each other. This is consistent with the fact that at low temperature the QCD topological transition rate is strongly suppressed and thus the CME is attenuated; when the temperature is so low that the chiral symmetry is spontaneously broken, the large mass effect will strongly diminish the axial chemical potential $\m_A$ and thus the CME. Note that the magnitude of the magnetic field is roughly proportional to $\sqrt{s}$, thus one may think that at lower energy the magnetic field is less efficient to transport charges. However, this is not likely the case, as the lifetime (as defined in \sect{sec:early}) of the magnetic field due to the spectators is inversely proportional to $\sqrt{s}$ which compensates the reduction of its magnitude when $\sqrt{s}$ decreases. Recently, the ALICE collaboration at LHC published their measurement of $\g_{\a\b}$ for Pb + Pb collisions at $\sqrt{s}=2.76$ TeV~\cite{Abelev:2012pa}. The result shows that the magnitude of the observed correlation $\g_{\a\b}$ is very close to that at RHIC in spite that the magnetic field at LHC is about much larger than at RHIC.

Let us briefly summarize the current experimental results for the charge-dependent correlation $\g_{\a\b}$ that are consistent with the CME interpretation.\\
${\bullet}$ In high-energy collisions (RHIC @ $\sqrt{s}=200$ GeV and LHC @ $\sqrt{s}=2.76$ TeV):\\
(1) Positive opposite-sign correlation and negative same-sign correlation for nonzero centrality are clearly seen.\\
(2) The observed magnitude of the charge-dependent correlation is at the same order as CME predication.\\
(3) The correlation grows as the centrality increases.\\
(4) The magnitude of the opposite-sign correlation is relatively smaller that that of the same-sign correlation at the same centrality.\\
(5) The magnitude of the correlation at given centrality is nearly independent of the collision energy.\\
${\bullet}$ In STAR beam-energy scan, $\sqrt{s}=7.7-200$ GeV:\\
(6) The difference between same-sign and opposite-sign correlations decreases when collision energy drops and finally nearly disappears for energy less than $\sim20$ GeV.\\

{\bf The background effects. ---} In the STAR publications~\cite{Abelev:2009ac,Abelev:2009ad,Adamczyk:2013hsi,Adamczyk:2014mzf}, they also reported results for another, reaction-plane independent, charge-dependent correlation,
\begin{eqnarray}
\label{delta}
\d_{\a\b}&\equiv&\lan\cos(\f_\a-\f_\b)\ran.
\end{eqnarray}
If the CME is the only source for the charge-dependent correlations, then one would expect that $\d_{++}=\d_{--}>0$ while $\d_{+-}<0$ which is easy to understand from the out-of-plane charge separation picture. However, the data shows the opposite: $\d_{++}\approx \d_{--}<0$ while $\d_{+-}>0$. This tells us that the measured charge-dependent correlations contain contributions from background effects other than CME and these background contributions are likely at the same order of magnitude of (or even larger than) the CME contribution~\cite{Bzdak:2009fc}. A number of studies have been conducted aiming to identify and quantify possible background effects~\cite{Wang:2009kd,Millo:2009ar,Asakawa:2010bu,Muller:2010jd,Pratt:2010gy,Bzdak:2010fd,Liao:2010nv,Schlichting:2010qia,Schlichting:2010na,Pratt:2010zn,Bzdak:2011np,Toneev:2011aa,Ma:2011uma,Hori:2012kp,Toneev:2012zx,Deng:2014uja,Voloshin:2014gja,Yin:2015fca}, but so far there still lack a quantitative and reliable way to estimate and subtract those backgrounds. Detailed discussions on possible background effects can be found in the review papers~\cite{Bzdak:2012ia,Liao:2014ava}. We here will just briefly discuss two possible background effects, namely, the transverse momentum conservation~\cite{Pratt:2010gy,Pratt:2010zn,Bzdak:2010fd} and the local charge conservation~\cite{Pratt:2010gy,Schlichting:2010na,Schlichting:2010qia}; they are considered to comprise the most important backgrounds.

\begin{figure}[!htb]
\begin{center}
\includegraphics[width=5.3cm]{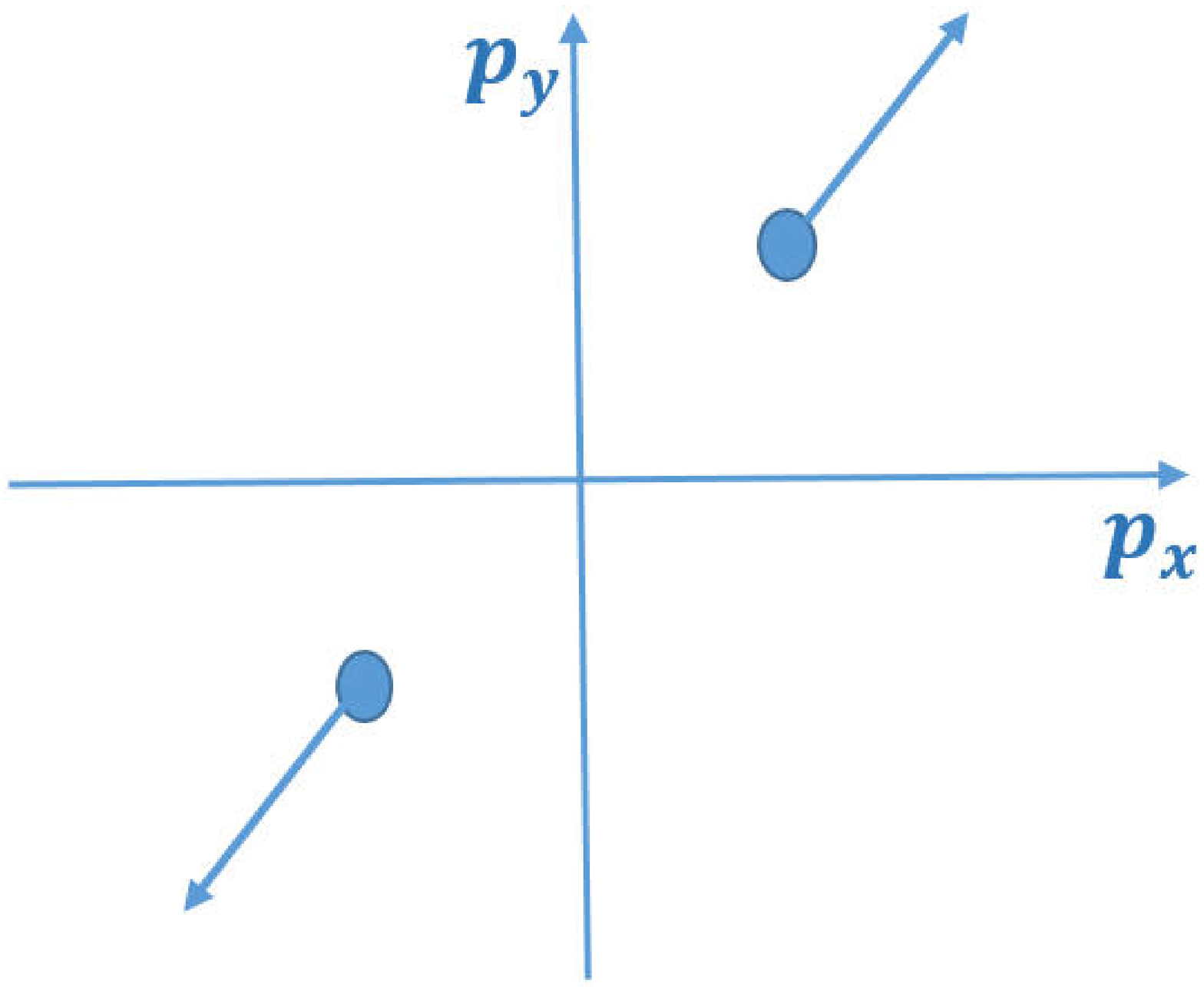}\;\;\;\;
\includegraphics[width=7cm]{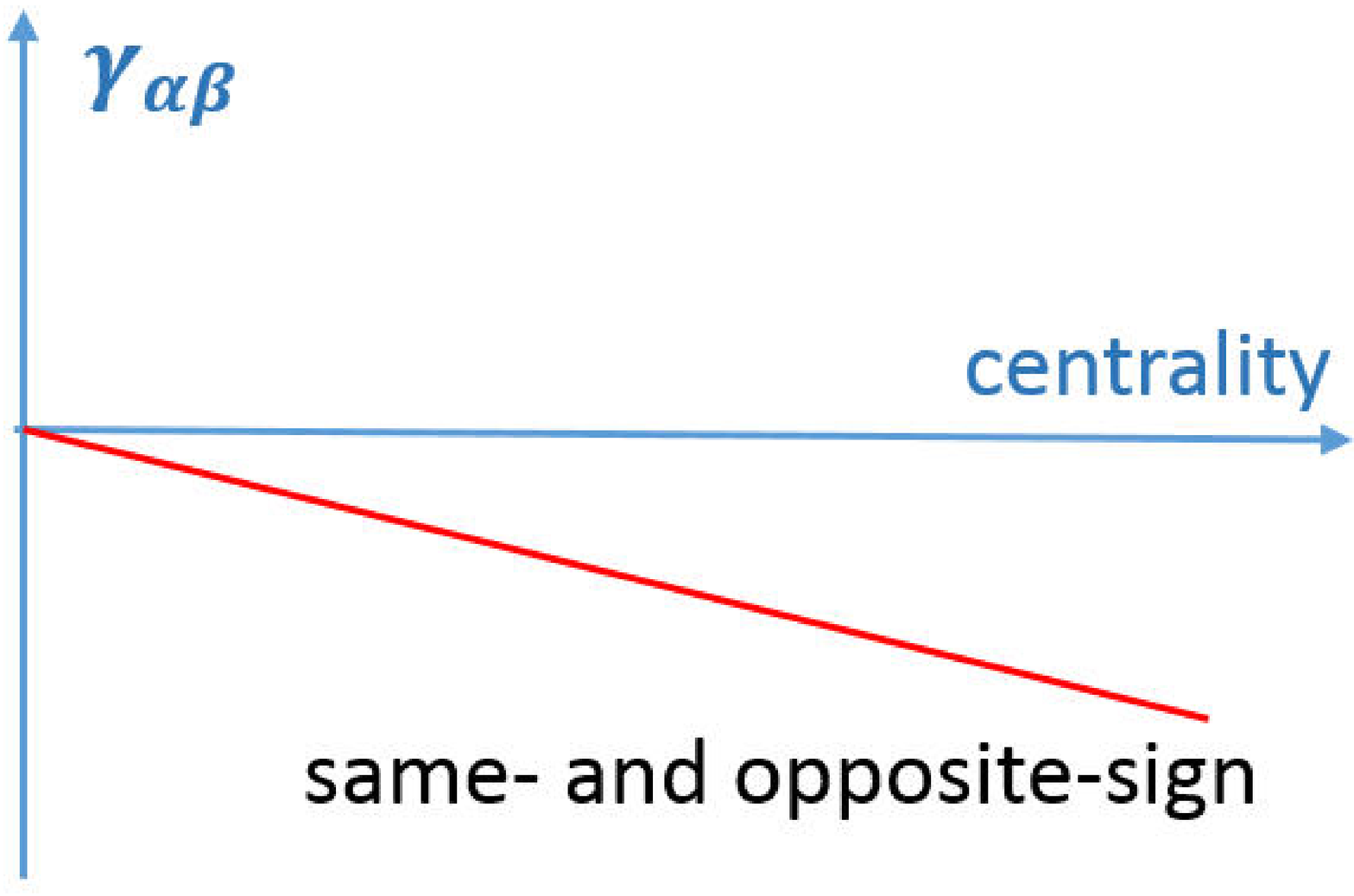}
\caption{Illustration for the transverse momentum conservation and its contribution to $\g_{\a\b}$.}
\label{figtmc}
\end{center}
\end{figure}
The momentum conservation enforces the sum of the transverse momenta of all the particles to be zero at any time during the evolution of the system (here ``transverse momentum" means the momentum component transverse to the beam direction). Thus if we observe a particle of transverse momentum $\bp_\perp$, the transverse momentum conservation (TMC) requires all the other particle must have a total transverse momentum $-\bp_\perp$; see \fig{figtmc} (left-panel) for an illustration. This builds an intrinsic back-to-back two-particle correlation (here ``intrinsic" means that this correlation is irreducible in the sense that it cannot be described by single-particle distribution) and it is evidently charge-blind. Suppose on average each particle has a transverse momentum of the same magnitude $p_\perp$, and the total multiplicity is $N\gg 1$. Then the TMC requires that $\sum_{i=1}^N\cos(\f_i-\J_\rp)=\sum_{i=1}^N\sin(\f_i-\J_\rp)=0$ for each event where $\f_i$ is the azimuthal angle of particle $i$. Then we can obtain that
\begin{eqnarray}
\g^{\rm TMC}_{\rm SS}&=&\g^{\rm TMC}_{\rm OS}=\lan\frac{\sum_{i\neq j}\cos(\f_i+\f_j-2\J_\rp)}{\sum_{i\neq j}}\ran\non
&=&\lan\frac{[\sum_{i}\cos(\f_i-\J_\rp)]^2-[\sum_{i}\sin(\f_i-\J_\rp)]^2-\sum_{i}\cos(2\f_i-2\J_\rp)}{\sum_{i\neq j}}\ran\non
&=&-\frac{1}{N-1}\lan\cos(2\f_i-2\J_\rp)\ran\approx-\frac{v_2}{N},\\
\d^{\rm TMC}_{\rm SS}&=&\d^{\rm TMC}_{\rm OS}=\lan\frac{\sum_{i\neq j}\cos(\f_i-\f_j)}{\sum_{i\neq j}}\ran\non
&=&\lan\frac{[\sum_{i}\cos(\f_i-\J_\rp)]^2+[\sum_{i}\sin(\f_i-\J_\rp)]^2-\sum_{i}}{\sum_{i\neq j}}\ran\non
&=&-\frac{1}{N-1}\approx-\frac{1}{N},
\end{eqnarray}
where SS and OS denote ``same-sign" and ``opposite-sign" and we have used the assumption that $N\gg 1$. Although the TMC itself is reaction-plane independent, the elliptic flow $v_2$ knows the reaction plane; furthermore the elliptic flow also ``measures" the centrality. The TMC always gives negative contributions to both $\g$ and $\d$, and it becomes more important for more peripheral collisions because the multiplicity decreases and $v_2$ increases as the centrality decreases.

\begin{figure}[!htb]
\begin{center}
\includegraphics[width=4.8cm]{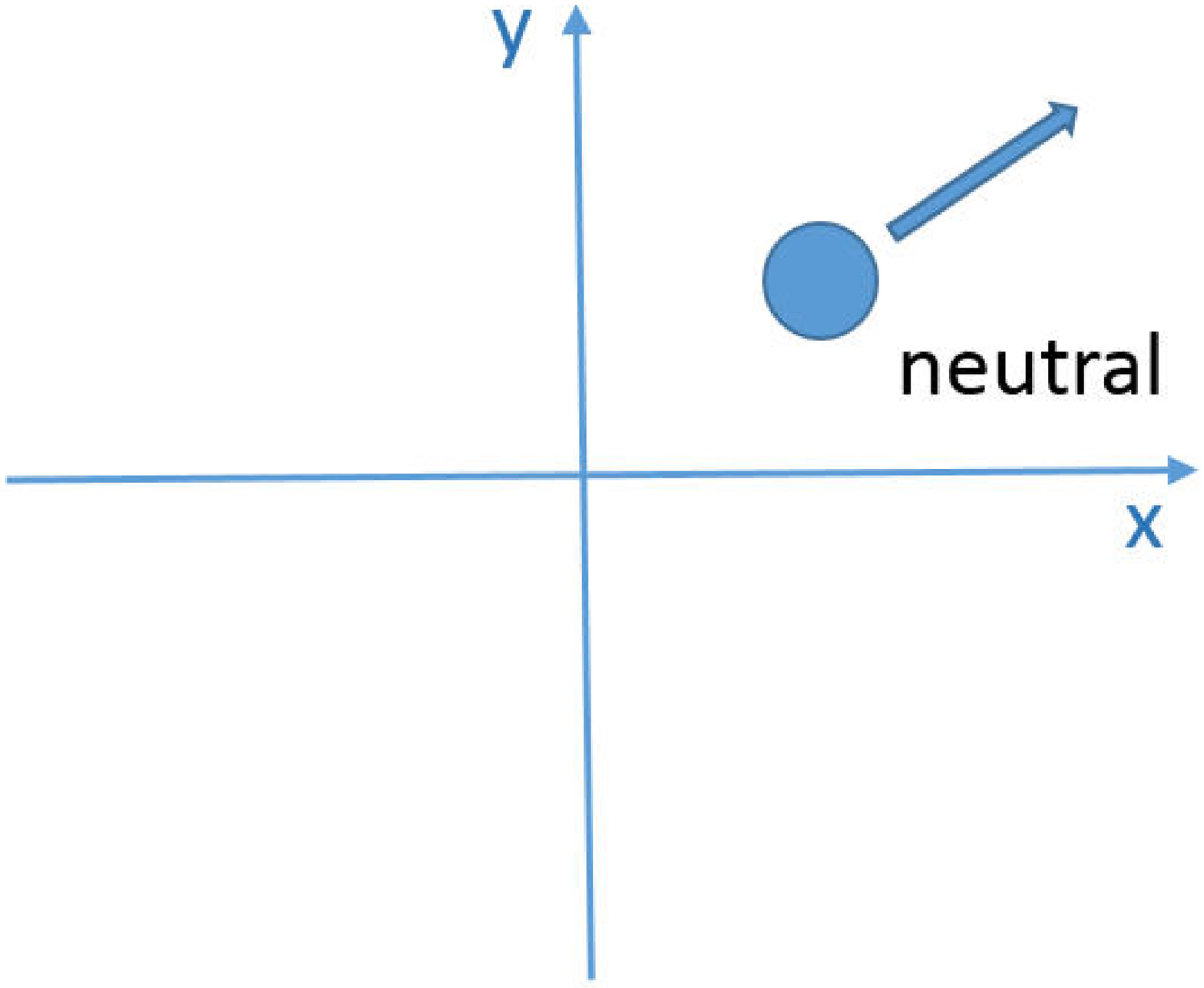}\;\;\;\;
\includegraphics[width=6.5cm]{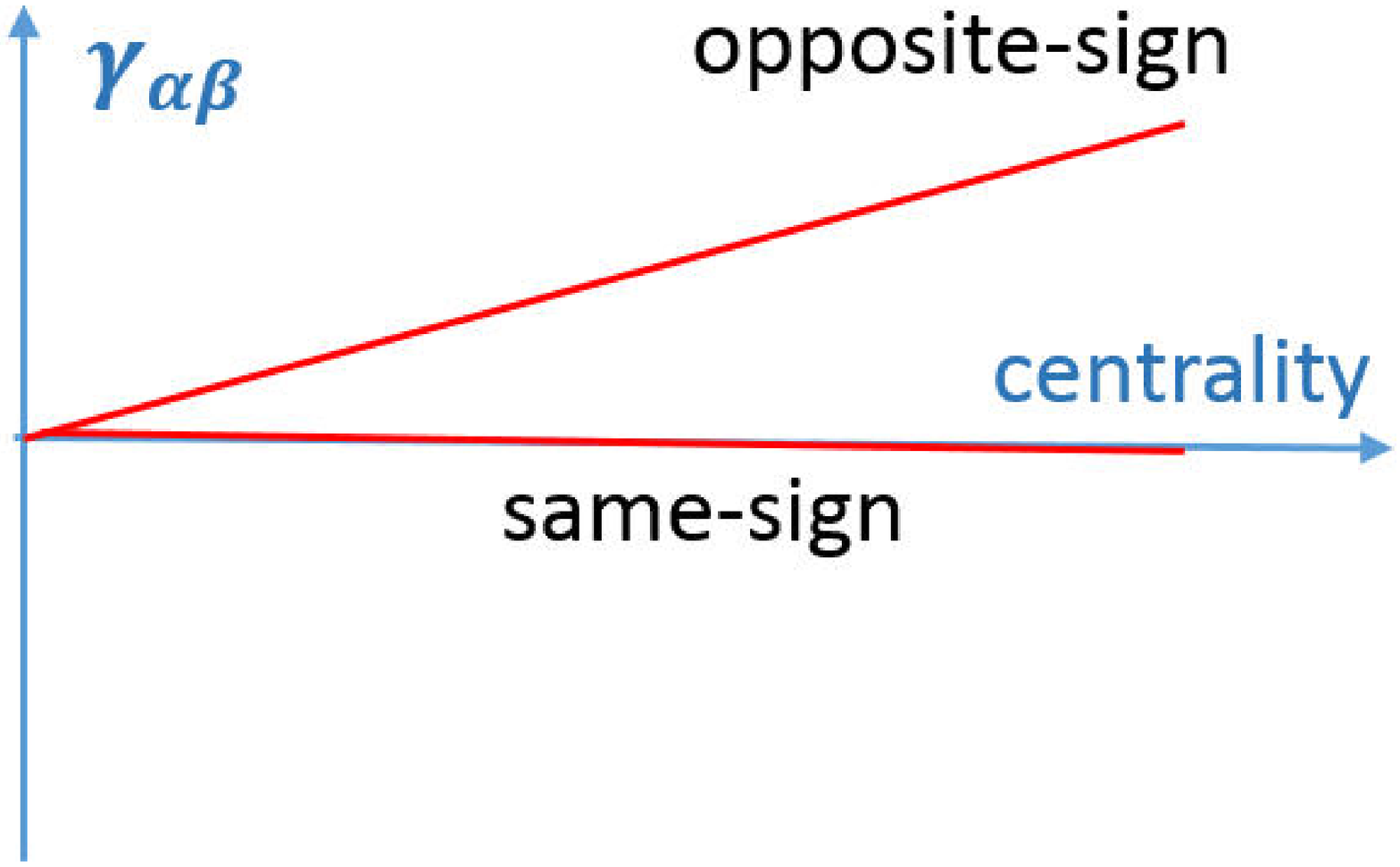}
\caption{Illustration for the local charge conservation and its contribution to $\g_{\a\b}$.}
\label{figlcc}
\end{center}
\end{figure}
The TMC can give a large amount of contribution to $\g_{\a\b}$ and $\d_{\a\b}$ as shown by AMPT simulation~\cite{Ma:2011uma,Shou:2014zsa} (This can also be roughly estimated by choosing $v_2\sim 0.1$ and $N\sim 1000$ for moderate centrality bins at RHIC which gives $\g^{\rm TMC}\sim 10^{-4}$ and $\d^{\rm TMC}\sim 10^{-3}$ being at the same order of magnitude as the STAR results~\cite{Abelev:2009ac,Abelev:2009ad}). In order to evidence the contributions of CME, it is useful to subtract the TMC contribution from $\g$. This can be implemented by taking the difference between opposite-sign and same-sign correlations,
\begin{eqnarray}
\D\g&\equiv&\frac{1}{2}\lb\g_{+-}-\g_{++}-\g_{--}\rb.
\end{eqnarray}
Although this correlation does not receive contribution from TMC, it gets contribution from the local charge conservation (LCC) which is charge dependent. The schematic cartoon for LCC is depicted in \fig{figlcc}. Suppose that in the coordinate space the charge is conserved locally in the sense that a cell of certain size $\sim R$ contains zero total charge. If we pick up one such cell which moves along certain azimuthal direction, then upon freeze-out, the hadrons emitted from this cell will contain equal numbers of positive and negative charges and they collectively move along the same direction as the original cell. In such a way, the LCC induces a near-side opposite-sign correlation but no same-sign correlation. Let us now make a rough estimate of the LCC contribution in an idealized setup. Suppose at the freeze-out the system is consist of $N_c$ non-overlapping cells with each containing $M/2$ positive and negative hadrons, thus $N=M N_c$ with $N$ the total charged hadron multiplicity. Let the cell $c$ move along the azimuthal angle $\f_c$ and assume all hadrons inside $c$ emit along the same direction $\f_c$. We can then express the opposite-sign correlations as (the same-sign correlations due to LCC are zero)
\begin{eqnarray}
\g_{+-}^{\rm LCC}&=&\lan\frac{\sum_{i, q_i=+}\sum_{j, q_j=-}\cos(\f_i+\f_j-2\J_\rp)}{N_+N_-}\ran\non
&\approx&\lan\frac{\sum_c\sum_{i\in c, q_i=+}\sum_{j\in c, q_j=-}\cos(\f_i+\f_j-2\J_\rp)}{N_+N_-}\ran\non
&\approx&\lan\frac{M^2\sum_c\cos(2\f_c-2\J_\rp)}{4N_+N_-}\ran\non
&\approx&\frac{M}{N}v_2,\\
\d_{+-}^{\rm LCC}&=&\lan\frac{\sum_{i, q_i=+}\sum_{j, q_j=-}\cos(\f_i-\f_j)}{N_+N_-}\ran\non
&\approx&\lan\frac{\sum_c\sum_{i\in c, q_i=+}\sum_{j\in c, q_j=-}}{N_+N_-}\ran\non
&\approx&\frac{M}{N},
\end{eqnarray}
where $N_+$ and $N_-$ are the multiplicities of positively and negatively charged hadrons and we have used $N_+\approx N_-=N/2\gg1$ and $N_c\gg1$. More thorough  analysis can be found in Refs.~\cite{Pratt:2010gy,Schlichting:2010na,Schlichting:2010qia} in which the effects of finite size of the cell and the finite emission angle (the hadrons may emit from the cell $c$ within a small relative angle $(\f_c-\D\f/2, \f_c+\D\f/2)$) are more accurately addressed by using the charge balance function. The LCC contribution is inversely proportional to the total multiplicity and proportional to the volume of the neutralized cell. Comparing to the TMC we see that LCC is very strong for both $\g$ and $\d$ as it is proportional to $M$ which is larger than 1.

Clearly, to fully understand the experimental data and to determine the CME-induced signals, one has to build a realistic modeling of the charge-dependent correlations in which all the strong backgrounds and the CME should be taken into account on the same footing. By assuming that the dominant backgrounds are those driven by elliptic flow, like TMC and LCC, it is plausible to make a two-component modeling of the two-particle correlations as~\cite{Bzdak:2012ia,Bloczynski:2013mca},
\begin{eqnarray}
\label{twocomp}
\g_{\a\b}&=&\k v_2 F_{\a\b}-H_{\a\b},\\
\d_{\a\b}&=&F_{\a\b}+H_{\a\b},
\end{eqnarray}
where the $F$ and $H$ represent the strengths of background effects and CME, respectively. The constant $\k$ is an ``uncertainty factor" accounting for some practical issues such as finite kinematic selections in real experiments that may drive $\k$ to deviate from unity~\cite{Adamczyk:2014mzf}. By adopting these relations, one may use the data for $\g_{\a\b}$, $\d_{\a\b}$, and $v_2$ to extract the strength factors $F$ and $H$ as functions of the centrality. Recently, the STAR collaboration reported the measurement of $\D H=H_{\rm SS}-H_{\rm OS}$~\cite{Adamczyk:2014mzf} and found it is finite, increasing with the centrality, and insensitive to collision energy: features consistent with the expectation of CME.

{\bf The U + U and Cu + Au collisions. ---}
It was proposed that the U + U collisions can serve as a potential mean to distinguish the flow-driven effects from the CME in $\g_{\a\b}$~\cite{Voloshin:2010ut}, and such collisions were recently done at RHIC with some preliminary results reported in~\cite{Wang:2012qs}. The main point is that, unlike the gold or lead nuclei, the uranium nucleus is highly deformed from spherical shape and even in the most central U + U collisions there can be sizable elliptic flow while the magnetic field is expected to be very tiny~\cite{Bloczynski:2013mca}. Therefore, if $\g_{\a\b}$ is dominated by CME, the signal would disappear in the most central U + U collisions; while if $\g_{\a\b}$ is dominated by flow-driven backgrounds like TMC and LCC, there will still be sizable signal even for very central collisions. In Ref.~\cite{Bloczynski:2013mca}, by substituting the $F$ and $H$ factors extracted from the Au + Au data, the computed magnetic field, as well as the $v_2$ data for U + U collisions~\cite{Wang:2012qs} to \eq{twocomp}, the authors are able to make prediction for $\D\g$ in U + U collisions. Although there is still quantitative discrepancy between the prediction and the preliminary experimental data~\cite{Wang:2012qs}, the two show consistent magnitudes and trends. This suggests that the two-component scenario may be a good approximation toward a more reliable modeling of the charge dependent correlations. But more experimental releases and more quantitative simulations are definitely needed, see for example Refs.~\cite{Chatterjee:2014sea,Shou:2014cua} for recent attempts.

\begin{figure}[!htb]
\begin{center}
\includegraphics[width=7cm]{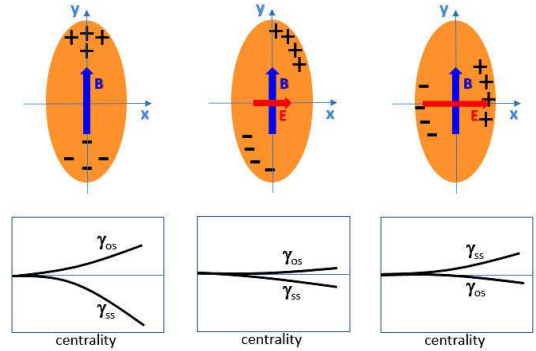}\;\;\;\;
\caption{Illustration of the charge dipole induced by CME and electric field in Cu + Au collisions. (Left): No in-plane electric field. The dipole is driven by CME to be perpendicular to the reaction plane. (Middle): Weak electric field. The in-plane electric field drives the dipole to skew from the magnetic field direction. Both $\g_{\rm OS}$ and $\g_{\rm SS}$ should be reduced. (Right): Strong electric field. The in-plane electric field drives a large in-plane component of the charge dipole, and thus the signs of $\g_{\rm OS}$ and $\g_{\rm SS}$ will be reversed. This figure is from Ref.~\cite{Deng:2014uja}}
\label{figillca}
\end{center}
\end{figure}

In Ref.~\cite{Deng:2014uja}, a new possible way by using the Cu + Au collisions to test the CME was proposed. The main idea is the following (see \fig{figillca} for an demonstration). As we have seen in \sect{sec:other} that Cu + Au collisions can generate a in-plane electric field as well as the out-of-plane magnetic field. This in-plane electric field can drive a in-plane charge dipole in addition to the out-of-plane charge dipole due to CME. As numerically shown in Ref.~\cite{Deng:2014uja}, the appearance of the in-plane dipole makes both $\g_{\rm SS}$ and $\g_{\rm OS}$ descend or get reversed if the electric field is much stronger than the magnetic field. However, this picture is based on the assumption that $\g_{\a\b}$ is dominated by EM-field induced effect; if it is dominated by $v_2$-driven effects rather than EM-field effect, we do not expect that $\g_{\a\b}$ changes too much from Au + Au collisions to Cu + Au collisions. (A plausible guess would be that $\D\g$ as a function of centrality in Cu + Au collisions lie between that in Cu + Cu and Au + Au collisions.) Thus, measuring $\g_{\a\b}$ in Cu + Au collisions may give useful insight into the different mechanisms underlying $\g_{\a\b}$.

\subsection {The charge-dependent elliptic flow and chiral magnetic wave}\label{sec:expcmw}
\begin{figure}[!htb]
\begin{center}
\includegraphics[width=10cm]{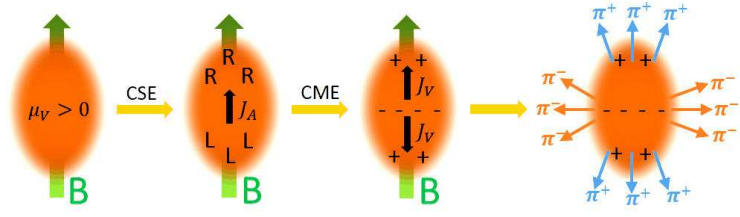}\;\;\;\;
\caption{The CMW-induced electric quadrupole in QGP and charge-dependent elliptic flow of pions.}
\label{figcmwv2}
\end{center}
\end{figure}

As we have seen from \sect{sec:cse}, the CMWs are able to transport both chirality and electic charge. Phenomenologically, they can induce an electric quadrupole in the QGP~\cite{Gorbar:2011ya,Burnier:2011bf}. This can be schematically seen from \fig{figcmwv2}. In the heavy-ion collision, the overlapping region contains a small amount of vector density (inheriting from the the colliding nuclei) so that $\m_V>0$~\footnote{Taking into account the event-by-event fluctuations, there can be events with overall $\m_V<0$ in the overlapping region. The induced electric quadruple will then flip.}. Therefore the $\bB$ field can induce an axial current along it via the CSE (\ref{cse}) and leads to a chirality separation with respect to the reaction plane with $\m_A>0$ in one tip of the overlapping region and $\m_A<0$ in another tip. The CME in turn induces two vector currents which transport positive charges toward the two tips and negative charges toward the equator of the overlapping region and eventually form an electric quadrupole. Qualitatively, the strength of the electric quadrupole is proportional to $\m_V (e\bB)^2$.

Then the next question is: what is the observable signal for this electric quadrupole? It was proposed in Ref.~\cite{Burnier:2011bf} that such a signal could be the splitting between the elliptic flows of $\p^+$ and $\p^-$; see Refs.~\cite{Burnier:2012ae,Taghavi:2013ena,Yee:2013cya,Hirono:2014oda} for more advanced numerical implementation of this proposal. As illustrated in \fig{figcmwv2}, owing to the strong in-plane pressure gradient, the electric quadrupole in coordinate space can be converted into a specific pattern in the momentum-space distribution of the final-state hadrons after freeze-out: more negatively charged hadrons move along the in-plane direction while more positively charged hadrons move along the out-of-plane direction. As a result, the elliptic flow of $\p^-$ will be larger than that of $\p^+$. This is true for events with $\m_V>0$; For a event with $\m_V<0$, the result will be that $v_2(\p^+)>v_2(\p^-)$.
To quantify this $v_2$ splitting, one can consider the single-particle azimuthal distribution $f_q(\f), q=\pm$, per rapidity. The appearance of the electric quadrupole modifies the net charge distribution to the following form:
\begin{eqnarray}
\label{chargedis}
\r(\f)=f_+(\f)-f_-(\f)=\frac{N_+-N_-}{2\p}\ls 1+2 v_2\cos(2\f-2\J_\rp)- r \cos(2\f-2\J_\rp)\rs,
\end{eqnarray}
where $r \propto(e\bB)^2$ is a parameter quantifying the strength of the electric quadrupole at given $N_+-N_-\propto\m_V$,
\begin{eqnarray}
r=-\frac{2}{N_+-N_-}\int_0^{2\p} d\f \cos(2\f-2\J_\rp)\ls\r(\f)-\r(\f)|_{\bB=0}\rs,
\end{eqnarray}
$v_2$ is the elliptic flow when the electric quadrupole is absent, and $N_+, N_-$ are the multiplicities of positively and negatively charged hadrons. Suppose the quadrupole is small so that the total distribution is unchanged,
\begin{eqnarray}
\label{totaldis}
f_+(\f)+f_-(\f)=\frac{N_++N_-}{2\p}\ls 1+2 v_2\cos(2\f-2\J_\rp)\rs.
\end{eqnarray}
From \eq{chargedis} and \eq{totaldis}, we can obtain the single particle distribution of charge $q$ as
\begin{eqnarray}
\label{spdis}
f_q(\f)\approx\frac{N_q}{2\p}\ls 1+2 v_2\cos(2\f-2\J_\rp)-q A_{\rm ch} r (2\f-2\J_\rp)\rs,
\end{eqnarray}
where $A_{\rm ch}=(N_+-N_-)/(N_++N_-)$ is the net charge asymmetry parameter and we have kept only terms up to linear order in $A_{\rm ch}$. From \eq{spdis}, we immediately see that the elliptic flow becomes charge dependent:
\begin{eqnarray}
\label{cmwpred}
v_2(\p^{\pm})=v_2\mp \frac{rA_{\rm ch}}{2}.
\end{eqnarray}
We give two comments here. (1) Obviously, the presence of the electric quadrupole in the QGP should affect not only the elliptic flow of charged pions but also the elliptic flows of other charged hadrons. The reason why we finally take into account only pions, according to the argument in Ref.~\cite{Burnier:2011bf}, is because other abundant charged hadrons, the proton, antiproton, kaons, etc, suffer from large difference in the absorption cross section in hadronic matter at finite baryon density which may strongly mask or even reverse the the effect of the electric quadrupole. (2) We should also notice that in \eq{chargedis} we assumed that the quadrupole moment is perpendicular to the reaction plane. This is not exactly true because, as we saw in \sect{sec:azimu}, $\bB$ orientation fluctuates event-by-event; by taking into account this azimuthal fluctuation of the $\bB$, the parameter $r$ should be proportional to $\lan(e\bB)^2\cos[2(\j_\bB-\J_\rp)]\ran$ rather than $\lan(e\bB)^2\ran$~\cite{Bloczynski:2012en}.

\begin{figure}[!htb]
\begin{center}
\includegraphics[height=4cm]{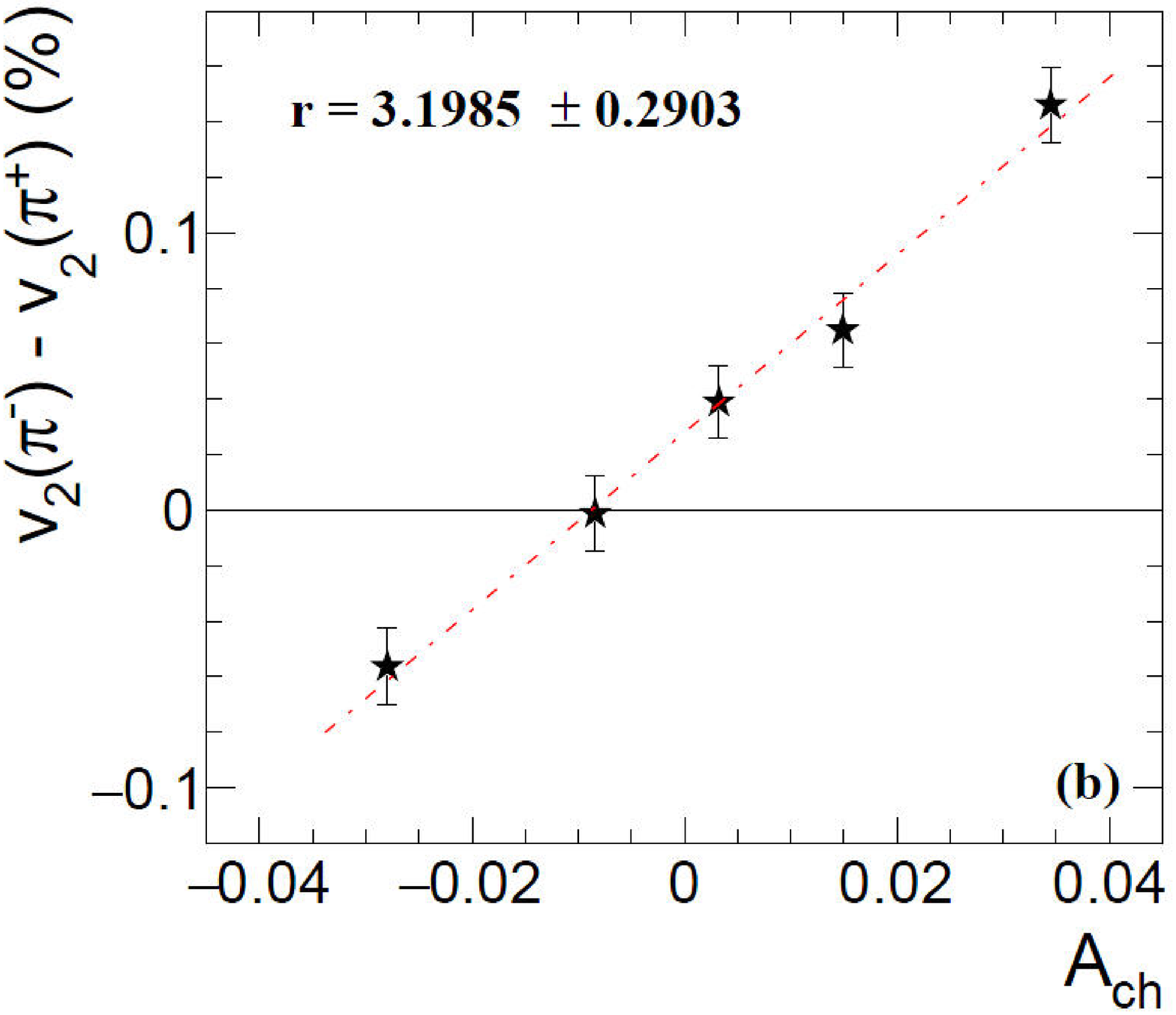}\;\;\;\;\;\;\;\;
\includegraphics[height=4cm]{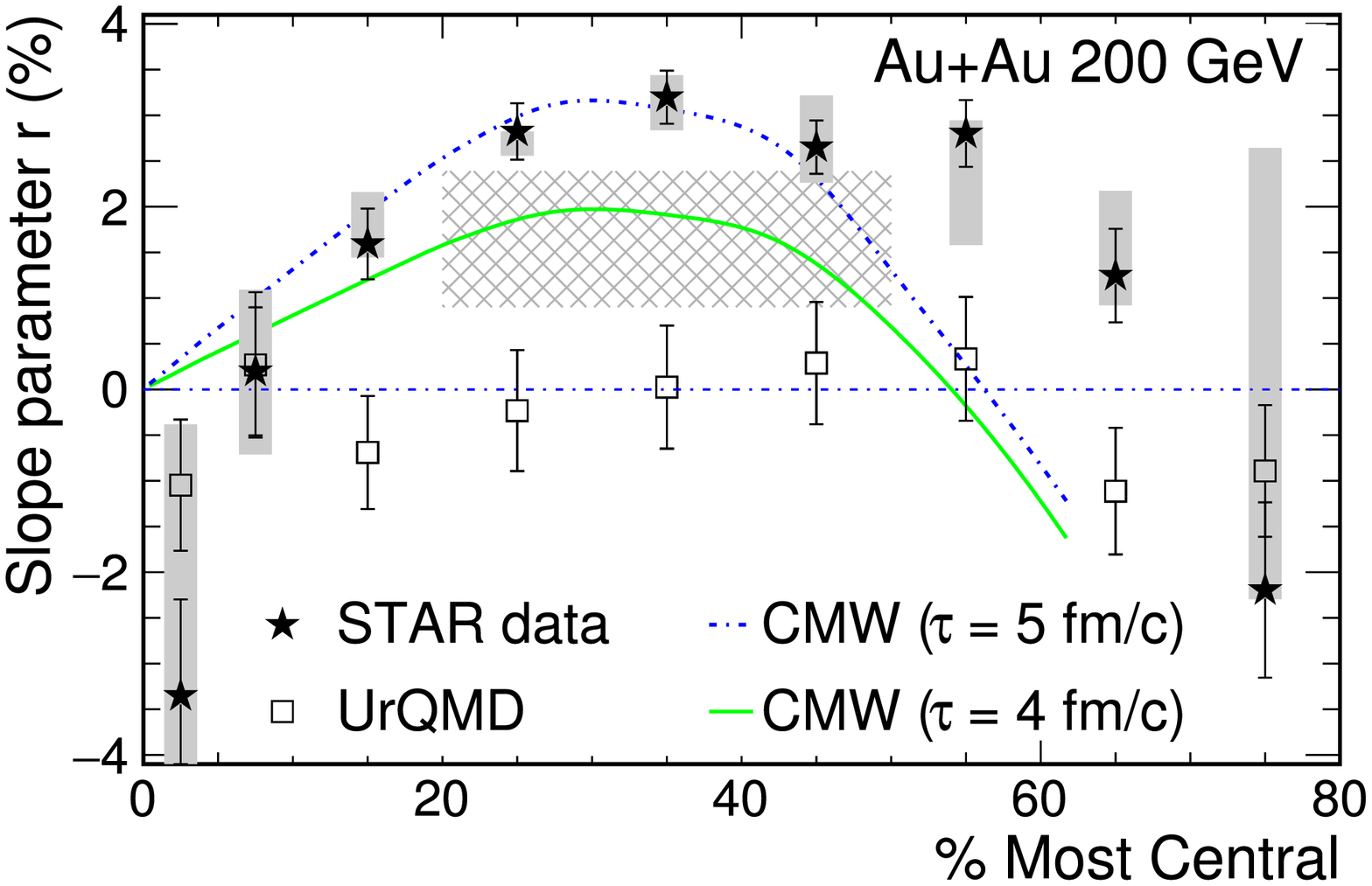}
\caption{(Left) The $v_2$ splitting between $\p^-$ and $\p^+$ as a function of charge asymmetry. (Right) The measurement of the slop parameter $r$ by STAR Collaboration as well as the UrQMD and CMW predictions. Figures are from Ref.~\cite{Adamczyk:2015eqo}.}
\label{figcmwexp}
\end{center}
\end{figure}

Recently, the $v_2$ splitting of charged pions was measured by STAR Collaboration~\cite{Wang:2012qs,Ke:2012qb,Adamczyk:2015eqo} and the data is in agreement with the CMW predictions in several aspects (See \fig{figcmwexp} for the experimental result for the centrality dependence of the slope parameter $r$ at $\sqrt{s}=200$ GeV):\\ (1) The measured $v_2$ difference $v_2(\pi^-)-v_2(\pi^+)$ is linear in $A_{\rm ch}$ with a positive slope as predicted by CMW; see \fig{figcmwexp}. \\(2) The magnitude of the measured slope $r$ versus centrality can be fitted by the CMW estimation by adopting the magnetic field simulated from the event-by-event calculations with field lifetime $\t_B\sim 4-5$ fm.\\ (3) The data for the slope parameter $r$ as a function of centrality shows a maximum in midcentrality or midperipheral collisions which is also consistent with the CMW calculation~\cite{Adamczyk:2015eqo}. \\(4) The slope parameter $r$ displays no obvious trend of the beam energy dependence for $10-60\%$ centrality at $\sqrt{s}=20-200$ GeV. This is consistent with the fact that the ability of the magnetic-field in transporting charges in a tiny rapidity interval is approximately energy independent because the strength of the magnetic field due to spectators is proportional to $\sqrt{s}$ while the its lifetime is proportional to $1/\sqrt{s}$. \\(5) At $\sqrt{s}=11.5$ and $5.5$ GeV the data shows that the slop parameters are consistent with zero --- a feature in agreement with CMW picture because the arising of CMW requires chiral symmetry restoration which is not expected to be the case at very low $\sqrt{s}$.

Although the experimental data shows very encouraging features in line with the CMW expectations, one cannot exclude the possibility that other effects may contribute to the observed $v_2$ splitting pattern as well. Indeed, the data (see the left panel of \fig{figcmwexp}) shows clearly a positive $v_2(\pi^-)-v_2(\pi^+)$ at $A_{\rm ch}=0$ which is not understood within (but also not in contradiction with) the CMW picture. A variety of possible background effects were proposed, for instances, the out-of-plane electric-field can induce out-of-plane electric currents which can produce a charge quadrupole similar with that produced by CMW (Note that this electric-field induced effect exists even at $A_{\rm ch}=0$)~\cite{Deng:2012pc,Stephanov:2013tga}, the electromagnetic chirality $\bE\cdot\bB$ distributes in the transverse plane like a chiral dipole which when combined with triangle anomaly and CME can also induce a charge quadrupole~\cite{Stephanov:2013tga}, the mean-field potential from the hadronic as well as partonic phases acts differently on particles and anti-particles and produces a $v_2$ difference between $\p^-$ and $\p^+$~\cite{Xu:2012gf,Song:2012cd}, the effect of baryon stopping, when assuming that the $v_2$
of transported quarks is larger than that of produced ones, can induce a larger $v_2$ of $\p^-$ than $\p^+$~\cite{Dunlop:2011cf}, the effect due to the rapidity cut in the data analysis when combined with the local charge conservation in rapidity can lead to $v_2$ splitting of charged pions~\cite{Bzdak:2013yla}, the isospin effect combined with viscous hydrodynamics may also show a $v_2$ splitting~\cite{Hatta:2015hca}. It should be noted that none of the above backgrounds can successfully explain all the features of the experimental data. It is quite plausible that different effects are entangled to produce the measured pattern. To distinguish different effects and separate the CMW contribution, we certainly need more detailed measurements such as the measurement of the $v_2$ splitting versus transverse momentum and rapidity, the measurement for other hadrons like kaons, and the measurement of splitting in other harmonic flows like $v_3$.

\subsection {The possible observables for chiral electric separation effect}\label{sec:expcese}
Because the CESE is driven by electric field, in order to observe the CESE in heavy-ion collisions one must have a collision system where the electric field is generated. As seen in \sect{sec:impac}, owing to the event-by-event fluctuations of the proton locations inside the nuclei, heavy-ion collisions do generate strong electric field even in the center of the collision region. However, as we discussed in \sect{sec:azimu}, so-generated electric field is not correlated to the matter geometry (the reaction plane or participant plane), and thus invisible in any reaction plane dependent observable. To have an electric field that correlated to the matter geometry, we need to consider asymmetric collisions. The Cu + Au collision is such a collision where a persistent electric field in generated with the direction from Au nucleus to Cu nucleus. Thus let us consider the possibility of detection of the CESE in the Cu + Au collisions~\cite{Huang:2013iia}.

\begin{figure}[!htb]
\begin{center}
\includegraphics[width=7cm]{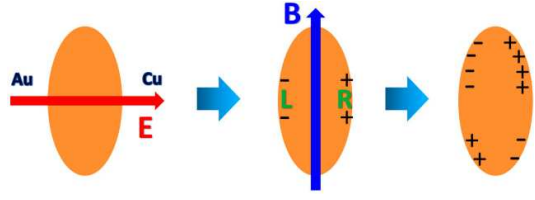}
\caption{Illustration of the charge configuration due to the in-plane charge separation effect due to usual electric conduction, CESE, and CME in Cu + Au collisions. The figure is from Ref.~\cite{Huang:2013iia}.}
\label{figillucese}
\end{center}
\end{figure}
In Cu + Au collisions, supposing that the created hot QGP contains both vector and axial charge densities from fluctuations and topological transitions, then the $\vec E$ field will lead to both an in-plane charge separation via usual Ohm's conduction and an in-plane chirality separation via CESE. The resulting in-plane axial dipole will then further separate charges via CME along the magnetic field which is in the out-of-plane direction, and cause an approximate quadrupole at certain angle $\psi_c$ in between in- and out-of-plane, see \fig{figillucese} for demonstration. After the collective motion and freeze out of the system, this spacial distribution of charges can cause corresponding single-particle distribution of charged hadrons in momentum space, which can be expressed as
\begin{eqnarray}
\label{singleparticle}
f_q(\f)&=&\frac{N_q}{2\p}\{ 1+2 v_1 \cos (\f-\J_1) + 2 v_2\cos[2(\f-\J_2)]+ 2q d_E\cos(\f-\j_\bE)\non&& + 2\c q d_B\cos(\f-\j_\bB) +2\c q h_B\cos[2(\f-\j_c)] + {\rm higher\;\; harmonics}\},
\end{eqnarray}
where $q=\pm$ is the charge of the particle, $N_q$ is the number of hadrons with charge $q$ per rapidity, $v_{1,2}$ are the usual directed and elliptic flows at $\bB=\bE={\vec0}$, $\J_{1,2}$ are corresponding harmonic angles, $\j_c$ is the CESE angle (as shown in \fig{figillucese}, positive/negative charges fly out at the angle $\pm\j_c$ and $\pm\j_c+\p$) which lies in $(0,\p/2)$, $d_E$ and $d_B$ characterize the strength
of the dipole induced by $\bE$ and $\bB$, respectively, and $\c= \pm$ is a random variable accounting
for the fact that in a given event there may be a sphaleron
or an anti-sphaleron transition resulting in $\m_A>0$ or $\m_A<0$ via triangle anomaly. The event average
of $\c$ should be zero. Equation (\ref{singleparticle}) represents four kinds of
¡°flows¡±: the normal directed flow and elliptic flow (the second and third terms), and the normal electric conduction along $\j_{\bE}$ (the fourth term), the
pure CME along $\j_\bB$ (the fifth term), and the combined CESE and CME (the sixth term, which we will simply call CESE term).

From the single-particle distribution (\ref{singleparticle}), we first observe that the directed flow of charge $q$ acquires a contribution from the electric field:
\begin{eqnarray}
v_1(q)&=&v_1 +q \lan  d_E \cos(\j_\bE-\J_1)\ran
\end{eqnarray}
where $\lan\cdots\ran$ denotes event average. Note that the CME term does not contribute to $v_1(q)$ because the event average of $\c$ is zero.
Thus in Cu + Au
collisions, the measured directed flow is likely to be charge
dependent. This may provide us a possibility to extract the
electric conductivity of the produced matter by measuring the
difference between the directed flows of positive and negative
hadrons because $d_E$ is proportional to the electric conductivity.
This idea has been exploited in Refs.~\cite{Hirono:2012rt,Voronyuk:2014rna}.

In Ref.~\cite{Ma:2015isa}, two possible observables for CESE were propoesed. The first one is the charge-dependent two-particle correlation
\begin{eqnarray}
\label{zet}
\zeta_{\alpha\beta}=\langle\cos[2(\phi_{\alpha}+\phi_{\beta}-2\Psi_{\rm RP})]\rangle,
\end{eqnarray}
where $\f_\a$ stands for the azimuthal angle of hadrons of charge $\a$ with $\a=\pm$ and the average is taken over events.
We will set $\J_{\rm RP}=0$ for simplicity. It is easy to find from \eq{singleparticle} and the configuration shown in \fig{figillucese} that the same-sign correlation $\zeta_{\rm SS}\sim \langle\cos(4\psi_{c})\rangle$ should be in general different from the opposite-sign correlation $\zeta_{\rm OS}\sim 1$ while the purely dipolar charge distribution (like the in-plane charge separation effect (in-plane CSE) due to usual conduction by $\bE$ or CME, the fourth and fifth terms in \eq{singleparticle}) does not contribute to $\zeta_{\a\b}$. Thus the correlation (\ref{zet}) carries information of the quadrupolar distribution due to CESE. However, in real experiments, $\z_{\a\b}$ may receive contributions from background effects related to the elliptic flow $v_2$ because, if we turn off all the anomalous effects, $\z_{\a\b}\sim v_2^2$. Another background may be the transverse momentum conservation~\cite{Pratt:2010gy,Pratt:2010zn,Bzdak:2010fd} which causes a contribution $\z_{\a\b}\propto v_4/N$ with $v_4$ the fourth harmonic flow and $N$ the multiplicity. Thus, just like what has been done for CME observeble $\g_{\a\b}$, we can take the difference $\Delta\zeta=\zeta_{\rm OS}-\zeta_{SS}$ to subtract these charge-blind backgrounds. But there may remain another backgrounds, for example, due to the local charge conservation~\cite{Pratt:2010gy,Schlichting:2010na,Schlichting:2010qia} (which leads to $\z_{\rm OS}\propto v_4/N$ and $\z_{\rm SS}\sim 0$ and thus $\D\z\propto v_4/N$) or due to the chiral magnetic wave induced quadrupole. These backgrounds again need to be more carefully explored and subtracted in any real experimental measurements. The AMPT simulation for $\D\z$ is shown in Fig.~5 of Ref.~\cite{Ma:2015isa} for Cu + Au collisions from which one can clearly see that when the CESE is present the increasing trend of $\D\zeta$ from central to peripheral collision are present.

The second observable for CESE is the charge-dependent event-plane angles. Define $\D\Psi=\langle|\Psi_{2}^{+}-\Psi_{2}^{-}|\rangle$, where $\Psi_{2}^{+}$ and $\Psi_{2}^{-}$ are the second harmonic angles of the event planes reconstructed by final positively charged hadrons and negatively charged hadrons. Then from \fig{figillucese} one can find that $\D\J\sim 2\langle\psi_{c}\rangle$. The AMPT simulation for $\D\J$ is given in Fig.~6 of Ref.~\cite{Ma:2015isa}: As expected, once the CESE effect happens, a visible $\D\Psi$ is present which shows a linear dependence on the centrality while if for the initial settings that do not include CESE, the nonzero $\D\J$ is absent for the whole centrality window, making $\D\J$ a good observable for CESE.

\section {Discussions}\label{sec:discu}
In this review, we have discussed the recent progresses in understanding the properties of the strong electromagnetic (EM) fields generated in heavy-ion collisions and the anomalous transport phenomena induced by these EM fields. The discussions have shown that heavy-ion collisions, the only mean on earth to create the quark-gluon plasma, can also generate extremely strong EM fields. These EM fields provide us the possible monitor of the non-trivial topological structure of the quantum chromodynamics (QCD) via the anomalous transport phenomena. We have focused on several types of the anomalous transports, namely, the chiral magnetic effect (CME), chiral separation effect (CSE), chiral electric separation effect (CESE), and the corresponding collective modes, e.g., the chiral magnetic wave (CMW), emerging from the mutual induction of the vector and axial vector charges and currents through them.

We have also discussed the ideas and the current status of the experimental searches of these anomalous transports. Hitherto, the experimental measurements show very encouraging features that are qualitatively consistent with the expectations of the CME and CMW, although many background effects may give significant contributions to the observables as well, making the interpretation of the experimental measurements very challenging. More efforts are certainly needed before the final conclusions can be made. On the theoretical side it is very desirable to make realistic studies of the CME and other anomalous transports in heavy-ion collisions based on kinetic simulations or (magneto)hydrodynamics by consistently encoding the time evolution of the EM fields, the event-by-event generation of the axial chemical potential and its spatial distribution and time evolution, etc.

We keep all the discussions as pedagogical as possible so that this review can be accessed by readers who are not familiar with the topics that we have covered. We give literature at proper places where deeper and more technical treatments can be found. This review will be potentially useful for graduate students who want to begin their study in the area of EM-field induced anomalous transport phenomena in heavy-ion collisions.

There are a number of interesting topics that are untouched so far but are closely related to the subjects covered by this article. Here we briefly address three of them which have attracted remarkable attention recently:

(1) The noncentral heavy-ion collisions can generate fluid vorticity in the interaction region~\cite{Liang:2004ph,Becattini:2007sr,Huang:2011ru,Csernai:2013bqa,Becattini:2015ska,Jiang:2016woz,Deng:2016gyh}. The fluid vorticity can also induce anomalous transports, generally called the chiral vortical effect (CVE)~\cite{Erdmenger:2008rm,Banerjee:2008th,Son:2009tf} (earlier suggestion is given in Ref.~\cite{Kharzeev:2007tn}), which are the vortical analogues of the chiral magnetic and separation effects. The corresponding collective modes owing to the CVE are the chiral vortical waves (CVW)~\cite{Jiang:2015cva} just like that the chiral magnetic waves are owing to CME and CSE. Phenomenologically, the CVE can induce baryon number separation with respect to the reaction plane~\cite{Kharzeev:2010gr} which can be tested by using the baryon number dependent correlation $\w_{\a\b}=\lan\cos(\f_\a+\f_b-2\J_{\rm RP})\ran$ with here $\f_\a$ the azimuthal angle of hadrons of baryon number $\a$ (=baryon or antibaryon) and the CVW can induce a $v_2$ splitting between $\L$ and $\bar{\L}$ in a way very similar with how the CMW induce $v_2$ splitting between $\p^-$ and $\p^+$. Recently, the STAR Collaboration reported the first measurement of $\w_{\a\b}$ which shows evident splitting between the same-sign (i.e., baryon-baryon or antibaryon-antibaryon) correlation and the opposite-sign (i.e., baryon-antibaryon) correlation for large centrality\cite{Zhao:2014aja}. This is consistent with the expectation of CVE. However, just like the case of the CME detections, the experimental signal is possibly masked by contributions from other effects like the transverse momentum conservation and the local baryon number conservation, and more efforts are certainly needed to understand the data.\\

(2) Quite recently, the triangle anomaly was reported to be realized in special condensed matter systems called Weyl semimetals~\cite{Xu:weyl2015,Lu:weyl2015,Lv:weyl2015,Shekhar:weyl2015,Xu2:weyl2015}. These materials are (3+1)-dimensional analogues of (2+1)-dimensional graphene. Their band structure permits the existence of the band-touching points at which the dispersion relation of the quasiparticles is approximately linear and thus effectively ``relativistic". Near a band-touching point the equation of motion of the quasiparticle can be described by Weyl equation and hence the name Weyl semimetal; the corresponding band-touching point is called Weyl point. The Nielsen-Ninomiya theorem~\cite{Nielsen:1980rz,Nielsen:1981xu} requires that the Weyl points in the Brillouin zone of a Weyl semimetal must come in pairs with opposite chirality. If two Weyl points with opposite chiralities happen to overlap with each other, the effective equation of motion becomes Dirac type and such materials are called Dirac semimetals~\cite{Neupane:dirac2014,Jeon:dirac2014,Liu:dirac2014,Liu2:dirac2014,Zhang:dirac2015}. The Dirac semimetal can be effectively transmuted into a Weyl semimetal by applying parallel electric and magnetic fields. Actually, in such a way, the authors of Ref.~\cite{Li:cme2014} reported the first observation of the chiral magnetic effect in Dirac semimetal $ZrTe_5$. The discovery of the Weyl and Dirac semimetals opens a new era for the studying of the topological matter and we can expect that other anomalous transport phenomena, like CMW and CESE, could also be realized in Weyl and Dirac semimetals in future.

(3) In addition to the Weyl and Dirac semimetals, the spin-orbit coupled atomic gases may provide another possibility to realize the anomalous transports in condensed matter systems. The synthetic spin-orbit coupling (SOC) was first successfully generated for Bose gas of $^{87}Rb$ atoms in which two hyperfine states (referred to as (pseudo)spin-up and -down states) of the atoms are coupled to their orbital motion in a manner described by Hamiltonian $\D H\sim\l \s_x k_y$ --- the so-called Rashba-Dresselhaus SOC Hamiltonian--- where $\s_x$ is a Pauli matrix, $k_y$ is $y$ component of the momentum, and $\l$ is the SOC strength~\cite{Lin:soc2011}. The same type of SOC was also realized in Fermi gases of $^6Li$~\cite{Cheuk:soc2012} and $^{40}K$~\cite{Wang:soc2012}, and the pure Rashba SOC was generated in Fermi gas of $^{40}K$~\cite{Huang:soc2015}. Very promisingly, the Weyl SOC which can be described by the Hamiltonian $H=\bk^2/(2m)-\l{\bm\s}\cdot\bk$ with $m$ the mass of atoms could also be realized~\cite{Anderson:soc2012,Anderson:soc2013,Li:soc2012}. In Ref.~\cite{Huang:2015mga}, based on the chiral kinetic theory, the author showed that the triangle anomaly can be realized in Weyl spin-orbit coupled atomic gases in a harmonic trap under rotation. In this case, the trapping potential provide an effective electric field and the rotation provides an effective magnetic field. As a consequence of this triangle anomaly, the currents of opposite helicities are generated in parallel or anti-parallel to the rotation axis which mimic the chiral magnetic effect and chiral separation effect. The potential experimental signature is that the chiral magnetic wave in this case can induce a mass quadrupole in the atomic cloud which may be detected by, e.g., light absorption images.

\emph{Acknowledgments}---
The author thanks J. Bloczynski, W.-T. Deng, Y. Jiang, J. Liao, G .L. Ma, X. Zhang for collaborations and A. Bzdak, G. Cao, M. Chernodub, K. Fukushima, D. Hou, D. Kharzeev, H. Warringa, M. Huang, S. Pu, A. Sedrakian, V. Skokov, S. Ozonder, A. Tang, G. Wang, Q. Wang, X.-N. Wang, N. Xu, Z. Xu, H.-U. Yee, Y. Yin, P. Zhuang for many useful discussions and communications.
The author is currently supported by Shanghai Natural Science Foundation (Grant No. 14ZR1403000), 1000 Young Talents Program of China, the Key Laboratory of Quark and Lepton Physics (MOE) of CCNU (Grant No. QLPL20122), and Scientific Research Foundation of State Education Ministry for Returned Scholars.

\end{document}